\documentclass[reprint,superscriptaddress,nofootinbib,amsmath,amssymb]{revtex4-1}
\usepackage{xcolor}
\usepackage{graphicx} % Include figure files

\newcommand{\nupos}{\nu_{\text{{p}}}}
\newcommand{\nuneg}{\nu_{\text{{e}}}}

\newcommand{\nuPLU}{\nu_{\text{{+}}}}

\newcommand{\Nhel}{N_{\text{{h}}}}
\newcommand{\Npos}{N_{\text{{p}}}}
\newcommand{\Nneg}{N_{\text{{e}}}}

\newcommand{\ppos}{p_{\text{{p}}}}
\newcommand{\pneg}{p_{\text{{e}}}}

\newcommand{\Pf}{P_{\text{{f}}}}

\newcommand{\Apos}{A_{\text{{p}}}}
\newcommand{\Aneg}{A_{\text{{e}}}}
\newcommand{\Bpos}{B_{\text{{p}}}}
\newcommand{\Bneg}{B_{\text{{e}}}}

\newcommand{\upos}{\upsilon_{\text{{p}}}}
\newcommand{\uneg}{\upsilon_{\text{{e}}}}
\newcommand{\uHe}{\upsilon_{\text{{h}}}}

\newcommand{\chipos}{\chi_{\text{{p}}}}
\newcommand{\chineg}{\chi_{\text{{e}}}}

\newcommand{\vkpos}{\varkappa_{\text{{p}}}}
\newcommand{\vkneg}{\varkappa_{\text{{e}}}}

\newcommand{\alphaS}{\alpha_{\text{\tiny{S}}}}

\newcommand{\veps}{\varepsilon}

\newcommand{\drm}{{d}}

\newcommand{\Ddr}{\frac{\drm\phantom{s}}{\drm r}}
\newcommand{\Ddrho}{\frac{\drm\phantom{s}}{\drm \rho}}

\newcommand{\ups}{\upsilon}

\newcommand{\refeq}[1]{(\ref{#1})}

\newcommand{\mEL}{m_{\text{e}}}
\newcommand{\mPR}{m_{\text{p}}}
\newcommand{\mHe}{m_{\text{h}}}

\newcommand{\vect}[1] {\boldsymbol{{ #1}} }

\newcommand{\Rset}{\mathbb{R}}

\newcommand{\sV}{{\vect{s}}}            % position 3-vector
\renewcommand{\leq}{\leqslant}
\renewcommand{\geq}{\geqslant}

\newcommand{\cN}{{\cal N}}
\newcommand{\kB}{k_{\mbox{\tiny{B}}}}

\begin{document}
	
% \title{On electrically non-neutral ground states of stars}

\title{Electrically non-neutral ground states of stars}
	
\author{Parker Hund$^1$ and Michael K.-H. Kiessling}

\affiliation{Department of Mathematics, Rutgers University,
                110 Frelinghuysen Rd., Piscataway, NJ 08854, USA}\vspace{-10pt}
\email{\copyright (2020) The authors.\vspace{-40pt}}

\begin{abstract}
\noindent 
 To approximately compute the non-relativistic ground state of an electrically non-neutral star,
an exactly solvable model was recently introduced, and partly solved, in \cite{KNY}.
 The model generalizes the well-known Lane--Emden equation of a polytropic gas ball of index $n=1$ 
to a two-fluid setting.
 Here its complete solution is presented in terms of simple elementary functions;
it is also generalized to a more-than-two-fluid setting where it remains exactly solvable.
 It is shown that, given the number of nuclei, a maximal negatively and a maximal positively charged solution exists, 
plus a continuous family of solutions which interpolates between these extremes.
 Numerical comparisons show that this exactly solvable model captures the qualitative behavior 
of the more physical model it is supposed to approximate.
 Furthermore, it correctly answers the question: how non-neutral can the star be?
 The answer is independent of the speed of light $c$ and the Planck quantum $h$.
 It supports Penrose's weak cosmic censorship hypothesis, in the sense that the 
bounds on the excess charge are compatible with the bound on the charge of a 
Reissner--Weyl--Nordstr\"om black hole.
\end{abstract}

%\begin{description}
	%\item[PACS numbers] % ??? 
%\end{description}
$\phantom{xi}$\hfill 1 
\maketitle

 $\phantom{nix}$\vspace{-1.5truecm}

%%%%%%%%%%%%%%%%%%%%%%%%%%%%%%%%%%%%%%%%%%%%%%%%%%%%%%%%%%%%%%
%%%%%%%%%%%%%%%%%%%%%%%%%%%%%%%%%%%%%%%%%%%%%%%%%%%%%%%%%%%%%%
%%%%%%%%%%%%%%%%%%%%%%%%%%%%%%%%%%%%%%%%%%%%%%%%%%%%%%%%%%%%%%%%%%%%
\section{Introduction}  \vspace{-10pt}
%%%%%%%%%%%%%%%%%%%%%%%%%%%%%%%%%%%%%%%%%%%%%%%%%%%%%%%%%%%%%%
%%%%%%%%%%%%%%%%%%%%%%%%%%%%%%%%%%%%%%%%%%%%%%%%%%%%%%%%%%%%%%
%%%%%%%%%%%%%%%%%%%%%%%%%%%%%%%%%%%%%%%%%%%%%%%%%%%%%%%%%%%%%%%%%%%% 
% a common plausible assumption is that zero-angular momentum stars are spherically symmetric}. This simplifies 
%
% is the radial distance from the star's center.

{In the theory of stellar structure \cite{Emden}, \cite{Chandra}, \cite{KippenhahnWeigert}, 
it is common practice to work with an effective two-fluid approximation, and to reduce it further to an
effective single-density  model by invoking a local neutrality approximation.
 The two-fluid approximation simply means that charge and mass densities are computed with the density function 
$\nuneg(\sV)$ of the electrons and a single effective density function $\nuPLU(\sV)$ for all the species of
positively charged nuclei, with $\nuneg(\sV)$ and $\nuPLU(\sV)$ normalized to the number of electrons, $\Nneg$, and 
nucleons, $N_n$, in the star; here, $\sV$ is the space point at which the densities are considered.
 The charge density is then given by $\sigma(\sV) = -e\nuneg(\sV) + e\overline{z}\nuPLU(\sV)$, where $\overline{z}$ denotes
the average number of elementary charges per nucleon in the star.
 Neglecting small differences between the proton mass and the average mass per nucleon,
the mass density is essentially given by $\mu(\sV) = \mEL\nuneg(\sV) + \mPR \nuPLU(\sV)$
(the mass $\mEL$ of the electrons is usually neglected here, due to its smallness relative to the nuclear masses).
 Imposing on this the local-neutrality approximation $\sigma(\sV)= 0\;\forall\,\sV$, based on the argument that the electrical 
coupling between electron and proton is about $10^{39}$ times stronger than their gravitational coupling so that any 
local electric imbalance must be negligible for the purpose of computing the overall mass density function $\mu(\sV)$,
one eliminates $\nuPLU(\sV)$ in favor of $\nuneg(\sV)$, say. 
 Further arguments are still needed to obtain a closed equation for $\nuneg(\sV)$.}

 For example, we recall Chandrasekhar's theory of non-rotating white dwarfs \cite{Chandra}. 
 Based on Fowler's insight \cite{Fowler} that white dwarfs are stabilized against their gravitational inward pull by the
Pauli principle for electrons \cite{LiebSeiringer}, modeled in form of the gradients of the pressure $\pneg(\sV)$ of a 
degenerate ideal gas of electrons,
Chandrasekhar computed first the non-relativistic and subsequently also the special-relativistic relationship between $\nuneg(\sV)$
and $\pneg(\sV)$ for a \emph{completely} degenerate Fermi gas, expected to characterize the fate of the electrons in a 
white dwarf after it has radiated away all its available energy and settled into a black dwarf. 
(Chandrasekhar himself often spoke of a models for black dwarfs rather than white dwarfs.)
 In the locally neutral approximation to the two-fluid approximation, a star whose angular momentum vanishes
the hydrostatic force balance reads $-\mu(\sV)\nabla\phi_N(\sV)- \nabla\pneg(\sV)=0$, with $\phi_N(\sV)$ 
the gravitational Newton potential; it satisfies Poisson's equation $\Delta\phi_N(\sV) = 4\pi G\mu(\sV)$. 
 With $\pneg(\sV)$ given in terms of $\nuneg(\sV)$, and with $\mu(\sV)$ also given in terms of $\nuneg(\sV)$ through the
locally neutral approximation to the two-fluid model, it is clear that a closed equation for $\nuneg(\sV)$ ensues.
 In the non-relativistic setting it has a radially symmetric solution for any finite mass $M>0$, but in the 
special-relativistic setting this is only true if $M<M_{\mbox{\tiny{Ch}}} = C \sqrt{3\pi} (\Nneg/N_n)^2(\hbar c/G)^{3/2}/\mPR^2$,
the critical mass discovered by Chandrasekhar, with $C\approx 1.01$; cf. \cite{LiebYau}.
 When $M\to M_{\mbox{\tiny{Ch}}}$, the mass density function degenerates into a Dirac $\delta$ function concentrated at a point.

 In general relativity it is not possible to continuously shrink a mass density to a $\delta$ function 
with finite mass $M>0$; before that could happen, a dynamical instability sets in and causes the collapse of the 
so-modelled star, forming a black hole in the process. 
 The critical mass $M_{\mbox{\tiny{GR}}}< M_{\mbox{\tiny{Ch}}}$, but not by much; it corresponds to a smallest non-zero
radius which such a star could have. 

 The critical mass is affected also by the finite size of the nucleons, and their strong and weak interactions.
 In particular, inverse $\beta$ decay causes electrons to be absorbed by nuclei (converting their protons 
into neutrons), when the central density exceeds a critical value.

 All these investigations have not challenged the local neutrality approximation, which seems to have been perceived 
as so compelling (see \cite{HS}; cf.\cite{Balian}, chpt.16, sect.9.5) that research into the large scale electric structure of 
stars has for a long time lived a life in the shadows, by comparison; one of the few early papers on the subject is \cite{OBb}.
 However, since the local neutrality approximation trivially implies global neutrality, $Q=0$,
it throws the baby out with the bath in regard to the following important problems, which 
in the past dozen or so years have rekindled the interest in non-neutral stars.
 Thus there is a desire to better understand the formation of charged black holes, such as the
Reissner--Weyl--Nordstr\"om black holes or, if angular momentum is included, the Kerr--Newman black holes,
through the collapse of charged stars \cite{RRb}, \cite{NRM}.
 In particular, since there is a limit as to how non-neutral a charged black hole can be, given its ADM
mass $M$ and angular momentum $aM$, it is important 
to find out whether this limit is also obeyed in models of non-neutral stars, or whether stars could be more
non-neutral, in which case Penrose's weak cosmic censorship hypothesis \cite{Penrose} could be in jeopardy.
 Also certain collapse-unrelated questions, concerning hypothetical quark and strange stars,
seem to require an understanding of their large-scale electrostatic fields for answers \cite{KNY}.
 Their charge densities then become interesting subjects of research, see \cite{RRb}, \cite{KNY}, \cite{RRa}.
 
 In this paper we pick up on the recent publication \cite{KNY} where a two-fluid model of a star was studied without 
invoking the local neutrality approximation. 
 One fluid component represents the electrons, the other fluid component represents the mix of positively 
charged nuclei in the star, in the spirit of the reasoning recalled above.
 For both fluids a pressure-density relation in form of a polytropic power law is assumed, with special 
emphasis on the power $\gamma=5/3$ which is predicted by non-relativistic quantum mechanics for a completely 
degenerate gas of spin-1/2 fermions, in the spirit of Chandrasekhar's pioneering calculations \cite{Chandra}.
 We note already that while the assumption of a polytropic $\gamma=5/3$ pressure-density relation is
of course compelling for the electrons, and inherited by the mix-of-nuclei fluid if the local neutrality
approximation is made in the two-fluid model, without the local neutrality approximation the assumption of a 
polytropic $\gamma=5/3$ pressure-density relation for the nuclei fluid would need to be 
justified separately; we will come back to this point below.
 But first, we summarize what is done about this two-fluid model in \cite{KNY}, and which new results our paper
contributes.
 
  The non-linear system of equations of this non-neutral stellar Thomas--Fermi model are more complicated than those of its 
neutral approximation, and so the authors of \cite{KNY} have looked for other approximations which facilitate the study of 
the non-neutral models.
 Since the ratio of gravitational to electrical coupling constants are fantastically tiny numbers, e.g.
 ${G\mEL^2}/{e^2}\approx 2.40\cdot 10^{-43}$ for two electrons, ${G\mPR\mEL}/{e^2}\approx 4.41\cdot 10^{-40}$
for an electron proton system, and ${G\mPR^2}/{e^2}\approx 8.09\cdot 10^{-37}$ for two protons, one approach has been to
utilize these small numbers for a first-order perturbative expansion in their powers to access, and assess, the non-neutral 
neighborhood of a neutral stellar model; cf.  \cite{KNY}.
 Unfortunately, as noted in \cite{KNY}, such a perturbation is singular: one effectively perturbs around the zero-gravity case, 
but without gravity there are no nontrivial stellar equilibrium configurations. 
 Thus, instead of simplifying matters, such a singular expansion introduces artificial new difficulties, which are absent from
the non-linear Thomas--Fermi equations.
 
 To gain further qualitative insights into their nonlinear two-fluid model, the authors of \cite{KNY} also 
introduced a linear proxy model; see section IV.B of \cite{KNY}. 
 {The approximation consists in changing the polytropic power $\gamma=5/3$
in the pressure-density relation of a non-relativistic, completely degenerate ideal Fermi gas to $\gamma=6/3$. 
 This alteration is small, but it has the advantage that the structure equations become 
exactly solvable in terms of elementary functions, as already noted in \cite{KNY}.
  The solutions of this ``6/3 model'' can then be compared with numerically computed solutions of the ``5/3 model'' 
which it is meant to approximate.}

 The discussion in \cite{KNY} is, however, confined to the structure in the interior (the ``bulk'') of the star where
both two fluid density functions are nonzero. 
 A complete understanding of the model  requires also a discussion
of what we call the ``atmospheric region'' where one or the other density function, but not both, vanishes.
 It is the interplay between the bulk region and the atmospheric region which selects the admissible solutions.

 In this paper we present the complete set of finite mass solutions of this exactly solvable 6/3
two-fluid model, covering bulk and atmosphere.
 In an appendix we explain that the model can be generalized to an arbitrary number of fluid components 
while remaining exactly solvable in terms of simple elementary functions.
 
 We compare our exact solutions for the 6/3 model with numerical evaluations of the 5/3 model
which it is meant to approximate, and also with numerical solutions of a Chandrasekhar-type special-relativistic model. 
 We also supply rigorous arguments to back up the numerical results.
 An important by-catch of our results is that the bounds on how non-neutral a star can be
support Penrose's weak cosmic censorship hypothesis \cite{Penrose}.

 Thus, the exactly solvable 6/3 model introduced in \cite{KNY} can serve several purposes.
 First of all, it provides insight into the qualitative structure of the possible solutions to the physically
more realistic 5/3 model.
 Second, it can serve as a test case for numerical methods designed to solve these physically 
more realistic sets of non-linear equations that cannot be solved in closed form. 
 Third, there seem to be some ``universal'' electrical facts that are largely independent of 
the details of the stellar ground state model, and these are readily reproduced by the exactly solvable 6/3 model.
 Finally, a power $2$ pressure-density relation is {of course} well known in the general theory 
of stellar structure \cite{Emden}, \cite{Chandra}, \cite{KippenhahnWeigert} and yields a polytrope of index $n=1$.
 Aside from being discussed in the astrophysical literature, polytropes also appear in pedagogical papers, e.g.
they are found in \cite{Simon}, \cite{SilbarReddy}, \cite{Garfinkle}, \cite{Pesnell}, and \cite{GjerlovPesnell}.
 In this vein we believe that this exactly solvable model could also be incorporated in a course on stellar structure.

 We have reached the point where we need to come back to the question of how realistic the two-fluid 5/3 model is, 
which is approximated by the 6/3 model. 
 We already recalled that the treatment of the electrons as a completely degenerate Fermi gas is justified for
stars in their ground state, and a reasonable approximation for white dwarf stars which are energetically 
near their ground state, an insight which goes back to Fowler. 
 On the other hand, each and every non-collapsed star in the heavens contains 
more than one species of nuclei, presumably, and with the electrons treated as one of the fluid components, a
two-fluid approximation for a star is a plausible approximation only as long as the various positive nuclei 
species are sufficiently mixed by convective and turbulent motions so that throughout the star any species of nuclei 
with $z$ elementary charges per nucleus and mass $A_z\mPR$, where $A_z$ is the mass number (for each $z$ we only take
the dominant isotope into account, for simplicity) has a number density in essentially constant proportion 
to the number density of free protons $\nupos(r)$, viz. $\nu_z(r) \approx C_{z}\nupos(r)$ (with ``$=$'' instead of
``$\approx$'' when $z=1$, and $C_1=1$).
 Since nuclei of type ${}_z$ carry positive integer multiples $z$ of the elementary charge $e$,
the positive charge density function 
$e\overline{z}\nuPLU(r):= e\sum_z z\nu_z(r)\approx e\left(\sum_z z C_{z} \right) \nupos(r)$,
and the two-fluid approximation consists in replacing ``$\approx$'' by ``$=$;''
so one can work with the electron density $\nuneg$, and either $\nuPLU$ or $\nupos$, as one pleases.
 For the mass density, one then has
$\mu(r):= \sum_z m_z\nu_z(r) +\mEL\nuneg(r) \approx \left(\sum_z m_z C_{z} \right) \nupos(r) +\mEL\nuneg(r)$
% $\approx \left(\sum_k [m_k +\mEL z_k] C_{k,\mathrm{p}} \right) \nupos(r)$ when $\nuneg=\overline{z}\nuPLU$ is set.
and ``$\approx$'' is replaced by ``$=$'' in the two-fluid approximation (and $\mEL$ may be neglected).

 However, white dwarfs are stars where nuclear fusion processes have expired,
and the distribution of nuclei is no longer mixed up by convection and other processes, featuring instead an onion layer
structure, as addressed long ago by Hamada and Salpeter \cite{HS}.
 Moreover, while the protons are fermions, the other important species in a low-to-medium mass white dwarf are
bosons (for instance, $\alpha$ particles, ${}^{12}$C, and ${}^{16}$O nuclei).
 Treating such a segregated ensemble of nuclei species as a single completely degenerate Fermi fluid with effective
mass and charge parameters is not a compelling approximation.

 The only special case of the two-fluid 5/3 model in \cite{KNY} which is not subject to the just
leveled criticism is an idealized model star with only one species of nuclei: protons, which \emph{are} fermions. 
  Even though no astronomer will presumably ever see a star made of only electrons and protons, in particular
not one that is in its energetic ground state, it is certainly not absurd to contemplate this model as a valid
simplifying approximation to a model for a first generation star with such a low mass that it failed to ignite 
(a ``failed star,'' like a brown dwarf \cite{Jill}) and essentially cooled down to its lowest energy state:
 a black dwarf with a very low mass, between $\approx 13$ and $\approx 80$ Jupiter masses.
 To emphasize that the star never ignited, we prefer to rather speak of a ``failed white dwarf.''
 As per the cosmological standard model \cite{Schramm}, on average such a star's nuclei composition would consist of
$\approx 92\%$ {free} protons and $\approx 8\%$ $\alpha$ particles,
     % mass would be $75\%$ due to {individual} protons, $25\%$ due to $\alpha$ particles,
and the suggestive approximation which leads to the 5/3 model of two completely degenerate Fermi gases
consists in replacing the $\approx 8\%$ $\alpha$ particles by protons. 
 By some statistical fluke there may well be regions in the early universe where 
the percentages are even more lopsided toward the protons.
 A non-relativistic model suffices, with Newton's gravity and Coulomb's electricity 
stabilized by the gradients of the degeneracy pressures.
 The latter are then approximated by replacing $\gamma=5/3$ with $\gamma=6/3 \, (=2)$.

 Furthermore, as we pointed out in \cite{HK} already, though without giving any details,
the 6/3 model gives the same answer 
to the question how many electrons per proton fit on a failed white dwarf as does the 5/3 model,
and it allows one to demonstrate by explicitly writing down the solution pairs, that the bounds on $\Nneg/\Npos$ can be saturated.
 To come to the same conclusions in the 5/3 model requires more work.
 In the special-relativistic model one has to stay away from the critical Chandrasekhar mass, but this of course is 
implicitly understood because the critical mass beyond which degeneracy pressure no longer stabilize against
collapse is far greater than the critical mass beyond which a star's nuclear fuel ignites. 

 In section II we recall the Thomas--Fermi equations of a failed white dwarf star  made of
protons and electrons that are treated as ideal Fermi gases of spin-$\frac12$ particles.
 This section is essentially identical with {Sect.IV} of \cite{HK}.

 Then, in section III, we will apply the $5/3\to 6/3$ approximation to this two-species
model. We solve the approximate model explicitly in terms of elementary functions and
compute the allowed interval of $\Nneg/\Npos$ values.

 Section IV explains that {the allowed interval of $\Nneg/\Npos$ values is the same
also in more} physical models, cf. \cite{HK}.

 In section V we explain that in the more physical models the electron and proton numbers, $\Nneg$ and $\Npos$,
can be computed in terms of the zeros of the particle densities and the derivatives of the densities at the zeros.

 In section VI we illustrate our findings and compare the 6/3 model with the physical 5/3 model, and also with 
the Chandrasekhar-type special-relativistic model, for which we have carried out numerical evaluations.

 Section VII has the Kepler problem of charged binaries.

 In section VIII we convert the bounds on $\Nneg/\Npos$ into bounds on the total charge $Q$ which imply
a bound on $Q^2$ proportional to $M^2$, valid also for Reissner--Weyl--Nordstr\"om black holes. 
 In this sense our results support Penrose's weak cosmic censorship hypothesis. 

The conclusions are presented in section IX.

In an appendix we formulate the generalization of the exactly solvable two-species model to more than two species.
 We leave its solution to some future work.

 In another appendix we present some exact solutions to the two-species polytropic $n=5$ and the isothermal model, and
also for the $5/3$ model.

 In yet another appendix we invoke the usual local neutrality approximation which yields the single-density model
discussed in \cite{Emden}, \cite{Chandra}, \cite{KippenhahnWeigert}.
 We will take the opportunity, in a subsection in that appendix, to explain our $5/3\to 6/3$ approximation
for the locally neutral single-density model, which produces the Lane--Emden polytrope of index $n=1$ and its
elementary solution.

\vspace{-.5truecm}
%%%%%%%%%%%%%%%%%%%%%%%%%%%%%%%%%%%%%%%%%%%%%%%%%%%%%%%%%%%%%%
%%%%%%%%%%%%%%%%%%%%%%%%%%%%%%%%%%%%%%%%%%%%%%%%%%%%%%%%%%%%%%
%%%%%%%%%%%%%%%%%%%%%%%%%%%%%%%%%%%%%%%%%%%%%%%%%%%%%%%%%%%%%%%%%%%%
\section{The Thomas--Fermi equations of \\ a failed white dwarf star}\vspace{-5pt}
%%%%%%%%%%%%%%%%%%%%%%%%%%%%%%%%%%%%%%%%%%%%%%%%%%%%%%%%%%%%%%
%%%%%%%%%%%%%%%%%%%%%%%%%%%%%%%%%%%%%%%%%%%%%%%%%%%%%%%%%%%%%%
%%%%%%%%%%%%%%%%%%%%%%%%%%%%%%%%%%%%%%%%%%%%%%%%%%%%%%%%%%%%%%%%%%%%

 The basic equations of structure of a non-rotating
white dwarf star composed of electrons and nuclei can be found in Chandrasekhar's
original publications composed into his classic book \cite{Chandra}, in \cite{KippenhahnWeigert}, and also in
\cite{SilbarReddy}, \cite{Garfinkle}, for instance. 
  For a non-rotating star one may assume spherical symmetry, so all
the basic structure functions are then functions only of the radial distance $r$ from the star's center,
and the differential equations involved in the discussion reduce to the ordinary type.

 We specialize the discussion to a failed star composed only of protons and electrons,
both of which are spin-$\frac12$ fermions.
 Each species is treated as an ideal Fermi gas.
 The number density functions $\nupos(r)\geq 0$ and $\nuneg(r)\geq 0$ are
assumed to integrate to the total number of protons, respectively electrons, viz.
\begin{eqnarray}
\int_{\Rset^3} \nupos(r) d^3r = \Npos,
 \label{eq:Npos}\\
\int_{\Rset^3} \nuneg(r) d^3r = \Nneg.
 \label{eq:Nneg}
\end{eqnarray}
 The protons have rest mass $\mPR$ and charge $+e$, the electrons have rest mass $\mEL$ and charge $-e$.
 Thus the mass density of the star is given by
\begin{equation}
\mu(r) = \mPR \nupos(r) + \mEL\nuneg(r)
 \label{eq:massdensity}
\end{equation}
and its charge density by
\begin{equation}
\sigma(r) = e \nupos(r) - e \nuneg(r).
 \label{eq:chargedensity}
\end{equation}
 The star is overall neutral if $\Npos=\Nneg$, otherwise it carries an excess charge which may have
either sign.

 The electrons and protons jointly produce a Newtonian gravitational potential $\phi_N^{}(r)$ and 
an electric Coulomb potential $\phi_C^{}(r)$.
 The Newton potential $\phi_N^{}$ is related to the mass density $\mu$ by a radial Poisson equation,
\begin{equation}
\left(r^2\phi_N^{\prime}(r)\right)^\prime
  = 4\pi G \mu(r)r^2,
 \label{eq:PoissonN}
\end{equation}
where $G$ is Newton's constant of universal gravitation.
 Similarly, the Coulomb  potential $\phi_C^{}$ is related to the charge density $\sigma$ by a radial Poisson equation,
\begin{equation}
-\left(r^2\phi_C^{\prime}(r)\right)^\prime = 4\pi \sigma(r)r^2.
 \label{eq:PoissonC}
\end{equation}
 As usual, the primes in (\ref{eq:PoissonN}) and (\ref{eq:PoissonC})
mean derivative with respect to the displayed argument, in this case $r$.

 Each species, the electrons and the protons, satisfies an Euler-type mechanical force balance equation,
\begin{eqnarray}
\nupos(r)\left[-\mPR \phi_N^{\prime}(r) - e \phi_C^{\prime}(r)\right] -  \ppos^\prime(r) =0,
 \label{eq:forceBALANCEpos}\\
\nuneg(r)\left[-\mEL \phi_N^{\prime}(r) + e \phi_C^{\prime}(r)\right] -  \pneg^\prime(r) =0.\;
 \label{eq:forceBALANCEneg}
\end{eqnarray}
 Here, $\ppos$ and $\pneg$ are the degeneracy pressures of the ideal proton and electron gases, respectively.
 For a non-relativistic gas of spin-$\frac12$ fermions (subscript ${}_f$) of mass $m_{\text{f}}^{}$ and number 
density $\nu_{\text{f}}^{}$ one has (see, e.g. \cite{Chandra}, \cite{Balian}, \cite{Balescu})\vspace{-10pt}
\begin{equation}
p_{\text{f}}(r) = \frac{\hbar^2}{m_{\text{f}}}\frac{(3\pi^2)^{2/3}}{5}\nu^{5/3}_{\text{f}}(r);
 \label{eq:pDEG}
\end{equation}
here, ${}_f$ stands for either ${}_{\mathrm{p}}$ or ${}_e$, and $\hbar$ is the reduced Planck constant.
 We remark that (\ref{eq:pDEG}) is of the type $p = K_\gamma\nu^\gamma$ for some constant $K_\gamma$,
called a \emph{polytropic} law of power $\gamma$, here with $\gamma=5/3$.
 Associated to $\gamma$ is a polytropic index $n=1/(\gamma-1)$; here $n=3/2$.

 The system of structure equations can be reduced to a closed system of equations for the densites 
$\nupos(r)$ and $\nuneg(r)$ alone, but one needs to distinguish three regions:
\smallskip

(a) $\nupos(r)>0$ and $\nuneg(r)>0$ (the bulk region),

(b) $\nupos(r)>0$ and $\nuneg(r)=0$ (positive atmosphere),

(c) $\nupos(r)=0$ and $\nuneg(r)>0$ (negative atmosphere).

 We begin with the bulk region, where both Eqs.(\ref{eq:forceBALANCEpos}) and (\ref{eq:forceBALANCEneg}) are nontrivial. 
 We use (\ref{eq:massdensity}) and (\ref{eq:chargedensity}) to express $\mu$ and $\sigma$ in terms of 
$\nupos$ and $\nuneg$ in  (\ref{eq:PoissonN}) and (\ref{eq:PoissonC});
 now we multiply (\ref{eq:PoissonN}) by $-\mPR$ and (\ref{eq:PoissonC}) by $e$ and add the resulting two equations,
then use (\ref{eq:forceBALANCEpos}) to replace $-\mPR \phi_N^{\prime}(r) - e \phi_C^{\prime}(r)$ in terms
of $\nupos(r)$ %, $\nuneg(r)$, and $\ppos^\prime(r)$;
next we use (\ref{eq:pDEG}) to express $\ppos$ in terms of $\nupos$.
 Similarly, we multiply (\ref{eq:PoissonN}) by $-\mEL$ and (\ref{eq:PoissonC}) by $-e$ and also add these equations, 
then use (\ref{eq:forceBALANCEneg}) to replace $-\mEL \phi_N^{\prime}(r) + e \phi_C^{\prime}(r)$ in terms of
$\nuneg(r)$ and $\pneg^\prime(r)$;
next we use (\ref{eq:pDEG}) to express $\pneg$ in terms of $\nuneg$.
 This yields 
\begin{widetext}
\begin{eqnarray}
- \veps \zeta \frac{1}{r^2}\Ddr\left(r^2\Ddr \nupos^{2/3}(r)\right)
&\,\  =  - \left(1 -\frac{G\mPR^2}{e^2}\right) \nupos(r) + \left(1 +\frac{G\mPR\mEL}{e^2}\right) \nuneg(r),
 \label{eq:PoissonMUpTHREEhalf}\\
- \zeta \frac{1}{r^2}\Ddr\left(r^2\Ddr \nuneg^{2/3}(r)\right)
&\!\!  = \left(1 + \frac{G\mPR\mEL}{e^2}\right) \nupos(r) - \left(1 - \frac{G\mEL^2}{e^2}\right) \nuneg(r),
 \label{eq:PoissonMUeTHREEhalf}
\end{eqnarray}\vspace{-.6truecm}
\end{widetext}
a system of nonlinear second-order differential equations 
valid wherever both  $\nupos(r)>0$ and $\nuneg(r)>0$; 
here, $\veps:= {\mEL}/{\mPR}$ and  % and $K := {2(3{\pi^2})^{\frac23}}/{5}$.
$\zeta:= (3^{2/3}\pi^{1/3}/8)\, {\hbar^2} / {\mEL e^2}$ % $\approx  52.1846 \lambdabar_{\mathrm{C}}$.
is approximately 50 reduced Compton wavelengths of the electron.

 Coming to the atmospheric regimes, a positive atmosphere is governed by
(\ref{eq:PoissonMUpTHREEhalf}) with $\nuneg(r)=0$, while a negative atmosphere is governed by
(\ref{eq:PoissonMUeTHREEhalf}) with $\nupos(r)=0$. 

 Each equation is of second order and requires two initial conditions.
 At the bulk-atmosphere interface at $r=r_0^{}$ the density of the species which forms the atmosphere
needs to be continuously differentiable.
 In the bulk, conditions are posed at $r=0$. 
 Naturally $\nupos^\prime(0)=0=\nuneg^\prime(0)$. 
 The values of $\nupos(0)$ and $\nuneg(0)$ are to be chosen such that 
Eqs.(\ref{eq:Npos}) and (\ref{eq:Nneg}) hold.

 This system of coupled differential equations for the density functions $\nupos$ and $\nuneg$ in bulk and
atmosphere regions generalizes the single Lane--Emden equation for the polytrope of index $n=\frac32$, 
which has only a bulk interior; see \cite{Chandra},  \cite{KippenhahnWeigert}, \cite{SilbarReddy}, \cite{Garfinkle}.

 As for the numerical values of the parameters,
$\veps \approx 1/1836\approx 5.54\cdot 10^{-4}$ and  % and $K := {2(3{\pi^2})^{\frac23}}/{5}$.
$\zeta\approx 52.185\frac{\hbar}{\mEL c}$, where $\frac{\hbar}{\mEL c}\approx 3.86\cdot 10^{-13}$m
is the electron's reduced Compton wave length. 
 The three ratios of gravitational-to-electrical coupling constants which appear in the coefficient matrix at the
right-hand sides of Eqs.(\ref{eq:PoissonMUpTHREEhalf}) and (\ref{eq:PoissonMUeTHREEhalf}) are fantastically tiny numbers, viz.
  ${G\mEL^2}/{e^2}\approx 2.40\cdot 10^{-43}$, ${G\mPR\mEL}/{e^2}\approx 4.41\cdot 10^{-40}$, and
${G\mPR^2}/{e^2}\approx 8.09\cdot 10^{-37}$.
 All the same, the three tiny ratios of coupling constants are the only places where Newton's constant of universal
gravitation, $G$, enters the equations, and since it is gravity, not electricity, which binds the ideal Fermi gases together
to form a star, one cannot neglect these tiny numbers versus 1 in the cofficients --- 
this would result in a singular coefficient matrix, and there would not be any nontrivial solution pair $\nupos,\nuneg$.

 The nonlinearity of Eqs.(\ref{eq:PoissonMUpTHREEhalf}) and (\ref{eq:PoissonMUeTHREEhalf}), coupled to each other and to
their atmospheric counterparts, stands in the way of solving them generally in closed form,
although one special elementary solution can be found (see further below).
 In principle one can evaluate them numerically on a computer, but the tiny ratios of the coupling constants do create
problems. 
 Also the small ratio of the masses, $\mEL/\mPR\approx 1/1836$, is a source of numerical trouble. 
 In this situation it definitely is prudent to look for a solvable model, to which we turn next.
\newpage
%\vspace{-.5truecm}
%%%%%%%%%%%%%%%%%%%%%%%%%%%%%%%%%%%%%%%%%%%%%%%%%%%%%%%%%%%%%%
%%%%%%%%%%%%%%%%%%%%%%%%%%%%%%%%%%%%%%%%%%%%%%%%%%%%%%%%%%%%%%
%%%%%%%%%%%%%%%%%%%%%%%%%%%%%%%%%%%%%%%%%%%%%%%%%%%%%%%%%%%%%%%%%%%%
\section{The $5/3\to 6/3$ approximation}\label{5363}%\vspace{-10pt}
%%%%%%%%%%%%%%%%%%%%%%%%%%%%%%%%%%%%%%%%%%%%%%%%%%%%%%%%%%%%%%
%%%%%%%%%%%%%%%%%%%%%%%%%%%%%%%%%%%%%%%%%%%%%%%%%%%%%%%%%%%%%%
%%%%%%%%%%%%%%%%%%%%%%%%%%%%%%%%%%%%%%%%%%%%%%%%%%%%%%%%%%%%%%%%%%%%

 Note that we cannot simply replace $\nu^{5/3}_{\text{f}}$ by $\nu^{6/3}_{\text{f}}$, for $\nu_{\text{f}}$ is not dimensionless. 
 This can be overcome by switching to dimensionless densities with the help of some reference density. 
 In the astrophysical literature one often finds the central density as reference density, a choice 
motivated by seeking a definite initial value problem for the numerical integration of the Lane--Emden equation on
a computer: the so-normalized dimensionless density takes the value 1 at $r=0$, and its derivative vanishes there.
 We will be able to solve the $6/3$ model equations explicitly, so we have no need for such a normalization. 
 Instead, since the fermionic degeneracy pressure already is expressed with the microscopic constants $\hbar,\mPR,\mEL$, 
we may as well now choose as reference length the electron's reduced Compton length $\hbar/\mEL c$,
where $c$ is the speed of light in vacuum.
 While this is somewhat unconventional, it is not unnatural and the resulting formulas are easy to interpret.
 Thus we set $r =: (\hbar/\mEL c)\rho$ and $\nu(r) =: (\mEL c/\hbar)^3\upsilon(\rho)$, and
we also set $\nupos(r) =: (\mEL c/\hbar)^3\upos(\rho)$ and  $\nuneg(r) =: (\mEL c/\hbar)^3\uneg(\rho)$. 
 Inserted into the formulas for the degeneracy pressures, we find $\ppos(r)\propto \upos(\rho)^{5/3}$ and 
$\pneg(r)\propto \uneg(\rho)^{5/3}$, and 
\emph{now} we can replace $\upos^{5/3}$ by $\upos^{6/3}$ and $\uneg^{5/3}$ by $\uneg^{6/3}$. 
 
 This hurdle cleared, we may for the sake of completeness also introduce dimensionless potential functions through
$\phi_N^{}(r) =: c^2 \psi_N^{}(\rho)$ and $\phi_C^{}(r) =: c^2\frac{\mEL}{e} \psi_C^{}(\rho)$, but we 
won't need this, given we already have the system of Eqs.(\ref{eq:PoissonMUpTHREEhalf}) and (\ref{eq:PoissonMUeTHREEhalf}), 
plus their atmospheric specializations.

 As in the 5/3 model we distinguish the regions:
\smallskip

(a) $\upos(\rho)>0$ and $\uneg(\rho)>0$ (the bulk region),

(b) $\upos(\rho)>0$ and $\uneg(\rho)=0$ (positive atmosphere),

(c) $\upos(\rho)=0$ and $\uneg(\rho)>0$ (negative atmosphere).
\vspace{-10pt}

%%%%%%%%%%%%%%%%%%%%%%%%%%%%%%%%%%%%%%%%%%%%%%%%%%%%%%%%%%%%%% 
%%%%%%%%%%%%%%%%%%%%%%%%%%%%%%%%%%%%%%%%%%%%%%%%%%%%%%%%%%%%%%
                \subsection{The bulk region}\vspace{-5pt}
%%%%%%%%%%%%%%%%%%%%%%%%%%%%%%%%%%%%%%%%%%%%%%%%%%%%%%%%%%%%%%
%%%%%%%%%%%%%%%%%%%%%%%%%%%%%%%%%%%%%%%%%%%%%%%%%%%%%%%%%%%%%%

 In the bulk region we now have the following coupled system of linear second-order differential equations 
for the density functions $\upos$ and $\uneg$, which generalizes the single Lane--Emden equation for the 
polytrope of index $n=1$, (\ref{eq:LaneEmdenONEa}), in the common interior of the charged gases where 
both $\upos(\rho)>0$ and $\uneg(\rho)>0$:
\begin{widetext}
\begin{eqnarray}
- \veps \varsigma \frac{1}{\rho^2}\left(\rho^2\upos^{\prime}(\rho)\right)^\prime
&\,\  =  - \left(1 -\tfrac{G\mPR^2}{e^2}\right) \upos(\rho) + \left(1 +\tfrac{G\mPR\mEL}{e^2}\right) \uneg(\rho),
 \label{eq:PoissonMUp}\\
- \varsigma \frac{1}{\rho^2}\left(\rho^2\uneg^{\prime}(\rho)\right)^\prime
&\!\!  = \left(1 + \tfrac{G\mPR\mEL}{e^2}\right) \upos(\rho) - \left(1 - \tfrac{G\mEL^2}{e^2}\right) \uneg(\rho),
 \label{eq:PoissonMUe}
\end{eqnarray}
\end{widetext}
 Here,  $\varsigma:= \frac{3^{2/3}\pi^{1/3}}{10}\tfrac{\hbar c} {e^2}\approx 41.74766$.
 
 We now solve the system of equations (\ref{eq:PoissonMUp}) and (\ref{eq:PoissonMUe}) explicitly. 
 A non-singular system of linear second-order differential equations has four linearly independent solutions, 
from which we have to select the ones compatible with our physical problem. 
 This is done as follows.

 We remark that similarly to the Lane--Emden equation for the polytrope of index $n=1$, (\ref{eq:LaneEmdenONEa}),
a change of dependent variables $\upos(\rho) \mapsto \rho\upos(\rho) =: \chipos(\rho)$ and
 $\uneg(\rho) \mapsto \rho\uneg(\rho) =: \chineg(\rho)$ transforms 
Eqs.(\ref{eq:PoissonMUp}) and (\ref{eq:PoissonMUe}) into a linear second-order system with constant coefficients
for $\chipos(\rho),\chineg(\rho)$,
and such a system (when not singular) can always
be solved by the ansatz $\chi_{\text{f}}(\rho) \propto \exp(\kappa \rho)$, with $_f$ standing for either $_{\mathrm{p}}$ or $_e$.
 In terms of $\upos,\uneg$ this means that the ansatz 
$\upos(\rho) = \Bpos \exp(\kappa \rho)/\rho$ and $\uneg(\rho) = \Bneg \exp(\kappa \rho)/\rho$, with the same $\kappa$, 
will transform the system of differential equations (\ref{eq:PoissonMUp}) and (\ref{eq:PoissonMUe}) into a linear
system of algebraic equations. 
 Indeed, away from $\rho=0$ we have 
\begin{equation}\label{Yukawa}
\frac{1}{\rho^2}\Ddrho\left(\rho^2\Ddrho \frac{\exp(\kappa \rho)}{\rho}\right) = \kappa^2 \frac{\exp(\kappa \rho)}{\rho},
\end{equation}
and so we obtain the matrix problem
\begin{equation}
\left(
\begin{array}{cc}
1 -\tfrac{G\mPR^2}{e^2} -  \kappa^2 \veps\varsigma \;; &  - 1 -\tfrac{G\mPR\mEL}{e^2} \\
- 1 -\tfrac{G\mPR\mEL}{e^2}\quad\; ; & 1 -\tfrac{G\mEL^2}{e^2} -  \kappa^2 \varsigma  
\end{array}
\right)\!\!
\left(
\begin{array}{c}
\!\Bpos\!  \\
\!\Bneg\!
\end{array}
\right)
=
\left(
\begin{array}{c}
\!0\!  \\
\!0\!
\end{array}
\right)\!\!;
 \label{eq:matrixPROBLEM}
\end{equation}
here we have placed semi-colons in the matrix to facilitate the identification of the matrix elements.
  The solvability condition for Eq.(\ref{eq:matrixPROBLEM}) is the characteristic equation
\begin{equation}
\det \left(
\begin{array}{cc}
1 -\tfrac{G\mPR^2}{e^2} -  \kappa^2 \veps\varsigma \;; &  - 1 -\tfrac{G\mPR\mEL}{e^2} \\
- 1 -\tfrac{G\mPR\mEL}{e^2}\quad\; ; & 1 -\tfrac{G\mEL^2}{e^2} -  \kappa^2 \varsigma  
\end{array}
\right)\!\!
=
0,
 \label{eq:matrixPROBLEMdet}
\end{equation}
which yields
\begin{equation}
\left(1 -\tfrac{G\mEL^2}{e^2} -  \kappa^2 \varsigma  \right)\!
\left(1 -\tfrac{G\mPR^2}{e^2} -  \kappa^2 \veps\varsigma \right)
-\left(1+\tfrac{G\mPR\mEL}{e^2}\right)^2\!\!
=
0,
 \label{eq:matrixPROBLEMdetEXPL}
\end{equation}
a quadratic problem in $\kappa^2$, viz. $a\kappa^4 +b\kappa^2 +c =0$, 
with $a=\veps\varsigma^2>0$,
$b= -\varsigma\left(1+\veps -{G(\veps\mEL^2+\mPR^2)}/{e^2}\right)<0$,
and $c= - {G\left(\mEL + \mPR\right)^2}/{e^2}<0$.
 By the quadratic formula we have two real solutions,
\begin{equation}
(\kappa^2)_\pm = -\tfrac{b}{2a}\left(1\pm\sqrt{1 - 4 \tfrac{ac}{b^2}}\right)
 \label{eq:kappaPMsquare}
\end{equation}
one of which is positive and the other one negative,
with $(\kappa^2)_+^{}\approx 44.0025$, and $(\kappa^2)_-^{}\approx - 1.94025\cdot 10^{-38}$.
 This now yields the \emph{hyperbolic} $\kappa_h^{} := \surd{(\kappa^2)_+}\approx  {6.63344}$
and the \emph{trigonometric} $\kappa_t^{} := |\surd{(\kappa^2)_-}| \approx { 1.3929\cdot 10^{-19}}$.
 The latter step obviously generates two imaginary $\kappa$ values. 
 Converted to real solutions by taking appropriate linear combinations, the set of four linear independent 
solutions consists of one exponentially growing mode, one exponentially decaying mode, one sine and one cosine 
mode, each of them divided by the independent variable $\rho$. 
 
 Next we recall the well-known fact that Newton's and Coulomb's $1/r$ potentials correspond to a point source at $r=0$, 
and this we need to rule out. 
 This means that the mode $\cos(\kappa \rho)/\rho$ is not admissible, while $\sin(\kappa\rho)/\rho$ is.
 Similarly, only the linear combination of the exponential modes into the hyperbolic $\sinh(\kappa\rho)/\rho$ mode is
admissible, while all other linear combinations are not, in particular the hyperbolic 
$\cosh(\kappa\rho)/\rho$ mode is not admissible.

 Thus, the physically admissible general solution of Eqs.(\ref{eq:PoissonMUp}) and (\ref{eq:PoissonMUe}) is of the form
 (cf. sect.IV.B in \cite{KNY})
\begin{eqnarray}\label{bulkPOS}
\upos(\rho) = \Bpos^h \frac{\sinh(\kappa_h^{}\rho)}{\rho} + \Bpos^t \frac{\sin(\kappa_t^{}\rho)}{\rho},\\
\label{bulkNEG}
\uneg(\rho) = \Bneg^h \frac{\sinh(\kappa_h^{}\rho)}{\rho} + \Bneg^t \frac{\sin(\kappa_t^{}\rho)}{\rho},\;
\end{eqnarray}
where we have added superscripts ${}^h$ and ${}^t$ at the bulk region coefficients $\Bpos$ and $\Bneg$ 
to match with the hyperbolic and trigonometic modes.
 Here, the pairs $(\Bpos^h,\Bneg^h)$ and $(\Bpos^t,\Bneg^t)$ are eigenvectors of the
coefficient matrix at the left-hand side of Eq.(\ref{eq:matrixPROBLEM}) for the corresponding
eigenvalues $(\kappa^2)_\pm^{}$, respectively, and so only two of the four bulk coefficients are independent
in the general physical solution. 
 Linear algebra yields the relationships between 
$\Bpos^h$ and $\Bneg^h$, respectively between $\Bpos^t$ and $\Bneg^t$, with the results
\begin{equation}\label{eq:BnegOverBposHYPO}
\frac{\Bneg^h}{\Bpos^h}
= 
\frac{1-\tfrac{G\mPR^2}{e^2} -\veps\varsigma \kappa^2_h}{1+\tfrac{G\mPR\mEL}{e^2} } \approx -{5.45\cdot 10^{-4}},
\quad
\end{equation}
\begin{equation}\label{eq:BnegOverBposTRIG}
\frac{\Bneg^t}{\Bpos^t}
= 
\frac{1-\tfrac{G\mPR^2}{e^2} +\veps\varsigma \kappa^2_t}{1+\tfrac{G\mPR\mEL}{e^2} } \approx {1- 8.09\times 10^{-37}}.
\end{equation}

 We pause for a moment to take in the results obtained.

 The trigonometric parts of the general solution obviously correspond to the $n=1$ polytrope of the 
Lane--Emden equation for the single-density approximation, with 
$\kappa_t^{}\approx \kappa$ given by Eq.(\ref{eq:LaneEmdenONEb}) to a high degree of accuracy, which in concert 
with Eq.(\ref{eq:BnegOverBposTRIG}) confirms that the positive and negative \emph{large scale} densities are 
{very well approximated by the single-density model almost all the way up to the bulk radius.}
 This confirms by explicit example what everyone knows already, 
that the locally neutral single-density approximation suffices to obtain the 
bulk structure of the white dwarf star. 

 In addition we now have information on  the charge separation effects, which are accounted for by
the hyperbolic parts of the general solution.
 These vary significantly on a very short scale by comparison, and so their amplitudes must be very tiny. 
 Interestingly, the hyperbolic modes of the positive and negative species have significantly different bulk amplitudes,
roughly corresponding in ratio to the ratio of the rest masses of electrons and protons. 

 The remaining two independent bulk amplitudes, say $\Bpos^h$ and $\Bpos^t$, 
cannot be fixed with the bulk densities alone; this requires also the atmospheric densities.
 By inspecting the general bulk solution formulas (\ref{bulkPOS}) and (\ref{bulkNEG})
it is easy to see, though, that $\upos(\rho)>0$ can only be achieved with $\Bpos^t>0$,
while $\Bpos^h$ can take either sign. 
 The analogous conclusion holds therefore for $\Bneg^t>0$ and $\Bneg^h$.
 As soon as one or the other density reaches zero, the system of equations changes to describe the atmospheric region,
unless it happens that both densities reach zero simultaneously (the case of no atmosphere; it will be addressed 
separately).
 An atmosphere can be populated either purely with protons or purely with electrons, 
yet either version is determined in a similar manner. 
 We next turn to these atmospheric cases.
\vspace{-10pt}

%%%%%%%%%%%%%%%%%%%%%%%%%%%%%%%%%%%%%%%%%%%%%%%%%%%%%%%%%%%%%% 
%%%%%%%%%%%%%%%%%%%%%%%%%%%%%%%%%%%%%%%%%%%%%%%%%%%%%%%%%%%%%%
                \subsection{The positive atmosphere}
%%%%%%%%%%%%%%%%%%%%%%%%%%%%%%%%%%%%%%%%%%%%%%%%%%%%%%%%%%%%%%
%%%%%%%%%%%%%%%%%%%%%%%%%%%%%%%%%%%%%%%%%%%%%%%%%%%%%%%%%%%%%%

 In the positive atmosphere the electron density vanishes, $\uneg(\rho)=0$, while the proton density is
still positive, $\upos(\rho)>0$, so the electrons' Eulerian force balance equations is trivially satisfied, 
while that for the protons now is given by (\ref{eq:PoissonMUp}) with $\upsilon_e=0$, viz.
\begin{eqnarray}
- \veps \varsigma \frac{1}{\rho^2}\left(\rho^2\upos^{\prime}(\rho)\right)^\prime
  =  - \biggl(1 -\frac{G\mPR^2}{e^2}\biggr) \upos(\rho),
 \label{eq:PoissonMUpNEGnull}
\end{eqnarray}
valid for $\rho>\rho_0$, where $\rho_0 = \sup\{\rho:\uneg(\rho)>0\}$ is the radius of the bulk region.
 If {${G\mPR^2}/{e^2}$} would be greater than 1, Eq.(\ref{eq:PoissonMUpNEGnull}) would be a Lane--Emden
equation of the $n=1$ polytrope (mathematically speaking).
 However, since {${G\mPR^2}/{e^2}$} is the tiny number it happens to be, Eq.(\ref{eq:PoissonMUpNEGnull}) differs
from this  Lane--Emden equation by the sign of its right-hand side. 
 Analogous to solving for the bulk region, the general solution of (\ref{eq:PoissonMUpNEGnull}) now reads
\begin{equation}\label{muposATMO}
\upos(\rho) = \Apos^+ \frac{\exp(\vkpos \rho)}{\rho} +\Apos^- \frac{\exp(-\vkpos \rho)}{\rho},
\end{equation}
where $\vkpos>0$ is the positive root of
\begin{equation}
{\vkpos^2
  = \frac{10}{3^{2/3}\pi^{1/3}}\frac{e^2}{\hbar c}\frac{\mPR}{\mEL} \biggl(1 - \frac{G\mPR^2}{e^2}\biggr).}
 \label{eq:kappaPOSatmo}
\end{equation}
 Note that in this expression one may approximate the last parenthetical factor by 1.
 Note furthermore that {$\vkpos\approx 6.63$} is essentially determined by the electrical coupling. 

 A few comments are in order right now.

 First, it could seem reasonable to throw out the exponentially growing mode, but 
note that a small negative $\Apos^+$ in concert with a large positive $\Apos^-$
will result in a $\upos(\rho)$ which rapidly goes to zero in the positive atmosphere region, 
so an exponentially growing mode is not a problem because it would be terminated as soon as the proton density vanishes.

 Second, since  $\rho>\rho_0$, there is no reason now to only allow the linear combination of the exponential modes 
into the hyperbolic sine, as was the case in the bulk region where there would otherwise be a problem at the origin $\rho=0$.
 Incidentally, equivalently to (\ref{muposATMO}) we may write the general solution of the positive atmosphere as
\begin{equation}\label{muposATMOalt}
\upos(\rho) = \Apos^+ \frac{\cosh(\vkpos \rho)}{\rho} +\Apos^- \frac{\sinh(\vkpos \rho)}{\rho}.
\end{equation}

 Third, the two atmospheric amplitudes $\Apos^+$ and $\Apos^-$ are constrained by the requirement that the 
proton density $\upos(\rho)$ be continuously differentiable at the boundary $\rho=\rho_0$ of the bulk region, 
so both are needed in general. 
 We will get to this shortly. 
\vspace{-10pt}

%%%%%%%%%%%%%%%%%%%%%%%%%%%%%%%%%%%%%%%%%%%%%%%%%%%%%%%%%%%%%% 
%%%%%%%%%%%%%%%%%%%%%%%%%%%%%%%%%%%%%%%%%%%%%%%%%%%%%%%%%%%%%%
                \subsection{The negative atmosphere}
%%%%%%%%%%%%%%%%%%%%%%%%%%%%%%%%%%%%%%%%%%%%%%%%%%%%%%%%%%%%%%
%%%%%%%%%%%%%%%%%%%%%%%%%%%%%%%%%%%%%%%%%%%%%%%%%%%%%%%%%%%%%%

 The discussion of the negative atmosphere region mirrors the one for the positive atmosphere region, so we may be 
brief.
 While $\upos(\rho)=0$, the structure equation for $\uneg(\rho)$ is given by (\ref{eq:PoissonMUe}) with $\upsilon_{\mathrm{p}}=0$, viz.
\begin{eqnarray}
- \varsigma \frac{1}{\rho^2}\left(\rho^2\uneg^{\prime}(\rho)\right)^\prime
  =  - \left(1 -\frac{G\mEL^2}{e^2}\right) \uneg(\rho),
 \label{eq:PoissonMUePOSnull}
\end{eqnarray}
valid for $\rho>\rho_0$, where now the radius of the bulk region is $\rho_0 = \sup\{\rho:\upos(\rho)>0\}$.

 The general solution of (\ref{eq:PoissonMUePOSnull}) reads
\begin{equation}\label{munegATMO}
\uneg(\rho) = \Aneg^+ \frac{\exp(\vkneg \rho)}{\rho} +\Aneg^- \frac{\exp(-\vkneg \rho)}{\rho},
\end{equation}
where $\vkneg>0$ is the positive root of
\begin{equation}
{\vkneg^2  = \frac{10}{3^{2/3}\pi^{1/3}}\frac{e^2}{\hbar c} \left(1 - \frac{G\mEL^2}{e^2}\right).}
 \label{eq:kappaNEGatmo}
\end{equation}
 Note that $\vkneg^2 \approx \frac{\mEL}{\mPR}\vkpos^2$, where the ``$\approx$'' is due to some 
slight differences beginning to show 36 decimal places after the leading digit.
 Again, also in (\ref{eq:kappaNEGatmo}) one may approximate the last parenthetical factor by 1.
 Note that also {$\vkneg\approx 0.155$} is essentially determined by the electrical coupling. 

 Of course, equivalently to (\ref{munegATMO}) we may also write the general solution of the negative atmosphere as
\begin{equation}\label{munegATMOalt}
\uneg(\rho) = \Aneg^+ \frac{\cosh(\vkneg \rho)}{\rho} +\Aneg^- \frac{\sinh(\vkneg \rho)}{\rho}.
\end{equation}

 The two atmospheric amplitudes $\Aneg^+$ and $\Aneg^-$ are constrained by the requirement that the 
electron density $\uneg(\rho)$ be continuously differentiable at the boundary $\rho=\rho_0$ of the bulk region.

 We will now address this matching of a positive or negative atmosphere to the bulk region.
\vspace{-10pt}

%%%%%%%%%%%%%%%%%%%%%%%%%%%%%%%%%%%%%%%%%%%%%%%%%%%%%%%%%%%%%% 
%%%%%%%%%%%%%%%%%%%%%%%%%%%%%%%%%%%%%%%%%%%%%%%%%%%%%%%%%%%%%%
                \subsection{The bulk-atmosphere interface}
%%%%%%%%%%%%%%%%%%%%%%%%%%%%%%%%%%%%%%%%%%%%%%%%%%%%%%%%%%%%%%
%%%%%%%%%%%%%%%%%%%%%%%%%%%%%%%%%%%%%%%%%%%%%%%%%%%%%%%%%%%%%%

 Having obtained the general physical solution type in the bulk region and the general physical solution type
in the atmosphere region, which can be either an electron or a proton atmosphere, we now match these
general solutions at their common bulk-atmosphere interface. 
 Both cases, positive and negative atmosphere, can be discussed in parallel.

 In the bulk region the two density functions together feature four amplitudes, but Eqs.(\ref{eq:BnegOverBposHYPO}) and 
(\ref{eq:BnegOverBposTRIG}) express the two electron amplitudes in terms of the two pertinent proton amplitudes, 
or the other way round. 
 The density function of the atmosphere-forming species features two further amplitudes in the atmosphere region.
 It has to vary continuously differentiably across the boundary $\rho_0$ of the bulk region, 
where the other density reaches zero. 
 In each case, whether the atmosphere consists of protons or of electrons, the requirement that the
atmosphere-forming density function $\upsilon_{\text{f}}(\rho)$ is continuously differentiable at the boundary $\rho=\rho_0$ 
of the bulk region allows us to express the two amplitudes of the density function $\upsilon_{\text{f}}(\rho)$
in the atmosphere region in terms of its two amplitudes in the bulk region.

 We explain the procedure using the positive atmosphere case. 
 The negative atmosphere case is completely analogous, and we will only state its final formulas. 

 The boundary $\rho_0^{}$ of the bulk region of a white dwarf star with positive atmosphere is determined by 
the vanishing of the right-hand side of (\ref{bulkNEG}), and cancelling $1/\rho_0^{}$ this yields 
\begin{equation}\label{xiNULLposATMO}
\Bneg^h \sinh(\kappa_h^{}\rho_0) + \Bneg^t \sin(\kappa_t^{}\rho_0) =0,
\end{equation}
where $\Bneg^h\propto \Bpos^h$ and  $\Bneg^t\propto \Bpos^t$; see Eqs.(\ref{eq:BnegOverBposHYPO}) and (\ref{eq:BnegOverBposTRIG}).
 This is an implicit equation for $\rho_0^{}$, given ${\Bneg^h}$ and ${\Bneg^t}$ (equivalently: given ${\Bpos^h}$ and ${\Bpos^t}$),
which can be easily solved numerically on a computer, but generally not in a closed form. 
 It should be noted, though, that Eq.(\ref{xiNULLposATMO}) permits $\Bneg^h$ to vanish (in which case also $\Bpos^h$
vanishes, by (\ref{eq:BnegOverBposHYPO})), given any $\Bneg^t>0$ (equivalently, given $\Bpos^t>0$), namely when
$\rho_0^{}=\pi/\kappa_t^{}$.
 This is perhaps the only case in which $\rho_0^{}$ is explicitly obtained from the bulk amplitudes, i.e. from $\Bneg^h=0$.
 We have already remarked earlier that only positive trigonometric bulk amplitudes are permitted, due to 
the requirement that the bulk densities must not be negative.

 At this point, a change of perspective will yield a decisive simplification:
From Eq.(\ref{xiNULLposATMO}) we obtain 
\begin{equation}\label{BratioASfctOFxiNULL}
\frac{\Bneg^h}{\Bneg^t} = -\frac{ \sin(\kappa_t^{}\rho_0)}{ \sinh(\kappa_h^{}\rho_0)}.
\end{equation}
 \emph{We will think of (\ref{BratioASfctOFxiNULL}) as yielding
the ratio ${\Bneg^h}/{\Bneg^t}$ (equivalently: ${\Bpos^h}/{\Bpos^t}$) explicitly as function of $\rho_0^{}$,
and hence treat the interface location  $\rho_0^{}$ as independent parameter.}

 Coming now to the matching of atmospheric amplitudes with the bulk amplitudes, we note that
for the protons we have, first of all, the continuity of their density function $\upos(\rho)$ at $\rho=\rho_0^{}$, 
which (after cancelling $1/\rho_0^{}$) yields
\begin{eqnarray}
\label{continuityA}
\Bpos^h \sinh(\kappa_h^{}\rho_0^{}) + \Bpos^t \sin(\kappa_t^{}\rho_0^{})= \\
\Apos^+ e^{\vkpos \rho_0^{}} +\Apos^- e^{-\vkpos \rho_0^{}},\quad \notag
\end{eqnarray}
equivalently,
\begin{eqnarray}
\label{continuityAA}
\Bpos^h \sinh(\kappa_h^{}\rho_0^{}) + \Bpos^t \sin(\kappa_t^{}\rho_0^{})= \\
\Apos^+ \cosh(\vkpos \rho_0^{}) +\Apos^- \sinh(\vkpos \rho_0^{}).\quad \notag
\end{eqnarray}
 Second, we need the continuity of the derivative of  their density function $\upos(\rho)$ at $\rho=\rho_0^{}$.
 By the product rule, the $\rho$-derivative of each term in the general solution is a sum of the $\rho$-derivative
of the numerator, divided by $\rho$, plus the numerator times the derivative of $1/\rho$.
 Yet all terms proportional to the derivative of $1/\rho$ can be grouped together and, 
with the help of (\ref{continuityA}), this group can be seen to vanish by itself.
 Thus, and after cancelling the remaining overall factor $1/\rho_0^{}$, continuity 
of the $\rho$-derivative of $\upos(\rho)$ at $\rho=\rho_0^{}$ yields
\begin{eqnarray}\label{continuityB}
\Bpos^h \kappa_h^{}\cosh(\kappa_h^{}\rho_0^{}) + \Bpos^t \kappa_t^{} \cos(\kappa_t^{}\rho_0^{})= \\
\Apos^+ \vkpos e^{\vkpos \rho_0^{}} - \Apos^- \vkpos e^{-\vkpos \rho_0^{}},\quad \notag
\end{eqnarray}
equivalently
\begin{eqnarray}\label{continuityBB}
\Bpos^h \kappa_h^{}\cosh(\kappa_h^{}\rho_0^{}) + \Bpos^t \kappa_t^{} \cos(\kappa_t^{}\rho_0^{})= \\
\Apos^+ \vkpos \sinh(\vkpos \rho_0^{}) + \Apos^- \vkpos \cosh(\vkpos \rho_0^{}).\quad \notag
\end{eqnarray}

 Using either the pair of equations (\ref{continuityA}), (\ref{continuityB}), or the pair 
(\ref{continuityAA}), (\ref{continuityBB}), we can write a linear transformation from the pair of $\Bpos$ amplitudes
to the pair of $\Apos$ amplitudes. 
 We choose the pair (\ref{continuityA}), (\ref{continuityB}) and obtain
\begin{eqnarray}
\left(\!\!
\begin{array}{cc}
\qquad \sinh(\kappa_h^{}\rho_0^{}) ; &\quad \sin(\kappa_t^{}\rho_0^{}) \\
\; \kappa_h^{}\cosh(\kappa_h^{}\rho_0^{})\; ; &\!  \kappa_t^{} \cos(\kappa_t^{}\rho_0^{})
\end{array}
\!\right)\!\! 
\left(
\begin{array}{c} \!\Bpos^h\!  \\ \!\Bpos^t\! \end{array} \right) 
= \label{eq:matrixPROBLEMb}
\ \\ 
\left( \!\!
\begin{array}{cc} 
\qquad\exp(\vkpos \rho_0^{})  ; &\quad\exp(-\vkpos \rho_0^{}) \\
\; \vkpos \exp(\vkpos \rho_0^{})\;; &\! -\vkpos \exp(-\vkpos \rho_0^{})
\end{array}
\!\right)\!\!
\left(
\begin{array}{c}\!\Apos^+\!  \\ \!\Apos^-\! 
\end{array}
\right).\notag\quad
\end{eqnarray}
 This linear transformation is valid as long as the left- (and therefore the right-)hand side of 
Eq.(\ref{continuityA}) is strictly positive, as required for having a positive atmosphere.

 We note that the determinant
of the coefficient matrix at the right-hand side of Eq.(\ref{eq:matrixPROBLEMb}) equals $-2\vkpos<0$, and
therefore the matrix is always invertible and the pair $(\Apos^+,\Apos^-)$ is uniquely given by (\ref{eq:matrixPROBLEMb}) 
in terms of the pair $(\Bpos^h,\Bpos^t)$, for any physically meaningful choice of $\rho_0^{}>0$.
 How to choose the physically meaningful $\rho_0^{}$ we work out in the next subsection.
 But first we list the analogous formulas for the case of a star with a negative atmosphere.

 The pertinent formulas are easily obtained 
from the formulas of the positive atmosphere setting.
 Thus, given ${\Bpos^h}/{\Bpos^t}$ (equivalently: given ${\Bneg^h}/{\Bneg^t}$), 
from the vanishing of the right-hand side of (\ref{bulkPOS}), and after cancelling $1/\rho_0^{}$, we obtain
\begin{equation}\label{BratioASfctOFxiNULLagain}
\frac{\Bpos^h}{\Bpos^t} = -\frac{ \sin(\kappa_t^{}\rho_0)}{ \sinh(\kappa_h^{}\rho_0)}.
\end{equation}
 Moreover, we now obtain the linear relationship 
\begin{eqnarray}
\left(\!\!
\begin{array}{cc}
\qquad \sinh(\kappa_h^{}\rho_0^{}) ; &\quad \sin(\kappa_t^{}\rho_0^{}) \\
\; \kappa_h^{}\cosh(\kappa_h^{}\rho_0^{})\; ; &\!  \kappa_t^{} \cos(\kappa_t^{}\rho_0^{})
\end{array}
\!\right)\!\!
\left(
\begin{array}{c}
\!\Bneg^h\!  \\
\!\Bneg^t\!
\end{array}
\right)
= \label{eq:matrixPROBLEMbb}
\ \\
\left(\!\!
\begin{array}{cc}
\qquad\exp(\vkneg \rho_0^{})  ; &\quad\exp(-\vkneg \rho_0^{}) \\
\; \vkneg \exp(\vkneg \rho_0^{})\;; &\! -\vkneg \exp(-\vkneg \rho_0^{})
\end{array}
\!\right)\!\!
\left(
\begin{array}{c}
\!\Aneg^+\!  \\
\!\Aneg^-\!
\end{array}
\right)\notag\quad
\end{eqnarray}
between the $\Bneg$ and $\Aneg$ amplitudes. 
 This linear transformation is valid as long as $\uneg(\rho_0^{})>0$, as required for having a negative atmosphere.
\vspace{-10pt}

%%%%%%%%%%%%%%%%%%%%%%%%%%%%%%%%%%%%%%%%%%%%%%%%%%%%%%%%%%%%%% 
%%%%%%%%%%%%%%%%%%%%%%%%%%%%%%%%%%%%%%%%%%%%%%%%%%%%%%%%%%%%%%
                \subsection{Two intervals of admissible $\rho_0^{}$ values}
%%%%%%%%%%%%%%%%%%%%%%%%%%%%%%%%%%%%%%%%%%%%%%%%%%%%%%%%%%%%%%
%%%%%%%%%%%%%%%%%%%%%%%%%%%%%%%%%%%%%%%%%%%%%%%%%%%%%%%%%%%%%%
 
By now we have determined the density functions $\upos(\rho)$ and $\uneg(\rho)$ of the
two-species $6/3$-model uniquely in terms of three parameters: (i) a choice of sign, as to whether the positive or negative
species defines the bulk {radius}, (ii) the location $\rho_0^{}$ of the interface between bulk region and atmosphere,
and (iii) the positive trigonometric bulk amplitude $B^t$ of the species defining the bulk {radius.}
 However, the resulting solution may not be integrable to yield finite total number of particles $\Npos$ and $\Nneg$. 
 The requirement that it should determines the physically allowed interval of $\rho_0^{}$ values in the positive and negative amplitude situation.
 We note that similarly to the $n=1$ polytropic single-density model, the value of the
 trigonometric amplitude $B^t>0$ is chosen independently of $\rho_0$.

 Again, having the answer worked out for the case of a star with a positive atmosphere, the answer for a star with 
a negative atmosphere will follow by dictionary.

 Therefore, assume that the star has a positive atmosphere. 
 Then $\rho_0^{}$ is the point where the electron bulk density $\uneg(\rho)$ has declined to zero. 
 We already know from our discussion that the trigonometric mode of the bulk regime essentially captures 
the density distribution, so $\Bneg^t>0$.
 Moreover, from  (\ref{BratioASfctOFxiNULL}) we see that $\Bneg^h<0$ if $\rho_0^{}<\pi/\kappa^{}_t$, and
$\Bneg^h>0$ if $\rho_0^{} >\pi/\kappa^{}_t$, with $\Bneg^h=0$ if $\rho_0^{}=\pi/\kappa^{}_t$. 
 By (\ref{eq:BnegOverBposHYPO}), (\ref{eq:BnegOverBposTRIG}), then also $\Bpos^t>0$,
while $\Bpos^h$ and  $\Bneg^h$ have opposite signs, except when both vanish.

 Of course, the case $\rho_0^{}=\pi/\kappa_t^{}$ which leads to $\Bneg^h=0=\Bpos^h$ is the case without 
atmosphere at all, and the bulk densities $\upos(\rho)$ and $\uneg(\rho)$ are then given by essentially the same 
Lane--Emden $n=1$ polytrope as in the single-density approximation, (\ref{eq:LaneEmdenONEsol}),
except for minute differences in the parameter values. 
 Therefore, to have a non-empty atmosphere we need to consider $\rho_0^{}\neq \pi/\kappa_t^{}$.
 In fact, we will need $\rho_0^{}< \pi/\kappa_t^{}$.

 Indeed, if $\rho_0^{}< \pi/\kappa_t^{}$, then since $\Bneg^t>0$ we have
$\Bneg^h<0$ by (\ref{BratioASfctOFxiNULL}), and therefore now both $\Bpos^t>0$ and $\Bpos^h>0$,
 by (\ref{eq:BnegOverBposHYPO}) and (\ref{eq:BnegOverBposTRIG}).
 Now, by assumption $\uneg(\rho_0^{})=0$, but $\upos(\rho_0^{})$ is the same linear combination of the 
$\Bpos$ amplitudes as $\uneg(\rho_0^{})$ is of the $\Bneg$ amplitudes, with $\Bpos^t\approx\Bneg^t>0$ yet 
$\Bpos^h>0$ while $\Bneg^h<0$, and so we conclude that $\upos(\rho_0^{}) >0$, as claimed. 

 Proceeding analogously when $\rho_0^{}>\pi/\kappa_t^{}$, we find that now both
$\Bneg^t>0$ and $\Bneg^h>0$ by (\ref{BratioASfctOFxiNULL}), and therefore now $\Bpos^t>0$ while $\Bpos^h<0$. 
 Thus, since {by assumption} $\uneg(\rho_0)=0$ with two positive amplitudes, the left-hand side 
of (\ref{bulkPOS}) with one positive and one negative amplitude evaluated at $\rho_0^{}$ is actually negative, 
in violation of the requirement that particle densities cannot be negative. 
 Thus a positive atmosphere is not possible with $\rho_0^{}>\pi/\kappa_t^{}$, which
cannot be a zero of $\uneg(\rho)$ in the bulk. 

 Next, since $\upos(\rho_0)>0$ in the case of a positive-atmosphere star, it is clear that 
$\Apos^+$ and $\Apos^-$ cannot both be {(strictly) positive or both be negative:
 two negative $\Apos$ amplitudes cannot produce a strictly positive particle atmospheric density.
 Two strictly positive $\Apos$ amplitudes do yield a positive particle density, but this density grows rapidly beyond
any upper bound and cannot integrate to a finite particle number.}
 On the other hand, the combination $A_{\text{f}}^+\leq 0$ and $A_{\text{f}}^->0$ is manifestly admissible, 
for it will always lead to an atmospheric density function $\upsilon^{}_{\text{f}}(\rho)$ which is integrable.
 
 We now rule out the combination $A_{\text{f}}^+ > 0$ and $A_{\text{f}}^- < 0$.
 It suffices to discuss one of these cases, for the other follows by analogy.

 Thus, consider the positive atmosphere. 
 Suppose $\Apos^+>0$ and $\Apos^-<0$. 
 Recall that $\rho>\rho_0^{}$ in the atmosphere, and that $\rho_0^{}\approx \pi/\kappa_t^{}\gg 1$ is huge. 
 Now {$\vkpos\approx 6.6$}, so $\vkpos\rho\gg 1$ is also huge, and therefore 
 the function $\rho\mapsto (\Apos^+ e^{\vkpos \rho} +\Apos^- e^{-\vkpos \rho})/\rho$ is increasing for
all $\rho>\rho_0^{}$. 
 Since it has to be positive at $\rho=\rho_0^{}$, it cannot be integrable over $\rho>\rho_0^{}$, 
which finishes the argument. 

 A similar reasoning rules out the combination $\Aneg^+>0$ and $\Aneg^-<0$. 
 Even though {$\vkneg\approx 0.155$} %0.23$ 
is smaller than 1, $\vkneg\rho$ is still so huge for $\rho>\rho_0^{}$  
that the function $\rho\mapsto (\Aneg^+ e^{\vkneg \rho} +\Aneg^- e^{-\vkneg \rho})/\rho$ is increasing for
all $\rho>\rho_0^{}$. 

 This proves that  $A_{\text{f}}^+ > 0$ \&\ $A_{\text{f}}^- < 0$ is not allowed.
\smallskip

 Thus the only possible combinations are $A_{\text{f}}^+\leq 0$ \&\ $A_{\text{f}}^->0$.
 The extremal case $\Apos^+ = 0$ and $\Apos^- >0$ defines the lower limit $\rho_0^{-}$ of the bulk boundary $\rho_0^{}$ 
if $\rho_0$ is {the zero of $\upsilon_e(\rho)$}. 
 It is straightforward to work out the equation defining $\rho_0^{-}$, and while it contains only simple
elementary functions, it is {transcendental and} cannot be solved in closed form.
 However, because of the fantastically tiny ratios of the gravitational to electric coupling constants, a
very accurate approximate expression for $\rho_0^{-}$ can be found in terms of simple elementary functions
{(see our subsection to this subsection below)}.
 It reads 
\begin{equation}\label{xiLOWERneg}
\rho_0^- \approx \frac{\pi}{\kappa_t^{}} - \frac{1}{(1-q)\vkpos -q\kappa_h},
\end{equation}
where
\begin{equation}\label{eq:q}
q
= 
\frac{1-\tfrac{G\mPR^2}{e^2} +\veps\varsigma \kappa^2_t}{1-\tfrac{G\mPR^2}{e^2} -\veps\varsigma \kappa^2_h}
\approx -1836.\quad
\end{equation}
 Note that $\kappa_t^{}\rho_0^+$ is just barely smaller than $\pi$.

 The discussion for a negative-atmosphere star mirrors the one for the positive-atmosphere star.
 Thus the only allowed combinations are $\Aneg^+\leq 0$ and $\Aneg^->0$. 
 Analogously to our computation in the positive atmosphere case we now find (see below)
\begin{equation}\label{xiLOWERpos}
\rho_0^+ \approx \frac{\pi}{\kappa_t^{}} - \frac{q}{(q-1)\vkneg -\kappa_h};
\end{equation}
also $\kappa_t^{}\rho_0^+$ is just barely smaller than $\pi$.
\smallskip

 We summarize: the bulk radii $\rho_0^\pm$ are defined as the smallest possible zeros of the positive,
 respectively negative species in a solution pair. 
 The ranges $[\rho_0^\pm,\pi/\kappa_t^{}]$ of possible bulk radii are very tiny intervals 
to the left of the no-atmosphere value $\rho_0=\pi/\kappa_t^{}$, relative to that value.
 No bulk radius is bigger than $\pi/\kappa_t^{}$.
 Since $\kappa_t^{}$ agrees nearly perfectly with the single-density model value $\kappa$ given by (\ref{eq:LaneEmdenONEb}),
the bulk radii of all the failed white dwarf stars in the $5/3\to 6/3$ approximation are essentially given by (\ref{R}).
 However, the atmosphere of a star can nevertheless be very extended.
 In particular, in the two extreme cases the atmosphere extends all the way out to infinity, yet with its density approaching
zero exponentially fast.  \vspace{-10pt}

%%%%%%%%%%%%%%%%%%%%%%%%%%%%%%%%%%%%%%%%%%%%%%%%%%%%%%%%%%%%%%%%%%%%%%%%%%%%%%%%%
%%%%%%%%%%%%%%%%%%%%%%%%%%%%%%%%%%%%%%%%%%%%%%%%%%%%%%%%%%%%%%%%%%%%%%%%%%%%%%%%%
\subsubsection{Computing $\rho_0^\pm$}\vspace{-5pt}
%%%%%%%%%%%%%%%%%%%%%%%%%%%%%%%%%%%%%%%%%%%%%%%%%%%%%%%%%%%%%%%%%%%%%%%%%%%%%%%%%
%%%%%%%%%%%%%%%%%%%%%%%%%%%%%%%%%%%%%%%%%%%%%%%%%%%%%%%%%%%%%%%%%%%%%%%%%%%%%%%%%

 In the case of an extreme negative atmosphere, $\rho_0^{+}<\pi/\kappa_t^{}$ is determined by the
matching of the bulk part of $\uneg(\rho)$ with its atmospheric part in the limiting case where $\Aneg^+=0$.
 So from (\ref{eq:matrixPROBLEMbb}) we obtain
\begin{equation}\label{eq:matrixPROBLEMbbEXTREME}
\left(\!\!\!
\begin{array}{cc}
\quad\, \sinh(\kappa_h^{}\rho_0^{})\; ; &\quad \sin(\kappa_t^{}\rho_0^{}) \\
\; \kappa_h^{}\!\cosh(\kappa_h^{}\rho_0^{}) ; &\!  \kappa_t^{}\! \cos(\kappa_t^{}\rho_0^{})
\end{array}
\!\!\right)\!\!
\left(
\begin{array}{c}
\!\!\Bneg^h\!\!  \\
\!\!\Bneg^t\!\!
\end{array}
\right)
= 
\Aneg^- e^{-\vkneg \rho_0^{}}\!
\left(
\begin{array}{c}
\!\! 1\! \!  \\
\!\!-\vkneg\! \!
\end{array}
\right)\!,
\end{equation}
and these are two different equations for $\Aneg^- \exp(-\vkneg \rho_0^{})$. 
 Elimination of $\Aneg^- \exp(-\vkneg \rho_0^{})$ now yields, after some simple manipulations,
\begin{eqnarray} \label{eq:bbEXreshuffle}
 \frac{\Bneg^h}{\Bneg^t} 
\sinh(\kappa_h^{}\rho_0^{}) + \sin(\kappa_t^{}\rho_0^{}) =\\
-\frac{\kappa_h^{}}{\vkneg}\frac{\Bneg^h}{\Bneg^t}\cosh(\kappa_h^{}\rho_0^{})- \frac{\kappa_t^{}}{\vkneg}\cos(\kappa_t^{}\rho_0^{}).
\notag
\end{eqnarray}
 With the help of Eqs.(\ref{eq:BnegOverBposHYPO}), (\ref{eq:BnegOverBposTRIG}), and 
{(\ref{BratioASfctOFxiNULLagain})} we find 
\begin{equation} \label{eq:Bratio}
 \frac{\Bneg^h}{\Bneg^t} = -\frac1q \frac{ \sin(\kappa_t^{}\rho_0)}{ \sinh(\kappa_h^{}\rho_0)},
\end{equation}
with $q$ given in (\ref{eq:q}).
{Note that (\ref{eq:Bratio}) is not in contradiction to (\ref{BratioASfctOFxiNULL}), for 
(\ref{eq:Bratio}) holds for the extreme negative atmosphere, while (\ref{BratioASfctOFxiNULL}) holds for any positive atmosphere.}
 Substituting (\ref{eq:Bratio}) in (\ref{eq:bbEXreshuffle}), dividing by $\sin(\kappa_t^{}\rho_0^{})$, and 
reshuffling now yields
\begin{equation} \label{eq:bbEXreshuffleB}
(q-1)\vkneg = \kappa_h^{}\coth(\kappa_h^{}\rho_0^{})- q\kappa_t^{}\cot(\kappa_t^{}\rho_0^{})
\end{equation}
for the lower limit $\rho_0^+<\pi/\kappa_t^{}$ of the zero of the bulk density $\upos(\rho)$.
 {Since $\coth$ is a monotonic decreasing function on the positive real line
and $\cot$ is a monotonic decreasing function on its first positive period, and since $q<0$, we see that 
the right-hand side of (\ref{eq:bbEXreshuffleB}) is a strictly monotonic decreasing function in
the interval $0<\kappa_t^{}\rho_0^{}<\pi$, thus it has a unique solution $\rho_0^+$.
  With the values of the parameters $q$, $\kappa_t^{}$, $\kappa_h^{}$, and $\vkneg$ as given, 
this solution is in the left vicinity of $\rho_0^{}=\pi/\kappa_t^{}$.}
 Recall that {$\kappa_h^{}\approx 6.6$ and $\kappa_t^{}\approx 2\cdot 10^{-19}$}. 
 Thus, if $\kappa_t^{}\rho_0^+\approx \pi$, then $\kappa_h^{}\rho_0^+\gg \pi$ is huge, and 
then $\coth(\kappa_h^{}\rho_0^{+})\approx 1$ asymptotically exact, with exponentially small corrections. 
 Moreover, in the left  vicinity of $\rho_0^{}=\pi/\kappa_t^{}$ we have
$\cot(\kappa_t^{}\rho_0^{})\approx 1/(\kappa_t^{}\rho_0^{}-\pi)<0$ asymptotically exact, and this
yields (\ref{xiLOWERpos}).

 Analogously  we handle the case of an extreme positive atmosphere, where
$\rho_0^{-} < \pi/\kappa_t^{}$ is determined by the
matching of the bulk part of $\upos(\rho)$ with its atmospheric part in the limiting case where $\Apos^+=0$.
{This time
\begin{equation} \label{eq:BratioPOS}
 \frac{\Bpos^h}{\Bpos^t} = - q \frac{ \sin(\kappa_t^{}\rho_0)}{ \sinh(\kappa_h^{}\rho_0)},
\end{equation}
with $q$ given in (\ref{eq:q}).
 Also (\ref{eq:BratioPOS}) is not in contradiction to (\ref{BratioASfctOFxiNULLagain}), for  
(\ref{eq:BratioPOS}) holds for the extreme positive atmosphere, while (\ref{BratioASfctOFxiNULLagain}) 
holds for any negative atmosphere.}
 We find 
\begin{equation} \label{eq:bbEXreshuffleC}
(1-q)\vkpos = q\kappa_h^{}\coth(\kappa_h^{}\rho_0^{})- \kappa_t^{}\cot(\kappa_t^{}\rho_0^{})
\end{equation}
for the lower limit $\rho_0^-<\pi/\kappa_t^{}$ of the zero of the bulk $\uneg(\rho)$.
 The right-hand side of (\ref{eq:bbEXreshuffleC}) is a strictly monotonic increasing 
function in $\rho_0$ {in the first positive period of the $\cot$ function}, with 
a solution in the left vicinity of $\rho_0^{}=\pi/\kappa_t^{}$.
 Using that $\kappa_h^{}\rho_0^{}\gg \pi$ is huge we again can set
$\coth(\kappa_h^{}\rho_0^{})\approx 1$ asymptotically exact, with exponentially small corrections. 
 Moreover, in the left vicinity of $\rho_0^{}=\pi/\kappa_t^{}$ we have
$\cot(\kappa_t^{}\rho_0^{})\approx 1/(\kappa_t^{}\rho_0^{}-\pi)<0$ asymptotically exact, and this now
yields (\ref{xiLOWERneg}).
\vspace{3pt} 

%%%%%%%%%%%%%%%%%%%%%%%%%%%%%%%%%%%%%%%%%%%%%%%%%%%%%%%%%%%%%%%%%%%%%%%%%%%%%%%%%
%%%%%%%%%%%%%%%%%%%%%%%%%%%%%%%%%%%%%%%%%%%%%%%%%%%%%%%%%%%%%%%%%%%%%%%%%%%%%%%%%
\subsection{Computing the ratio ${\Nneg}/{\Npos}$ as function of $\rho_0^{}$}\vspace{-10pt}
%%%%%%%%%%%%%%%%%%%%%%%%%%%%%%%%%%%%%%%%%%%%%%%%%%%%%%%%%%%%%%%%%%%%%%%%%%%%%%%%%
%%%%%%%%%%%%%%%%%%%%%%%%%%%%%%%%%%%%%%%%%%%%%%%%%%%%%%%%%%%%%%%%%%%%%%%%%%%%%%%%%

 As in any set of homogeneous linear equations, so also in the $6/3$ model there is an amplitude invariance,
i.e. if $(\nupos,\nuneg)$ is a solution pair, then so is $(\lambda\nupos,\lambda\nuneg)$ for any $\lambda$
 (with $\lambda >0$ to be meaningful). 
 Therefore there is no such thing as the number of protons $\Npos$ and the number of electrons $\Nneg$ associated
with a solution. 
 Incidentally, although any total number of particles is mathematically allowed in this linear model,
as explained in the introduction, a failed white dwarf is a low-mass star, and since the mass is 
essentially given by the number of protons, $\Npos$ should be restricted to about $1.5\cdot 10^{55}$ to $9\cdot 10^{55}$ 
protons to be physically meaningful.

 The ratio ${\Nneg}/{\Npos}$ is a well-defined quantity associated with any solution pair $(\nupos,\nuneg)$, though.
 Given the choice of either a positive or a negative atmosphere, the ratio $\Nneg/\Npos$ is uniquely 
determined by the allowed values of the bulk boundary location $\rho_0^{}$. 
 Its computation as a function of $\rho_0^{}$ can be effected 
by directly integrating the explicit solutions parameterized by $\rho_0^{}$.
 Yet it is easier to work directly with the differential equations.

 Starting with the case of a negative atmosphere, we multiply Eq.(\ref{eq:PoissonMUe}) 
with $4\pi\rho^2$ and integrate from $0$ to $\rho_e^{}$, obtaining
\begin{eqnarray}\hspace{-10pt}
 \varsigma 4\pi \rho_e^2\uneg^{\prime}(\rho_e^{})
 =  \left(1 - \tfrac{G\mEL^2}{e^2}\right) \Nneg - \left(1 + \tfrac{G\mPR\mEL}{e^2}\right) \Npos,
 \label{eq:PoissonMUeINTagainB}
\end{eqnarray}
where $\uneg^{\prime}(\rho_e^{})$ is the left-derivative of $\uneg(\rho)$ at $\rho=\rho_e^{}$.
 We next complement (\ref{eq:PoissonMUeINTagainB}) by deriving its counterpart for the positive species. 
 Thus we multiply Eq.(\ref{eq:PoissonMUp}) with $4\pi\rho^2$ and integrate from $0$ to $\rho_0^{}$, obtaining
\begin{widetext}
\begin{eqnarray}
- \veps \varsigma 4\pi \rho_0^2\upos^{\prime}(\rho_0^{})
  =  - \left(1 -\tfrac{G\mPR^2}{e^2}\right) \Npos
+ \left(1 +\tfrac{G\mPR\mEL}{e^2}\right) \displaystyle\int_0^{\rho_0}\uneg(\rho)4\pi\rho^2\drm\rho.
 \label{eq:PoissonMUpINTagain}
\end{eqnarray}
 Here, $\upos^{\prime}(\rho_0^{})$ is the left-derivative of $\upos(\rho)$ at $\rho=\rho_0^{}$.
 Noting that $\int_0^{\rho_0} \uneg(\rho)4\pi\rho^2\drm\rho + \int_{\rho_0}^{\rho_e}\uneg(\rho)4\pi\rho^2\drm\rho = \Nneg$,
we multiply Eq.(\ref{eq:PoissonMUePOSnull}) with $4\pi\rho^2$ and integrate from $\rho_0^{}$
to $\rho_e^{}$, the point where $\uneg(\rho)$ has decreased to zero.
 This yields
\begin{eqnarray}
4\pi \varsigma \left( \rho_0^2\uneg^{\prime}(\rho_0^{}) - \rho_e^2\uneg^{\prime}(\rho_e^{})\right)
  =  - {\textstyle\left(1 -\frac{G\mEL^2}{e^2}\right)} 
\int_{\rho_0}^{\rho_e}\uneg(\rho)4\pi\rho^2\drm\rho,
 \label{eq:PoissonMUePOSnullINTagain}
\end{eqnarray}
 Now we multiply (\ref{eq:PoissonMUePOSnullINTagain}) by 
$\left(1 +\tfrac{G\mPR\mEL}{e^2}\right)/\left(1 -\frac{G\mEL^2}{e^2}\right)$
and subtract the result from (\ref{eq:PoissonMUpINTagain}), which yields
\begin{eqnarray}
4\pi \varsigma \left[\left( \rho_0^2\uneg^{\prime}(\rho_0^{}) - \rho_e^2\uneg^{\prime}(\rho_e^{})\right)
\frac{1 +\tfrac{G\mPR\mEL}{e^2}}{1 -\frac{G\mEL^2}{e^2}}
+ \veps \rho_0^2\upos^{\prime}(\rho_0^{})\right]
 =   \left(1 -\tfrac{G\mPR^2}{e^2}\right) \Npos - \left(1 +\tfrac{G\mPR\mEL}{e^2}\right) \Nneg.
 \label{eq:PoissonMUpINTagainB}
\end{eqnarray}
 Eqs.(\ref{eq:PoissonMUeINTagainB}) and (\ref{eq:PoissonMUpINTagainB}) form a linear system for $\Npos$ and $\Nneg$ 
in terms of their coefficients and their left-hand sides. 
 This linear system is easily solved formally for $\Npos$ and $\Nneg$, from which we obtain $\Nneg/\Npos$.
 Symbolically, 
\begin{equation}
\left(
\begin{array}{c}
\!\Npos\!  \\
\!\Nneg\!
\end{array}
\right)
=
4\pi \varsigma 
\left(
\begin{array}{cc}
\quad 1 -\tfrac{G\mPR^2}{e^2}\quad; &  - 1 -\tfrac{G\mPR\mEL}{e^2} \\
- 1 -\tfrac{G\mPR\mEL}{e^2}\;\; ; & 1 -\tfrac{G\mEL^2}{e^2} 
\end{array}
\right)^{-1}\!\!
\left(
\begin{array}{c}
\!
\left( \rho_0^2\uneg^{\prime}(\rho_0^{}) - \rho_e^2\uneg^{\prime}(\rho_e^{})\right)
\frac{1 +\tfrac{G\mPR\mEL}{e^2}}{1 -\frac{G\mEL^2}{e^2}}
+ \veps \rho_0^2\upos^{\prime}(\rho_0^{})\!  \\
\! \rho_{e}^2\uneg^{\prime}(\rho_e^{})\!
\end{array}
\right)\!\!,
 \label{eq:matrixNposNnegSOL}
\end{equation}
and the inverse matrix is easily computed as
\begin{equation}
\left(
\begin{array}{cc}
\quad 1 -\tfrac{G\mPR^2}{e^2}\quad; &  - 1 -\tfrac{G\mPR\mEL}{e^2} \\
- 1 -\tfrac{G\mPR\mEL}{e^2}\;\; ; & 1 -\tfrac{G\mEL^2}{e^2} 
\end{array}
\right)^{-1}\!\!
=
\frac{e^2}{G(\mPR+\mEL)^2}
\left(
\begin{array}{cc}
\quad -1 +\tfrac{G\mEL^2}{e^2}\quad; &   -1 -\tfrac{G\mPR\mEL}{e^2} \\
- 1 -\tfrac{G\mPR\mEL}{e^2}\;\; ; & - 1 + \tfrac{G\mPR^2}{e^2} 
\end{array}
\right).
 \label{eq:matrixINV}
\end{equation}
  This gives $(\Npos,\Nneg)$ uniquely in terms of the zeros of the densities and the derivatives at the zeros.
 Recall, though, that a choice of the sign of the atmosphere (negative in this case) plus a choice of $\rho_0$ do not
uniquely determine a solution pair, by the linearity of the equations. 
 If $(\upos,\uneg)$ is a solution pair, then so is $(\lambda\upos,\lambda\uneg)$, and this changes the derivatives
$(\upos^\prime,\uneg^\prime)$ to $(\lambda\upos^\prime,\lambda\uneg^\prime)$ everywhere, and hence also $(\Npos,\Nneg)$ to 
$(\lambda\Npos,\lambda\Nneg)$.
 Therefore (\ref{eq:matrixNposNnegSOL}) does not yield $(\Npos,\Nneg)$ uniquely as function of $\rho_0^{}$ and the sign of
the atmosphere.
 However, the fraction $\Nneg/\Npos$ is scaling-invariant, and uniquely given as function of $\rho_0^{}$ and the sign of
the atmosphere. 
 It reads
\begin{equation}
\frac{\Nneg}{\Npos}
= \left[1 - 
 \frac{\rho_e^2\uneg^{\prime}\left(\rho_e^{}\right)\frac{G(\mPR+\mEL)^2}{e^2}}
{ {\rho_0^2}\uneg^{\prime}(\rho_0^{})\left(1 +\frac{G\mPR\mEL}{e^2}\right)^2 +
\veps{\rho_0^2}\upos^{\prime}(\rho_0^{})\left(1 -\frac{G\mEL^2}{e^2}\right)\left(1 +\frac{G\mPR\mEL}{e^2}\right)}
\right]\frac{1 +\tfrac{G\mPR\mEL}{e^2}}{1 -\frac{G\mEL^2}{e^2}} .
 \label{eq:NnegOverNposATMOneg}
\end{equation}

 Next we compute the pertinent derivatives at $\rho_0^{}$ and $\rho_e^{} = (1/2\vkneg)\ln(-\Aneg^-/\Aneg^+)$.
 We find
\begin{eqnarray}\label{munegPRIMExie}
\uneg^\prime(\rho_e^{}) &=& 
%\vkneg\left(\Aneg^+ \frac{\exp(\vkneg \rho_e)}{\rho_e} - \Aneg^- \frac{\exp(-\vkneg \rho_e)}{\rho_e}\right),
2\Aneg^+\vkneg  \frac{\exp(\vkneg \rho_e)}{\rho_e},\\
\label{munegPRIMExiNULL}
\uneg^\prime(\rho_0^{}) &= &
% \vkneg\left(\Aneg^+ \frac{\exp(\vkneg \rho_0)}{\rho_0} - \Aneg^- \frac{\exp(-\vkneg \rho_0)}{\rho_0}\right) -
% \left(\Aneg^+ \frac{\exp(\vkneg \rho_0)}{\rho_0^2} + \Aneg^- \frac{\exp(-\vkneg \rho_0)}{\rho_0^2}\right),
\left(\vkneg - \frac{1}{\rho_0}\right)\Aneg^+ \frac{\exp(\vkneg \rho_0)}{\rho_0} 
- 
\left(\vkneg + \frac{1}{\rho_0}\right)\Aneg^- \frac{\exp(-\vkneg \rho_0)}{\rho_0} , \\
\label{muposPRIMExiNULL}
\upos^\prime(\rho_0^{}) &=& 
\Bpos^h \kappa_h^{}\frac{\cosh(\kappa_h^{}\rho_0^{})}{\rho_0^{}} + \Bpos^t \kappa_t^{}\frac{\cos(\kappa_t^{}\rho_0^{})}{\rho_0^{}}.\end{eqnarray}
 Since the derivates of the densities enter linearly at the numerator and at the denominator of (\ref{eq:NnegOverNposATMOneg}), 
the expression (\ref{eq:NnegOverNposATMOneg}) is manifestly amplitude-scaling invariant.
 Thus  (\ref{eq:NnegOverNposATMOneg}) is an explicit formula for $\Nneg/\Npos$ as function of $\rho_0$ in the negative atmosphere
regime. 
 We note that the term in square parentheses is smaller than 1, yet converges upward to 1 when 
$\rho_e^2\uneg^\prime(\rho_e^{})\to0$ and $\rho_0^{}\searrow\rho_0^+$. 
 In that case $\Nneg/\Npos$ reaches its upper limit given by the right-hand side of (\ref{eq:NnegOverNposINTERVAL}).
 
 In a similar manner we can treat the case of a positive atmosphere and find
%\begin{equation}
%\frac{\Npos}{\Nneg}  = 
%\frac{\veps\left( \upos^{\prime}(\rho_0^{}) - \frac{\rho_{\mathrm{p}}^2}{\rho_0^2}\upos^{\prime}(\rho_{\mathrm{p}}^{})\right)
%\frac{1 +\tfrac{G\mPR\mEL}{e^2}}{1 -\frac{G\mPR^2}{e^2}} + \uneg^{\prime}(\rho_0^{})
%+ \veps \frac{\rho_{\mathrm{p}}^2}{\rho_0^2}\upos^{\prime}(\rho_{\mathrm{p}}^{})\frac{1 -\frac{G\mEL^2}{e^2}}{1 +\tfrac{G\mPR\mEL}{e^2}}}
%{\veps\upos^{\prime}(\rho_0^{}) +\uneg^{\prime}(\rho_0^{})\frac{1 -\frac{G\mPR^2}{e^2}}{1 +\tfrac{G\mPR\mEL}{e^2}}}.
% \label{eq:NnegOverNposATMOposOLD}
%\end{equation}
%\begin{equation}
%\frac{\Npos}{\Nneg}
%= \frac{\veps\upos^{\prime}(\rho_0^{}) +\uneg^{\prime}(\rho_0^{})\frac{1 -{G\mPR^2}/{e^2}}{1 +{G\mPR\mEL}/{e^2}}
%- \veps\frac{\rho_{\mathrm{p}}^2}{\rho_0^2}\upos^{\prime}\left(\rho_{\mathrm{p}}^{})\right)
% \frac{{G(\mPR+\mEL)^2}/{e^2}}{\left(1 +{G\mPR\mEL}/{e^2}\right)^2}}
%{\veps\upos^{\prime}(\rho_0^{}) +\uneg^{\prime}(\rho_0^{})\frac{1 -{G\mPR^2}/{e^2}}{1 +{G\mPR\mEL}/{e^2}}}
%\frac{1 +\tfrac{G\mPR\mEL}{e^2}}{1 -\frac{G\mPR^2}{e^2}}. \label{eq:NnegOverNposATMOpospos}
%\end{equation}
\begin{equation}
\frac{\Nneg}{\Npos}
= \left[1 - \frac{\veps\rho_{\mathrm{p}}^2 \upos^{\prime}\left(\rho_{\mathrm{p}}^{}\right)\frac{G(\mPR+\mEL)^2}{e^2}}
{ \veps{\rho_0^2}\upos^{\prime}(\rho_0^{})\left(1 +\frac{G\mPR\mEL}{e^2}\right)^2 +
{\rho_0^2}\uneg^{\prime}(\rho_0^{})\left(1 -\frac{G\mPR^2}{e^2}\right)\left(1 +\frac{G\mPR\mEL}{e^2}\right)}
\right]^{-1}\frac{1 -\frac{G\mPR^2}{e^2}}{1 +\tfrac{G\mPR\mEL}{e^2}}.
 \label{eq:NnegOverNposATMOpos}
\end{equation}
 The pertinent derivatives at $\rho_0^{}$ and $\rho_{\mathrm{p}}^{}= (1/2\vkpos)\ln(-\Apos^-/\Apos^+)$ read
\begin{eqnarray}\label{muposPRIMExip}
\upos^\prime(\rho_{\mathrm{p}}^{}) &=& 2 \Apos^+ \vkpos\frac{\exp(\vkpos \rho_{\mathrm{p}})}{\rho_{\mathrm{p}}},\\
\label{muposPRIMExiNULLagain}
\upos^\prime(\rho_0^{}) &=& 
\left(\vkpos - \frac{1}{\rho_0}\right)\Apos^+ \frac{\exp(\vkpos \rho_0)}{\rho_0} 
 - 
\left(\vkpos + \frac{1}{\rho_0}\right)\Apos^- \frac{\exp(-\vkpos \rho_0)}{\rho_0} , \\
\label{munegPRIMExiNULLagain}
\uneg^\prime(\rho_0^{}) &=& 
\Bneg^h \kappa_h^{}\frac{\cosh(\kappa_h^{}\rho_0^{})}{\rho_0^{}} + \Bneg^t \kappa_t^{}\frac{\cos(\kappa_t^{}\rho_0^{})}{\rho_0^{}}.
\end{eqnarray}
\end{widetext}
 Again all amplitudes are proportional to $\Bpos^t$ (equivalently, $\Bneg^t$), 
which actually cancels out from  (\ref{eq:NnegOverNposATMOpos}).
 Thus  (\ref{eq:NnegOverNposATMOpos}) is an explicit formula for $\Nneg/\Npos$ as function of $\rho_0$ in the positive atmosphere
regime.
 We note that the term in square parentheses is smaller than 1, yet converges upward to 1 when 
$\rho_{\mathrm{p}}^2\upos^\prime(\rho_{\mathrm{p}}^{})\to0$ and $\rho_0^{}\searrow\rho_0^-$.
 In that case $\Nneg/\Npos$ reaches its lower limit given by the left-hand side of (\ref{eq:NnegOverNposINTERVAL}).

 Consistency check: when $\rho_0^{}=\pi/\kappa_t^{}$,  then $\rho_e^{}=\rho_{\mathrm{p}}^{}=\rho_0^{}$, and both
(\ref{eq:NnegOverNposATMOpos}) and (\ref{eq:NnegOverNposATMOneg}) reduce to
\begin{equation}\hspace{-5pt}
\frac{\Nneg}{\Npos}
= 
\frac{\uneg^{\prime}(\frac{\pi}{\kappa_t^{}})
\left(1 -\frac{G\mPR^2}{e^2}\right)
+
\veps\upos^{\prime}(\frac{\pi}{\kappa_t^{}})\left(1 +\tfrac{G\mPR\mEL}{e^2}\right)}
{ \uneg^{\prime}(\frac{\pi}{\kappa_t^{}})\left(1 +\tfrac{G\mPR\mEL}{e^2}\right)+ 
\veps\upos^{\prime}(\frac{\pi}{\kappa_t^{}})\left(1 -\frac{G\mEL^2}{e^2}\right)},\hspace{-5pt}
 \label{eq:NnegOverNposNOatmo}
\end{equation}
with the derivatives reducing to $\uneg^{\prime}(\frac{\pi}{\kappa_t^{}}) = -\Bneg^t\kappa_t^2/\pi$ and 
$\upos^{\prime}(\frac{\pi}{\kappa_t^{}}) = -\Bpos^t\kappa_t^2/\pi$.
 Now factoring out $\uneg^{\prime}(\frac{\pi}{\kappa_t^{}})$ from both numerator and
denominator produces the ratio $\upos^{\prime}(\frac{\pi}{\kappa_t^{}})/\uneg^{\prime}(\frac{\pi}{\kappa_t^{}})
=\Bpos^t/\Bneg^t$, cf. (\ref{eq:NnegOverNposNOatmoEXPLICIT}), which can be read of from (\ref{eq:BnegOverBposTRIG}).
 From this expression one then finds that in this special no-atmosphere case the ratio $\Nneg/\Npos< 1$.

 Equations (\ref{eq:NnegOverNposATMOneg}) and (\ref{eq:NnegOverNposATMOpos}) are easy to implement on a computer. 
 We used them to generate Figures \ref{63rhoVsNeOverNp} and \ref{63reVSrp}. %use labels! 
\vspace{-10pt}

%%%%%%%%%%%%%%%%%%%%%%%%%%%%%%%%%%%%%%%%%%%%%%%%%%%%%%%%%%%%%%%%%%%%%%%%%%%%%%%%%
%%%%%%%%%%%%%%%%%%%%%%%%%%%%%%%%%%%%%%%%%%%%%%%%%%%%%%%%%%%%%%%%%%%%%%%%%%%%%%%%%
\subsection{The interval of allowed $\Nneg/\Npos$ ratios}
%%%%%%%%%%%%%%%%%%%%%%%%%%%%%%%%%%%%%%%%%%%%%%%%%%%%%%%%%%%%%%%%%%%%%%%%%%%%%%%%%
%%%%%%%%%%%%%%%%%%%%%%%%%%%%%%%%%%%%%%%%%%%%%%%%%%%%%%%%%%%%%%%%%%%%%%%%%%%%%%%%%

 We now ask: ``What does the $6/3$ model say about 
the possible numbers of electrons per proton, $\Nneg/\Npos$, in a failed white dwarf?'' (cf. \cite{HK}).

 Since the successful single-density models are based on the local neutrality approximation, which
implies $\Npos=\Nneg$, one should expect that any non-neutral pair $(\Npos,\Nneg)$ will have 
a ratio $\Nneg/\Npos\approx 1$ to a high degree of precision.

 It is clear that the extreme values of the ratio $\Nneg/\Npos$ 
will be obtained by inserting the extremal values $\rho_0^\pm$ for $\rho_0^{}$ into the 
$\Nneg/\Npos$ {formulas}, which we have already computed as an elementary function of $\rho_0^{}$.
 However, to obtain these extreme ratios we can resort to a simpler argument, cf. \cite{HK}, for which
we {here can} use that we have full knowledge of the solution family of the 6/3 model.

 Namely, consider the extreme case of a star with negative atmosphere, i.e. $\rho_0^{}=\rho_0^+$.
 We multiply Eq.(\ref{eq:PoissonMUe}) by $4\pi\rho^2$ and integrate over $\rho$ from $0$ to $\infty$.
 (Strictly speaking, (\ref{eq:PoissonMUe}) is a-priori only valid inside the bulk region, but comparison 
with the atmospheric equation (\ref{eq:PoissonMUePOSnull}) reveals that we can extend (\ref{eq:PoissonMUe})
to all $\rho$ by noting that $\upos(\rho)=0$ for $\rho\geq \rho_0^+$.)
 Using that $\int \upos(\rho)d^3\rho =\Npos$ and $\int \uneg(\rho)d^3\rho =\Nneg$, and using
that $\uneg^\prime(0)=0$ and that $\uneg^\prime(\rho)\to0$ exponentially fast when
${\rho\to\infty}$, we obtain
\begin{eqnarray}
0
 = \left(1 + \tfrac{G\mPR\mEL}{e^2}\right) \Npos^- - \left(1 - \tfrac{G\mEL^2}{e^2}\right) \Nneg^-,
 \label{eq:PoissonMUeINT}
\end{eqnarray}
where the negative superscript at $\Npos$ and $\Nneg$ indicates extreme negative atmosphere case.
 Similarly, consider the extreme case of a star with positive atmosphere, i.e. $\rho_0^{}=\rho_0^-$.
 We multiply Eq.(\ref{eq:PoissonMUp}) by $4\pi\rho^2$ and integrate over $\rho$ from $0$ to $\infty$;
again we extend also Eq.(\ref{eq:PoissonMUp}) to all $\rho$ by noting that $\uneg(\rho)=0$ for $\rho\geq \rho_0^-$.
 Using once again that $\int \upos(\rho)d^3\rho =\Npos$ and $\int \uneg(\rho)d^3\rho =\Nneg$, and using
now that $\upos^\prime(0)=0$ and that $\upos^\prime(\rho)\to0$ exponentially fast when
${\rho\to\infty}$, we obtain
\begin{eqnarray}
0
= 
 - \left(1 -\tfrac{G\mPR^2}{e^2}\right) \Npos^+
+ \left(1 +\tfrac{G\mPR\mEL}{e^2}\right) \Nneg^+,
 \label{eq:PoissonMUpINT}
\end{eqnarray}
where the positive superscript at $\Npos$ and $\Nneg$ indicates extreme positive atmosphere case.
 From Eqs.(\ref{eq:PoissonMUeINT}) and (\ref{eq:PoissonMUpINT}) we now obtain the allowed range
of ratios $\Nneg/\Npos$ in the 6/3 model as
\begin{eqnarray}
\boxed{
\frac{1 -\tfrac{G\mPR^2}{e^2}}{1 +\tfrac{G\mPR\mEL}{e^2}}
\leq 
\frac{\Nneg}{\Npos}
\leq 
\frac{1 + \tfrac{G\mPR\mEL}{e^2}}{1 - \tfrac{G\mEL^2}{e^2}}
      }\;.
 \label{eq:NnegOverNposINTERVAL}
\end{eqnarray}

 We have boxed formula (\ref{eq:NnegOverNposINTERVAL}), for it will turn out to be ``universal,'' in a sense
we will explain {next}.

\emph{Note that the bounds (\ref{eq:NnegOverNposINTERVAL}) are independent of $\hbar$ and $c$.} 

%%%%%%%%%%%%%%%%%%%%%%%%%%%%%%%%%%%%%%%%%%%%%%%%%%%%%%%%%%%%%%
%%%%%%%%%%%%%%%%%%%%%%%%%%%%%%%%%%%%%%%%%%%%%%%%%%%%%%%%%%%%%%
%%%%%%%%%%%%%%%%%%%%%%%%%%%%%%%%%%%%%%%%%%%%%%%%%%%%%%%%%%%%%%%%%%%%
\section{``Universality'' of the $\Nneg/\Npos$ bounds}\label{sec:U}\vspace{-5pt}
%%%%%%%%%%%%%%%%%%%%%%%%%%%%%%%%%%%%%%%%%%%%%%%%%%%%%%%%%%%%%%%%%%%%
%%%%%%%%%%%%%%%%%%%%%%%%%%%%%%%%%%%%%%%%%%%%%%%%%%%%%%%%%%%%%%%%%%%%
%%%%%%%%%%%%%%%%%%%%%%%%%%%%%%%%%%%%%%%%%%%%%%%%%%%%%%%%%%%%%%%%%%%%

 We will present a compelling argument for why 
{(\ref{eq:NnegOverNposINTERVAL}) is the correct $\Nneg/\Npos$
interval for a failed white dwarf star made of protons and electrons}, 
and not merely in the non-relativistic theory!

 In the 6/3 model the two extreme values of $\Nneg/\Npos$ are attained by the only two solutions which extend 
all the way out to spatial infinity, and the densities of the infinitely extended extremal atmospheres decay
faster than exponentially to zero when the radial variable goes to infinity.
 All other solutions have density function pairs $(\nupos(r),\nuneg(r))$ which have finite radial extent and
a $\Nneg/\Npos$ ratio sandwiched between the bounds in (\ref{eq:NnegOverNposINTERVAL}).
 This suggests that also among all the solutions of the structure equations of the physically more realistic models 
the solutions with an extreme surplus of charge
are those which have one of their two density functions extend to spatial infinity, 
approaching zero sufficiently rapidly together with its radial derivative so that some surface integrals vanish 
in the limit --- note that we cannot expect a decay to zero to be exponentially fast or even faster;
this is a model-specific detail.
 We will now confirm this. 
 The gist of the discussion can also be found in \cite{HK}. \vspace{-10pt}

%%%%%%%%%%%%%%%%%%%%%%%%%%%%%%%%%%%%%%%%%%%%%%%%%%%%%%%%%%%%%%%%%%%%%%%%%%%%%%%%%
%%%%%%%%%%%%%%%%%%%%%%%%%%%%%%%%%%%%%%%%%%%%%%%%%%%%%%%%%%%%%%%%%%%%%%%%%%%%%%%%%
\subsection{Proof that an atmospheric density has to reach zero with zero slope 
to saturate the bounds (\ref{eq:NnegOverNposINTERVAL})}
\vspace{-10pt}
%%%%%%%%%%%%%%%%%%%%%%%%%%%%%%%%%%%%%%%%%%%%%%%%%%%%%%%%%%%%%%%%%%%%%%%%%%%%%%%%%
%%%%%%%%%%%%%%%%%%%%%%%%%%%%%%%%%%%%%%%%%%%%%%%%%%%%%%%%%%%%%%%%%%%%%%%%%%%%%%%%%

 For simplicity we present the proof for the $5/3$ model, but it will be clear from the proof how to 
adjust it to also apply to the special-relativistic failed white dwarf model.

 Starting with the case of a negative atmosphere, we multiply Eq.(\ref{eq:PoissonMUeTHREEhalf}) 
with $4\pi r^2$ and integrate from $0$ to $ r_e^{}$, the point where the density $\nuneg( r)$ reaches $0$, 
obtaining
\begin{equation}
4\pi \zeta   r_e^2{\nuneg^{\frac23}}^{\prime}( r_e^{})
 = \left(1 - \tfrac{G\mEL^2}{e^2}\right)\Nneg - \left(1 + \tfrac{G\mPR\mEL}{e^2}\right) \Npos.
 \label{eq:PoissonMUeTHREEhalfINTagainB}
\end{equation}
 Since $\nuneg^{\prime}( r_e^{})$ is the left-derivative of $\nuneg( r)$ at the point $ r_e^{}$,
and since an otherwise positive function cannot reach $0$ with a positive slope, 
it follows that ${\nuneg^{\frac23}}^{\prime}( r_e^{})\leq 0$, and so
\begin{equation}
\frac{\Nneg}{\Npos}
\leq 
\frac{1 + \tfrac{G\mPR\mEL}{e^2}} {1 - \tfrac{G\mEL^2}{e^2}},
 \label{eq:NnegOverNposUPPERbound}
\end{equation}
which is the upper bound on $\Nneg/\Npos$ given in (\ref{eq:NnegOverNposINTERVAL}).
 We abbreviate the right-hand side of (\ref{eq:NnegOverNposUPPERbound}) by  {$(\Nneg/\Npos)^-$.}
 In the limit in which the upper bound is saturated,  from (\ref{eq:PoissonMUeTHREEhalfINTagainB}) we now obtain
\begin{equation}
\lim_{\frac{\Nneg} {\Npos}\nearrow  {\left(\frac{\Nneg}{\Npos}\right)^-}}
 r_e^2{\nuneg^{\frac23}}^{\prime}( r_e^{})=0,
 \label{eq:NnegOverNposSATURATED}
\end{equation}
where we consider $ r_e^{}$ as a function of $\Nneg/\Npos$.
 Since here $ r_e^{}> r_{\mathrm{p}}^{}>0$, it follows that ${\nuneg^{\frac23}}^{\prime}( r_e^{})\to 0$ in the limit.

 In a completely analogous manner we obtain in the case of a positive atmosphere that
\begin{equation}
4\pi \veps  \zeta   r_{\mathrm{p}}^2{\nupos^{\frac23}}^{\prime}( r_{\mathrm{p}}^{})
 = \left(1 - \tfrac{G\mPR^2}{e^2}\right)\Npos - \left(1 + \tfrac{G\mPR\mEL}{e^2}\right) \Nneg ,
 \label{eq:PoissonMUpTHREEhalfINTagainBnew}
\end{equation}
from which we deduce that
\begin{equation}
\frac{\Nneg} {\Npos}
\geq 
\frac{1 - \tfrac{G\mPR^2}{e^2}}{1 + \tfrac{G\mPR\mEL}{e^2}} ,
 \label{eq:NnegOverNposLOWERbound}
\end{equation}
which is the lower bound on $\Nneg/\Npos$ given in (\ref{eq:NnegOverNposINTERVAL}).
 Abbreviating the right-hand side of (\ref{eq:NnegOverNposLOWERbound}) by  {$(\Nneg/\Npos)^+$, }
in the limit in which the lower bound is saturated, from (\ref{eq:PoissonMUpTHREEhalfINTagainBnew}) we now obtain
\begin{equation}
\lim_{\frac{\Nneg}{\Npos} \searrow  {\left(\frac{\Nneg}{\Npos}\right)^+}}
 r_{\mathrm{p}}^2{\nupos^{\frac23}}^{\prime}( r_{\mathrm{p}}^{})=0,
 \label{eq:NnegOverNposSATURATEDagain}
\end{equation}
where we consider $ r_{\mathrm{p}}^{}$ as a function of $\Nneg/\Npos$.
 Since now $r_{\mathrm{p}}^{}> r_e^{}>0$, it follows that ${\nupos^{\frac23}}^{\prime}( r_{\mathrm{p}}^{})\to 0$ in the limit. 
%\vspace{-10pt}

%%%%%%%%%%%%%%%%%%%%%%%%%%%%%%%%%%%%%%%%%%%%%%%%%%%%%%%%%%%%%%%%%%%%%%%%%%%%%%%%%
%%%%%%%%%%%%%%%%%%%%%%%%%%%%%%%%%%%%%%%%%%%%%%%%%%%%%%%%%%%%%%%%%%%%%%%%%%%%%%%%%
\subsection{Proof that the atmosphere of an extremely surcharged solution is infinitely extended}
\vspace{-10pt}
%%%%%%%%%%%%%%%%%%%%%%%%%%%%%%%%%%%%%%%%%%%%%%%%%%%%%%%%%%%%%%%%%%%%%%%%%%%%%%%%%
%%%%%%%%%%%%%%%%%%%%%%%%%%%%%%%%%%%%%%%%%%%%%%%%%%%%%%%%%%%%%%%%%%%%%%%%%%%%%%%%%

 Consider an extremal solution with negative atmosphere.
 We have just seen that both $\nuneg(r_e^{})=0$ and $\nuneg^\prime(r_e^{})=0$.
 Now suppose $r_e^{}<\infty$.
 Then by a familiar uniqueness result for (\ref{eq:PoissonMUeTHREEhalf}), {and using that} $\nupos(r)=0$ for $r_{\mathrm{p}}<r<r_e$
in a negative-atmosphere star, it now follows that $\nuneg(r)=0$ for all $r>r_{\mathrm{p}}^{}$.
 But this violates the negative-atmosphere hypothesis which says that $\nuneg(r)$ is strictly 
positive for $r_{\mathrm{p}}^{}\leq r<r_{\text{e}}^{}$.
 Hence an extremal negative atmosphere extends to infinity.

 The analogous conclusion holds for an extremal positive atmosphere.
%\vspace{-10pt}

%%%%%%%%%%%%%%%%%%%%%%%%%%%%%%%%%%%%%%%%%%%%%%%%%%%%%%%%%%%%%%%%%%%%%%%%%%%%%%%%%
%%%%%%%%%%%%%%%%%%%%%%%%%%%%%%%%%%%%%%%%%%%%%%%%%%%%%%%%%%%%%%%%%%%%%%%%%%%%%%%%%
\subsection{Existence of extremely surcharged solutions}
\vspace{-10pt}
%%%%%%%%%%%%%%%%%%%%%%%%%%%%%%%%%%%%%%%%%%%%%%%%%%%%%%%%%%%%%%%%%%%%%%%%%%%%%%%%%
%%%%%%%%%%%%%%%%%%%%%%%%%%%%%%%%%%%%%%%%%%%%%%%%%%%%%%%%%%%%%%%%%%%%%%%%%%%%%%%%%

 The arguments presented in the previous two subsections establish that any extremely surcharged 
solution must be infinitely extended, and that in the limit where the radial variable goes to
infinity, the derivative of the atmospheric density must vanish very rapidly, see 
(\ref{eq:NnegOverNposSATURATED}) and (\ref{eq:NnegOverNposSATURATEDagain}).
 It remains to show that such extremely surcharged solutions do exist.
 Of course, we are only interested in solutions with finite total mass. 

 Here is the argument, which involves continuous dependence of solutions on the data, plus a-priori bounds.
 We already established the existence of the no-atmosphere solution, where
both densities go to zero at the same distance from the center. 
 Both densities, in this case, are rescaled standard $n=3/2$ polytropes.

 Note that the central electron density is smaller than the central proton density. 
 Now, keeping the central density of the protons fixed, start lowering the central density of
the electrons. 
 Considering the system of ordinary differential equations for the densities as initial value
problem at the origin, with vanishing slope, one can extract the information that the electron
density function decreases together with its central density, and so its zero 
now moves to the left, while the proton density increases and its zero moves to the right.
 Thus $\Nneg$ decreases and $\Npos$ increases. 
 The ratio of course can never violate the a-priori bounds in (\ref{eq:NnegOverNposINTERVAL}), and so,
given the opposite monotonicity of the particle numbers, $\Npos$ in particular cannot increase to infinity, 
and $\Nneg$ not decrease to zero. 
 How far can one push this? 
 Answer: as long as both densities hit zero with finite slope, one can continue into the neighborhood of the
solution to find a new solution. 
 This process can therefore be continued until the slope of the proton density vanishes when the proton 
density reaches zero. 
 As shown already, this can only happen if the proton density vanishes only at infinity.

 In a similar manner one can proceed keeping the central electron density fixed and lowering
the central proton density. 
 This leads to a sequence of decreasing $\Npos$ and increasing $\Nneg$, which can be continued until
the electron density extends all the way to infinity, with vanishing slope at infinity.

 We still need to show that in either of these borderline cases the density vanishes sufficiently rapidly 
so that (\ref{eq:NnegOverNposSATURATED}) and (\ref{eq:NnegOverNposSATURATEDagain}) hold. 
 So suppose this would not hold.
 Then (and assuming convergence here, for simplicity) in the case of the negative atmosphere we necessarily have
that $r^2\Ddr \nuneg^{2/3}(r)\to C<0$ for $r\to\infty$; this implies that 
$\nuneg(r)\sim 1/r^{3/2}$ for $r\to\infty$, but such a $\nuneg(r)$ is not integrable at $\infty$, in violation of
the fact that we know that $\Nneg<\infty$ and $\Npos < \infty$. 
 Similarly one can rule out that $r^2\Ddr \nupos^{2/3}(r)\to C<0$ for $r\to\infty$ for the positive atmosphere case.
 This shows that (\ref{eq:NnegOverNposSATURATED}) and (\ref{eq:NnegOverNposSATURATEDagain}) do hold. 

 This establishes the existence of two extremal atmosphere solutions in the $5/3$ model, satisfying 
(\ref{eq:NnegOverNposSATURATED}), respectively (\ref{eq:NnegOverNposSATURATEDagain}).

 But then we can multiply Eq.(\ref{eq:PoissonMUeTHREEhalf}) by $4\pi r^2$ and integrate over $r$ from $0$ to $\infty$,
with the understanding that $\nupos(r)=0$ for $r\geq r_{\mathrm{p}}^{}$. 
 Using that $r^2\Ddr \nuneg^{2/3}(r)\to 0$ for $r\to\infty$,
the result of this integration is again Eq.(\ref{eq:PoissonMUeINT}). 
 Similarly we can proceed in the case of an extreme positive atmosphere, and once again find Eq.(\ref{eq:PoissonMUpINT}).

 Thus our {bounds} (\ref{eq:NnegOverNposINTERVAL}) are also valid when working with 
the proper $5/3$ power law of the non-relativistic degeneracy pressures of the protons and the electrons, as claimed.
%\vspace{-10pt}
\newpage

%%%%%%%%%%%%%%%%%%%%%%%%%%%%%%%%%%%%%%%%%%%%%%%%%%%%%%%%%%%%%%%%%%%%%%%%%%%%%%%%%
%%%%%%%%%%%%%%%%%%%%%%%%%%%%%%%%%%%%%%%%%%%%%%%%%%%%%%%%%%%%%%%%%%%%%%%%%%%%%%%%%
\subsection{Relativity}
\vspace{-10pt}
%%%%%%%%%%%%%%%%%%%%%%%%%%%%%%%%%%%%%%%%%%%%%%%%%%%%%%%%%%%%%%%%%%%%%%%%%%%%%%%%%
%%%%%%%%%%%%%%%%%%%%%%%%%%%%%%%%%%%%%%%%%%%%%%%%%%%%%%%%%%%%%%%%%%%%%%%%%%%%%%%%%

Our discussion of the $5/3$ model can be adapted to the special-relativistic setting in the
manner done by Chandrasekhar \cite{Chandra} for the single-density model,
which in the structure equations (\ref{eq:PoissonMUpTHREEhalf}) and (\ref{eq:PoissonMUeTHREEhalf}) changes
the $\nu^{2/3}_f$ into some nonlinear function of $\nu_f$ that interpolates continuously between $\nu^{2/3}_f$ and $\nu^{1/3}_f$.
 Explicitly, introducing $\Pf=(1/24\pi^2)\, m_f^4c^5/ \hbar^3$          % $A_f=\frac{\pi m_f^4c^5}{3h^3}$ 
and $\ell_f=(3\pi^2)^{1/3}\hbar/m_fc$, and  $k_f = {m_f c^2}/{4\pi e^2}$,
the nonrelativistic pressure law (\ref{eq:pDEG}) gets replaced with
the somewhat intimidating expression
\begin{widetext}
\begin{equation}
    p_{\text{f}}(r)=\Pf \ell_f\nu_f^{1/3}(r)\left(2\ell_f^2\nu_f^{2/3}(r)-3\right)\sqrt{1+\ell_f^2\nu_f^{2/3}(r)}
+\sinh^{-1}\left({\ell_f\nu_f^{1/3}(r)}\right).
\end{equation}
 Wherever both  $\nupos(r)>0$ and $\nuneg(r)>0$
one can follow the same steps used in the derivation of (\ref{eq:PoissonMUpTHREEhalf}) and (\ref{eq:PoissonMUeTHREEhalf}) 
to get
\begin{eqnarray}
- k_{\mathrm{p}} \frac{1}{r^2}\Ddr\left(r^2\Ddr \sqrt{1+ \ell_{\mathrm{p}}^2\nupos^{2/3}(r)}\right)
&\,\  =  - \left(1 -\frac{G\mPR^2}{e^2}\right) \nupos(r) + \left(1 +\frac{G\mPR\mEL}{e^2}\right) \nuneg(r),
 \label{eq:PoissonMUpCHANDRA}\\
- k_{\mathrm{e}} \frac{1}{r^2}\Ddr\left(r^2\Ddr \sqrt{1+\ell^2_e\nuneg^{2/3}(r)}\right)
&\!\!  = \left(1 + \frac{G\mPR\mEL}{e^2}\right) \nupos(r) - \left(1 - \frac{G\mEL^2}{e^2}\right) \nuneg(r).
 \label{eq:PoissonMUeCHANDRA}
\end{eqnarray}
\end{widetext}
 All the same, the integration of the pertinent structure equations will always produce Eqs.(\ref{eq:PoissonMUeINT})
and (\ref{eq:PoissonMUpINT}), and therefore (\ref{eq:NnegOverNposINTERVAL}). 
 
 There is one caveat to what we just said, and that is that we have tacitly assumed that we stay away from the
Chandrasekhar mass  $\propto (\hbar c/G)^\frac32/\mPR^2$. 
 However, since we are only discussing failed white dwarfs, which
are low-mass stars, with $\Npos$ restricted to about $1.5\cdot 10^{55}$ -- $9\cdot 10^{55}$ protons, 
we certainly are on the safe side. 

 All our results so far are based on Newtonian gravity, though.
 We suspect that (\ref{eq:NnegOverNposINTERVAL}) also holds general-relativistically, again
for failed white dwarfs whose mass is far away from any critical mass beyond which no stellar 
equilibrium is possible in a general-relativistic setting.
 To show this in the detailed manner as done for the non-relativistic, and by analogy special-relativistic
models, is a more complicated problem which requires the discussion of the Einstein field equations
coupled with both the matter equations for the Fermi gases and the Maxwell equations of the electrostatic
field in curved spacetime; cf. \cite{OBb}, \cite{RRa}, \cite{RRb}.
 We plan to do this in a future work.
 Here we are content with the remark that
the key argument in our derivation of (\ref{eq:NnegOverNposINTERVAL}) is the behavior of
the atmospheric densities at spatial infinity, and in an asymptotically flat spacetime this is the 
region where the general-relativistic equations are expected to go over into the non-relativistic equations
of Newtonian physics --- hence the independence of $\hbar$ and $c$, and our conjecture that (\ref{eq:NnegOverNposINTERVAL}) 
is truly universally valid for \emph{failed white dwarfs} which never ignited, and assumed to consist
of electrons and protons.

 More realistic models of ground states of failed white dwarfs and white dwarfs (black dwarfs)
require other compositions of particles, not just electrons and protons, 
and this will of course change the bounds on the excess charge
in terms of nuclear-chemical composition.
 It is an interesting question whether they will be independent of $\hbar$ and $c$, all the way up to 
Chandrasekhar's critical mass, $\propto (\Nneg/N_n)^2(\hbar c/G)^\frac32/\mPR^2$, where
$N_n$ is the number of nucleons in the star.
\vspace{-20pt}

%%%%%%%%%%%%%%%%%%%%%%%%%%%%%%%%%%%%%%%%%%%%%%%%%%%%%%%%%%%%%%%%%%%%%%%%%%%%%%%%%
%%%%%%%%%%%%%%%%%%%%%%%%%%%%%%%%%%%%%%%%%%%%%%%%%%%%%%%%%%%%%%%%%%%%%%%%%%%%%%%%%
\subsection{Other pressure-density relations}
\vspace{-10pt}
%%%%%%%%%%%%%%%%%%%%%%%%%%%%%%%%%%%%%%%%%%%%%%%%%%%%%%%%%%%%%%%%%%%%%%%%%%%%%%%%%
%%%%%%%%%%%%%%%%%%%%%%%%%%%%%%%%%%%%%%%%%%%%%%%%%%%%%%%%%%%%%%%%%%%%%%%%%%%%%%%%%

 It is clear from our discussion so far that the key to (\ref{eq:NnegOverNposINTERVAL}) is the 
existence of infinitely extended density solutions which vanish rapidly at infinity such that 
(analogs of) (\ref{eq:NnegOverNposSATURATED}), respectively (\ref{eq:NnegOverNposSATURATEDagain}) hold,
where the $\nu_f^{2/3}$ at the left-hand side is replaced by some nonlinear function of $\nu_f$ obtained
from any pressure-density law which leads to solutions which are integrable.
 This is a large class of models which all lead to the same surcharge bounds for a failed star, which then
is not necessarily considered to be in the white dwarf stage already.

{There are also many laws which do not produce solutions with finite $\Npos$ and $\Nneg$;
e.g., polytropic laws with index $n>5$.
 Also the isothermal} pressure law with finite temperature will \emph{not} lead to integrable density functions.
{Yet there is an analog of (\ref{eq:NnegOverNposINTERVAL}); see Appendix B.}
\vspace{-10pt}

%%%%%%%%%%%%%%%%%%%%%%%%%%%%%%%%%%%%%%%%%%%%%%%%%%%%%%%%%%%%%%%%%%%%%%%%%%%%%%%%%
%%%%%%%%%%%%%%%%%%%%%%%%%%%%%%%%%%%%%%%%%%%%%%%%%%%%%%%%%%%%%%%%%%%%%%%%%%%%%%%%%
%%%%%%%%%%%%%%%%%%%%%%%%%%%%%%%%%%%%%%%%%%%%%%%%%%%%%%%%%%%%%%%%%%%%%%%%%%%%%%%%%
\section{Determining $\Npos$ and $\Nneg$ of non-extremal solutions}\vspace{-10pt}
%%%%%%%%%%%%%%%%%%%%%%%%%%%%%%%%%%%%%%%%%%%%%%%%%%%%%%%%%%%%%%%%%%%%%%%%%%%%%%%%%
%%%%%%%%%%%%%%%%%%%%%%%%%%%%%%%%%%%%%%%%%%%%%%%%%%%%%%%%%%%%%%%%%%%%%%%%%%%%%%%%%
%%%%%%%%%%%%%%%%%%%%%%%%%%%%%%%%%%%%%%%%%%%%%%%%%%%%%%%%%%%%%%%%%%%%%%%%%%%%%%%%%

 Having numerically computed a solution pair for the non-linear $5/3$ model or the special-relativistic model,
the number of protons $\Npos$ and electrons $\Nneg$ of the solution can of course be obtained by integrating
$4\pi r^2\nupos(r)$ over $r$ from $0$ to $r_{\mathrm{p}}$, and $4\pi r^2\nuneg(r)$ over $r$ from $0$ to $r_e$. 
 However, there is a simpler way to get to these numbers directly after integrating the differential equations,
in the manner done earlier for the $6/3$ model.

 Thus, for the $5/3$ model, we obtain $(\Npos,\Nneg)$ uniquely in terms of the zeros of the densities and the derivatives 
at the zeros by simply replacing $\upsilon_f^{\prime}(\rho)$ by 
$\frac54(\upsilon_f^{2/3})^{\prime}(\rho)$ in Eq.(\ref{eq:matrixNposNnegSOL}), ${}_f = {}_{\mathrm{p}}$ or ${}_e$,
where we have tacitly switched to the dimensionless variables of the $6/3$ model.
 So when solving the system of equations for $(\upos,\uneg)$ as an initial value problem with prescribed central densities 
and vanishing central radial derivatives, all one needs to compute are the zeros of the densities 
and their left derivatives at the zeros.
 This reduces the computational effort.

 In an analogous manner one can compute $(\Npos,\Nneg)$ uniquely in terms of the zeros of the densities and the derivatives 
at the zeros for the special-relativistic Chandrasekhar-type setup.\vspace{-.5truecm}

%%%%%%%%%%%%%%%%%%%%%%%%%%%%%%%%%%%%%%%%%%%%%%%%%%%%%%%%%%%%%%
%%%%%%%%%%%%%%%%%%%%%%%%%%%%%%%%%%%%%%%%%%%%%%%%%%%%%%%%%%%%%%
%%%%%%%%%%%%%%%%%%%%%%%%%%%%%%%%%%%%%%%%%%%%%%%%%%%%%%%%%%%%%%%%%%%%
\section{Comparison of the models}\label{check}\vspace{-.3truecm}
%%%%%%%%%%%%%%%%%%%%%%%%%%%%%%%%%%%%%%%%%%%%%%%%%%%%%%%%%%%%%%
%%%%%%%%%%%%%%%%%%%%%%%%%%%%%%%%%%%%%%%%%%%%%%%%%%%%%%%%%%%%%%
%%%%%%%%%%%%%%%%%%%%%%%%%%%%%%%%%%%%%%%%%%%%%%%%%%%%%%%%%%%%%%%%%%%%
%%%%%%%%%%%%%%%%%%%%%%%%%%%%%%%%%%%%%%%%%%%%%%%%%%%%%%%%%%%%%%

 Having a complete set of solution formulas for the $6/3$-model one can generate 
figures which illustrate the findings, and compare these with the results of numerical evaluations
of the $5/3$ model and also with the Chandrasekhar-type special-relativistic model.
 As emphasized earlier, the $6/3$ model serves also as a test case for the numerical 
algorithm, which has to reproduce the exact solutions to the degree of accuracy demanded.

 There are two compromises to be made, though.

 Namely, the fantastically tiny ratios of the gravitational to electrical coupling
constants between electron and proton are definitely a numerical problem, but also 
the small mass ratio $\mEL/\mPR\approx1/1836$ is a source of trouble. 
 Both these small numbers taken together make it sheer impossible to produce any useful graphs at all. 

 For instance, let us try to resolve the interval of the allowed values of the ratio $\Nneg/\Npos$.
 From (\ref{eq:NnegOverNposINTERVAL}) we see that $\Nneg/\Npos$ varies between about $1-8.1\cdot10^{-37}$
and about $1+4.4\cdot 10^{-40}$.
 This can be ameliorated a little bit by centering the $\Nneg/\Npos$ axis at $1$ and scaling up the units by 
a factor of $(1/4.4)\cdot 10^{40}$. 
 Incidentally, the construction just described is equivalent to using a rescaled $\ln(\Nneg/\Npos)$ as base variable. 
 This has eliminated the problems with the tininess of the coupling constant ratios!
 However, the small mass ratio $\mEL/\mPR$ still poses a hurdle, for 
the negatively charged stars will occupy about 1 positive unit in the allowed 
interval of $\ln(\Nneg/\Npos)$ and the positive stars $1836$ negative units.
 To resolve such a lopsided asymmetry graphically without introducing otherwise obscuring transformations is impossible.
 We therefore decided to work with the SciFi value $\mEL/\mPR = 1/10$. 

\begin{figure}[ht]
  \includegraphics[width = 7.2truecm,scale=2]{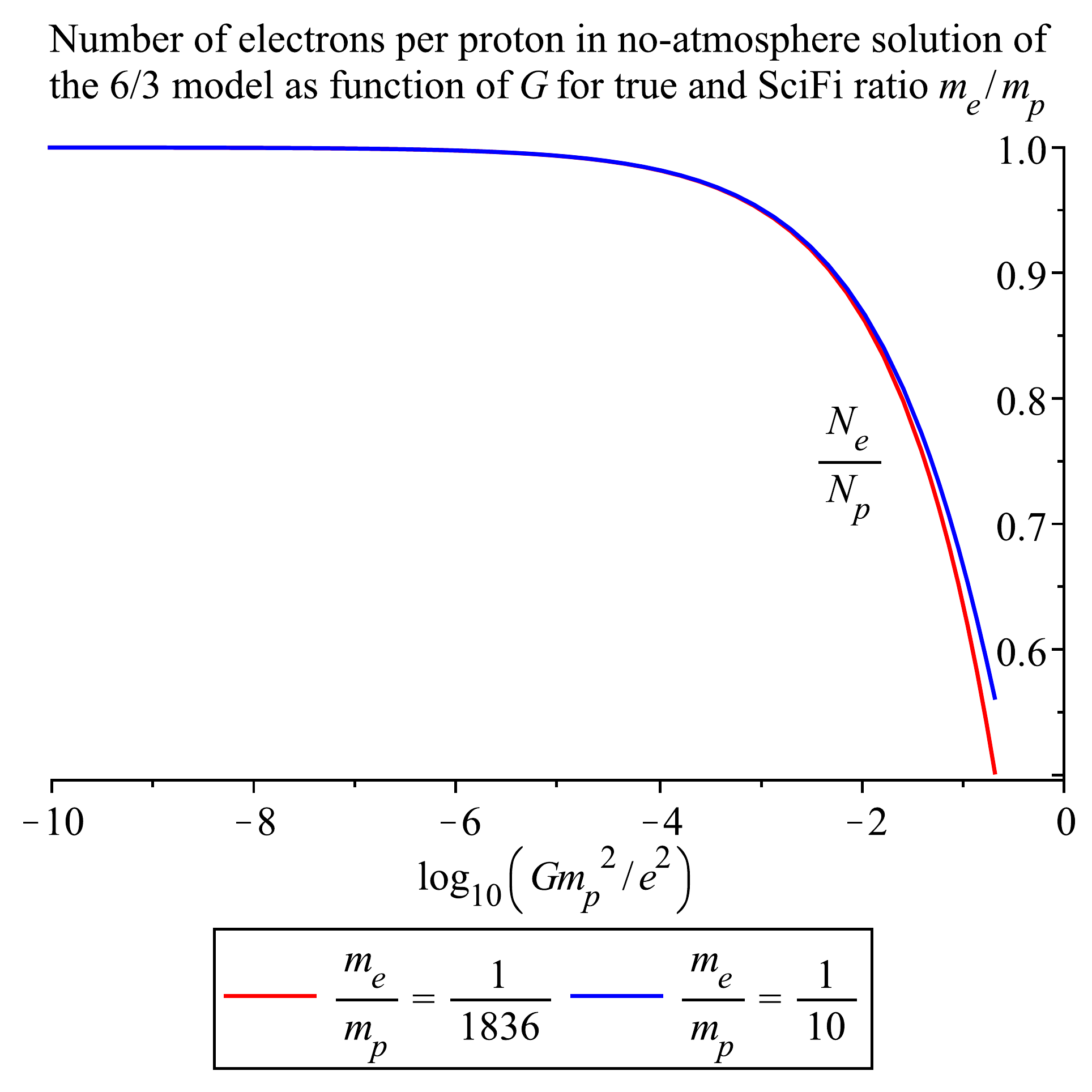} %\vspace{-6truecm}
\caption{Shown is $\Nneg/\Npos$ computed analytically as a function of $G\mPR^2/e^2$ for the no-atmosphere solutions of
the 6/3 model (formula \refeq{eq:NnegOverNposNOatmo}),
once for the physical ratio $\mEL/\mPR=1/1836$ and once for the science fiction ratio $\mEL/\mPR=1/10$. 
 Note that the physical value of $\log_{10}(G\mPR^2/e^2)$ is located on the horizontal axis at $\approx -36$.}   \vspace{-.5truecm}
\label{NeOverNpVsG63}
\end{figure}

 Furthermore, while the discussion just given shows that for certain questions the tiny ratios of the coupling
constants can be dealt with and only the small ratio of $\mEL/\mPR$ is a problem, when one wants to plot
both electron and proton density functions in one panel, they will appear indistinguishable when attempted with
the actual values of $G\mPR^2/e^2$, etc.
 To illustrate this, we plot $\Nneg/\Npos$ for the no-atmosphere solution of the 6/3 model,
computed analytically with formula \refeq{eq:NnegOverNposNOatmo},
versus $\log_{10} (G\mPR^2/e^2)$ for the actual $\mEL/\mPR=1/1836$ and for the SciFi value $\mEL/\mPR = 1/10$;
see Fig.~\ref{NeOverNpVsG63}.

 Fig.~\ref{NeOverNpVsG63} shows that to qualitatively visualize the difference of the particle densities one needs to
replace the actual value of $G\mPR^2/e^2$ with science fiction values. 
 In this vein, in the following we illustrate our findings for the SciFi values
$G\mPR^2/e^2=1/2$ and $\mEL/\mPR=1/10\;(=\veps)$; for consistency, therefore,
$G\mPR\mEL/e^2=\veps/2$ and $G\mEL^2/e^2=\veps^2 /2$.
 The other physical constant, $\alphaS= 1/137.036$.
 
\begin{figure}[ht]
  \includegraphics[width = 8.2truecm,scale=2]{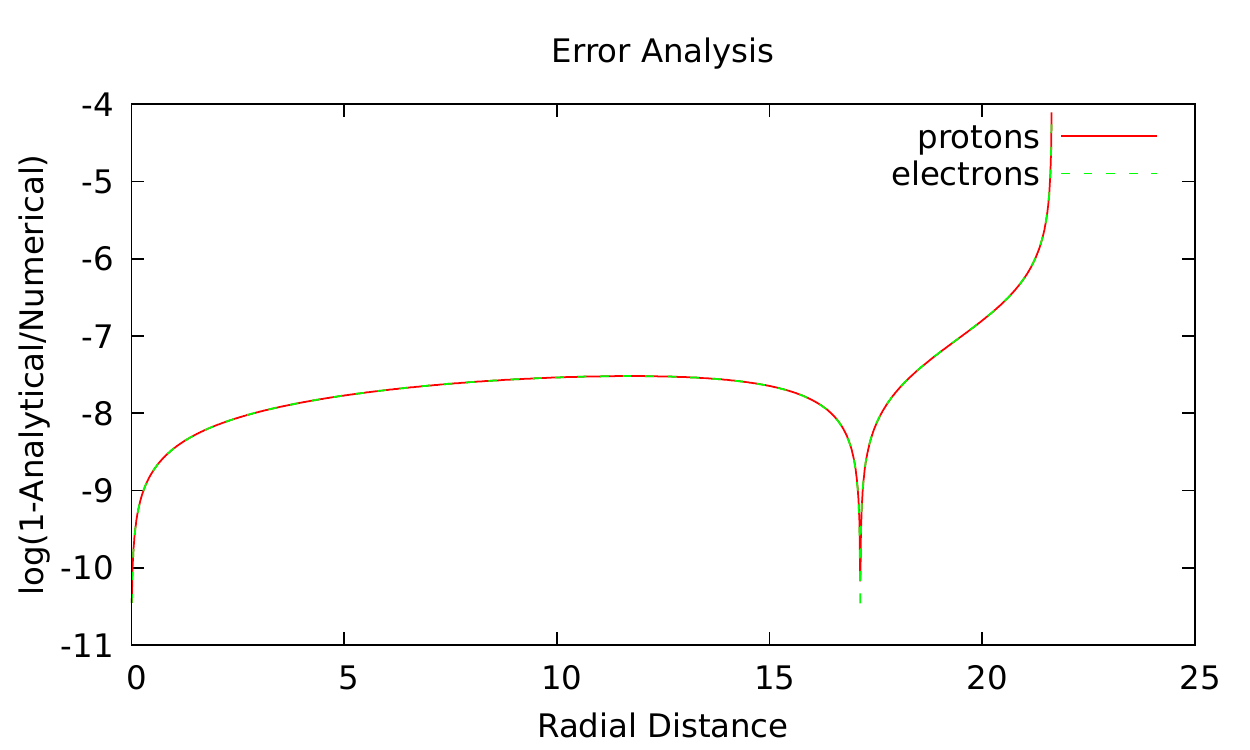} %\vspace{-6truecm}
\caption{Shown is log$_{10}|1-\ups_f^{(a)}/\ups_f^{(n)}|$, where $\ups_f^{(a)}$ and $\ups_f^{(n)}$
denote the analytically and the numerically computed scaled densities for the no-atmosphere solutions of the 6/3 model, 
for electrons (${}_f={}_e$) and for protons (${}_f={}_p$), 
 as functions of the scaled radial distance $\rho$ (for the units, see sect.III, first paragraph),
though for science fiction values $G\mPR^2/e^2=1/2$ and $\mEL/\mPR=1/10$. }   \vspace{-.5truecm}
\label{63densDIFF}
\end{figure}

 We have tested our numerical algorithm (essentially a Runge--Kutta 45 scheme) by comparing the plots of the
exact solution formulas with those produced by numerically solving the $6/3$ model
for the science fiction values of the constants; see Fig.~\ref{63densDIFF}  for a representative error plot.\vspace{-.75truecm}

%\newpage

 Fig.~\ref{63densDIFF}
 demonstrates that the numerical algorithm approximates the exact analytical solutions
with a relative error of less than $10^{-4}$, and even less than $10^{-7}$ over most of the bulk region.
 This indicates that our Runge--Kutta 45 scheme also computes the solutions to the physically more realistic
models accurately, where we do not have analytical solutions to compare. 
 
\begin{figure}[ht]
  \includegraphics[width = 10.2truecm,scale=2]{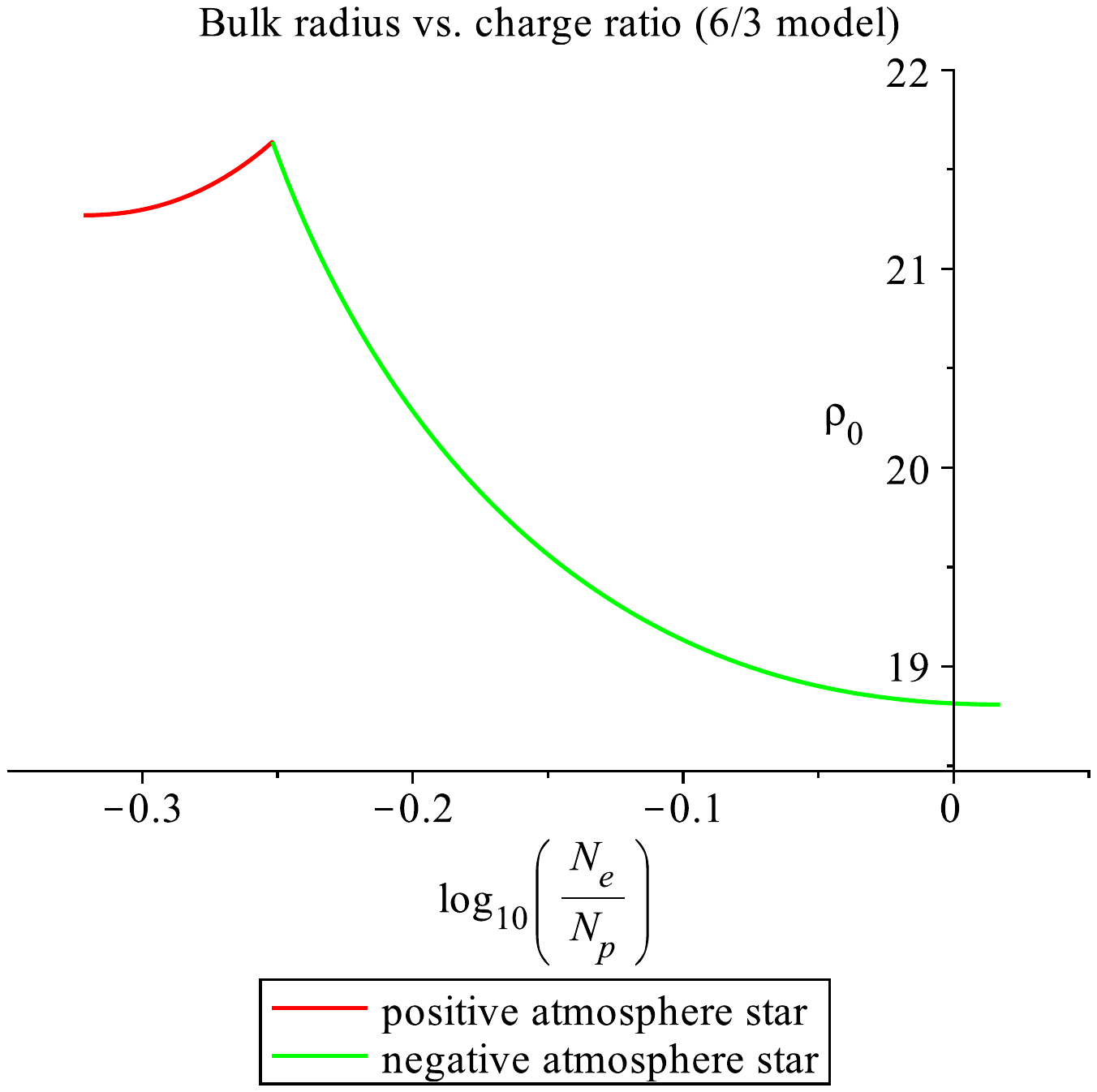} \vspace{-6truecm}
\caption{Shown is the bulk radius $\rho_0$ (units in sect.III, 1st paragraph) vs. $\log_{10}(\Nneg/\Npos)$, 
computed analytically, for the full range of allowed values, though for SciFi values $G\mPR^2/e^2=1/2$ and 
$\mEL/\mPR=1/10$. 
 Most solutions carry a positive surcharge, but only a small fraction of them has a positive atmosphere.
 This strong two-fold asymmetry is caused by $\mEL/\mPR \ll 1$.}  % \vspace{-1truecm}
\label{63rhoVsNeOverNp}
%\end{figure}
%\begin{figure}[ht]
  \includegraphics[width = 7truecm,height=6truecm]{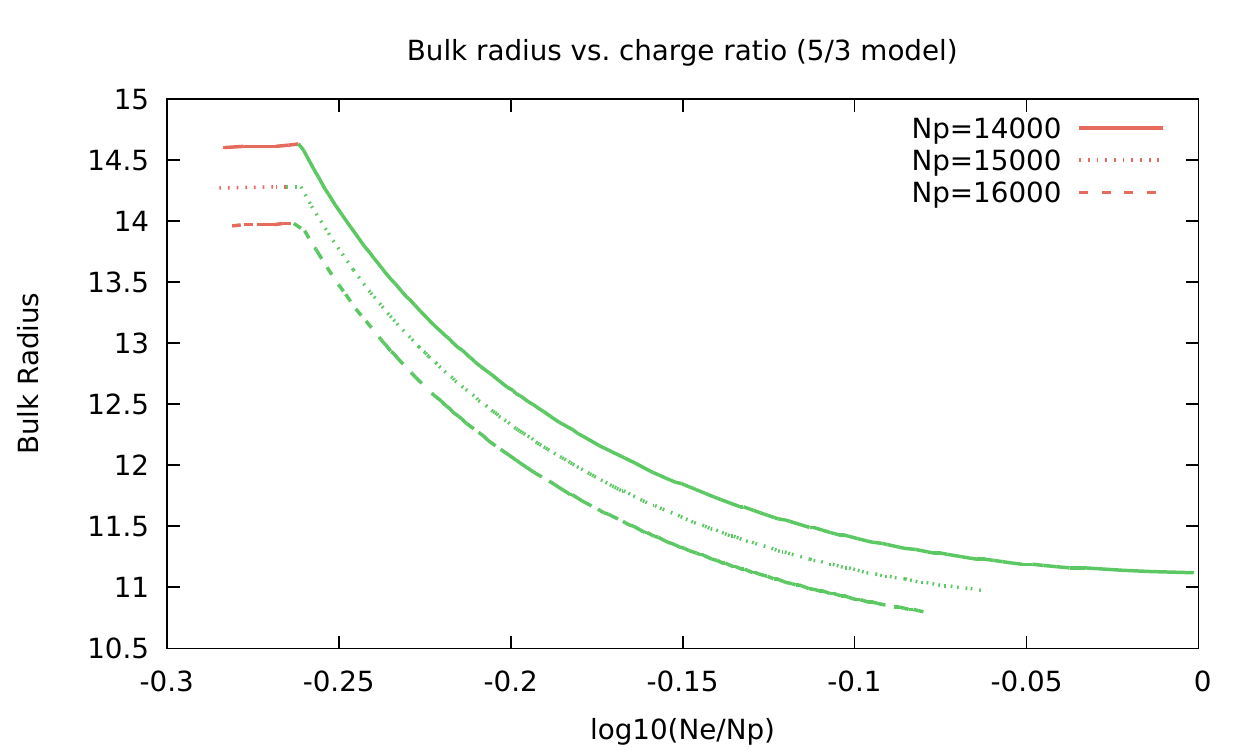} \vspace{-.5truecm}
\caption{Shown is the bulk radius $\rho_0$  (units in sect.III, 1st paragraph) vs. $\log_{10}(\Nneg/\Npos)$, for 
an {(not the full)} interval of allowed $\Nneg/\Npos$ values, computed numerically, for three different values of $\Npos$.
 Along each of the curves $\Npos$ is constant.
 We use science fiction values $G\mPR^2/e^2=1/2$ and $\mEL/\mPR=1/10$. }  \vspace{-0.3truecm}
\label{53rhoVsNeOverNp}
\end{figure}

 We next graph the bulk radius $\rho_0$ which the star adapts in response to $\Nneg/\Npos$, 
in Fig.~\ref{63rhoVsNeOverNp} for the 6/3 model (analytical)
and in Fig.~\ref{53rhoVsNeOverNp} for the 5/3 model (numerical).  \vspace{-1truecm}

 In the 6/3 model, thanks to the amplitude scaling invariance of its linear set of structure equations, there
is only one equal-$\Npos$ curve representing all solutions; this is of course a degenerate situation.
 Each point on the curve corresponds to a whole scaling family of solution pairs $(\nupos,\nuneg)$ with the
same ratio $\Nneg/\Npos$.
 The nonlinear set of structure equations of the 5/3 model breaks the amplitude scaling invariance of the
6/3 model. 
 To each $\Npos$ there now corresponds a separate curve representing solution pairs.
 On each such curve, every point belongs to a unique solution with the given $\Npos$ and an
associated $\Nneg$, which varies along the curve. %\vspace{-1truecm}
{It is however too time-consuming to push all the way to the extreme solutions, which is
noticeable by comparing the first two figures.}

  Next we show the sets of equal-$\Npos$ curves in the plane of radii,  
first for the analytical 6/3 solutions, then for the numerical 5/3 solutions. 
 The no-atmosphere solutions are situated on the diagonal in these two diagrams.  \vspace{-1.5truecm}
\begin{figure}[ht]
  \includegraphics[width = 12truecm]{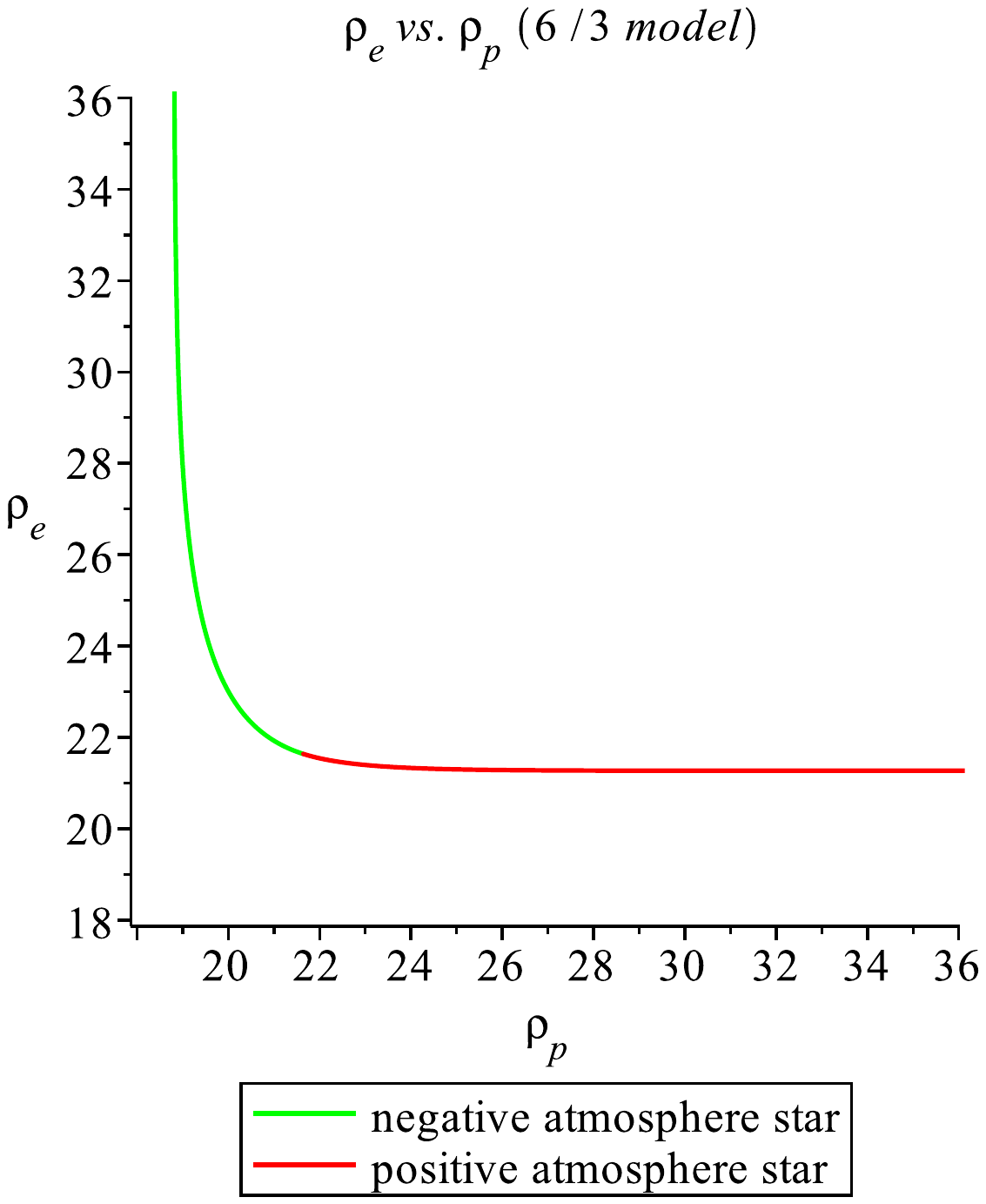} \vspace{-7truecm}
\caption{Shown is the radius $\rho_{\mathrm{e}}$  of the electron density vs. the radius $\rho_{\mathrm{p}}$ of the proton density
(for the units, see sect.III, first paragraph), computed analytically, 
for science fiction values $G\mPR^2/e^2=1/2$ and $\mEL/\mPR=1/10$. 
   Any point on the curve is a scaling family of solution pairs with fixed ratio $\Nneg/\Npos$.}  %\vspace{-1truecm}
\label{63reVSrp}
%\end{figure}
%\begin{figure}[ht]
\hspace{-2truecm}
  \includegraphics[width = 10.5truecm]{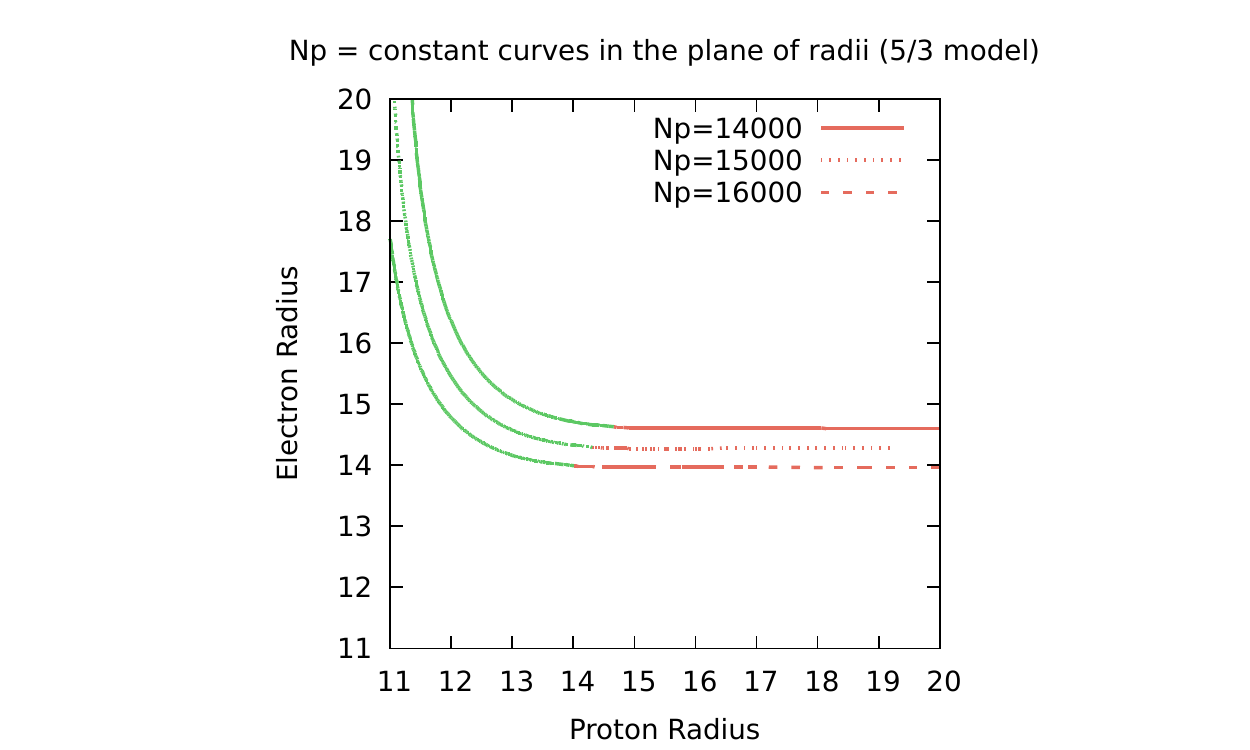} \vspace{-.5truecm}
\caption{Shown is the radius $\rho_{\mathrm{e}}$ of the electron density vs. the radius $\rho_{\mathrm{p}}$ of the proton density
(for the units, see sect.III, first paragraph), computed numerically,
for science fiction values $G\mPR^2/e^2=1/2$ and $\mEL/\mPR=1/10$. 
   Each point on a given curve represents a unique solution pair with same $\Npos$.} \vspace{-.5truecm}
\label{53reVSrp}
\end{figure}

%\newpage

 We next compare the numerically computed particle density functions of the 5/3 model with the analytically
computed ones of its 6/3 approximation, with the same SciFi values given to the physical constants.
 The central proton bulk density in the 5/3 model is the same in all examples.
 In the comparisons of  5/3 with pertinent 6/3 densities, the two solutions have the same $\Npos$.

\begin{figure}[ht]
  \includegraphics[width = 8.2truecm,scale=1.1]{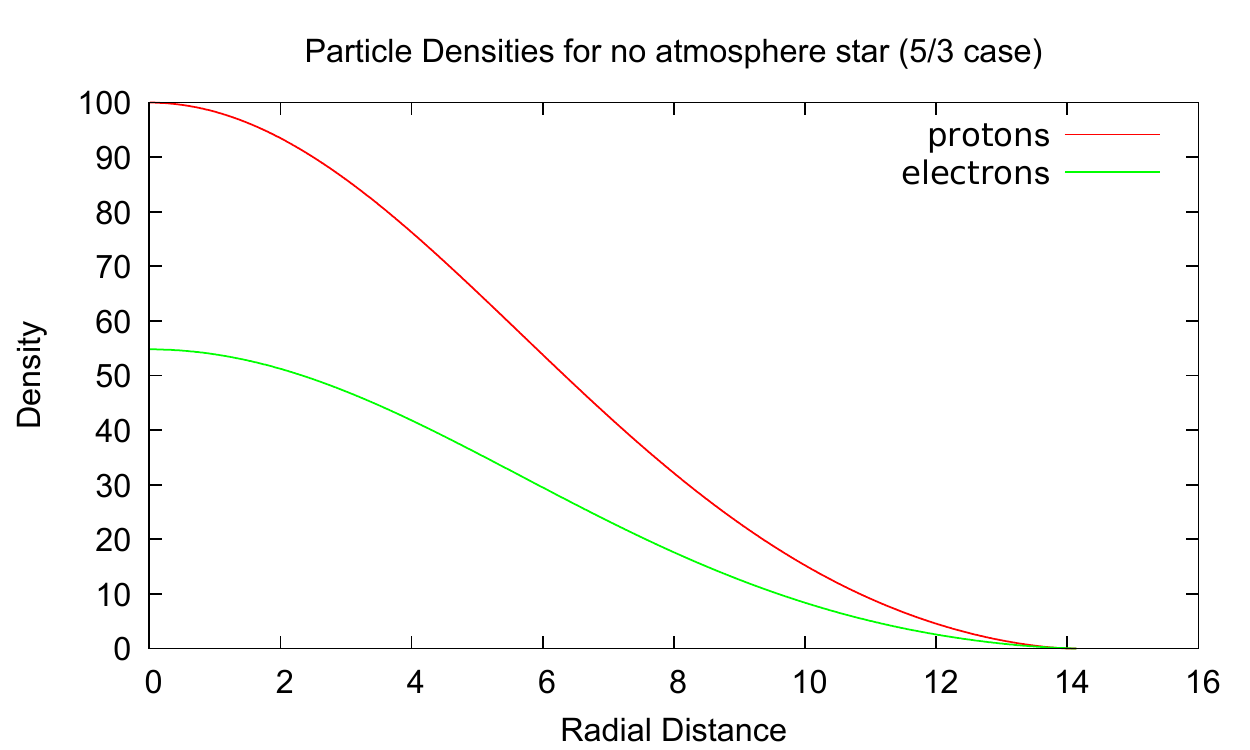} %{mpmeOFxiNOatmo.pdf} \vspace{-6truecm}
  \includegraphics[width = 8.2truecm,scale=1.1]{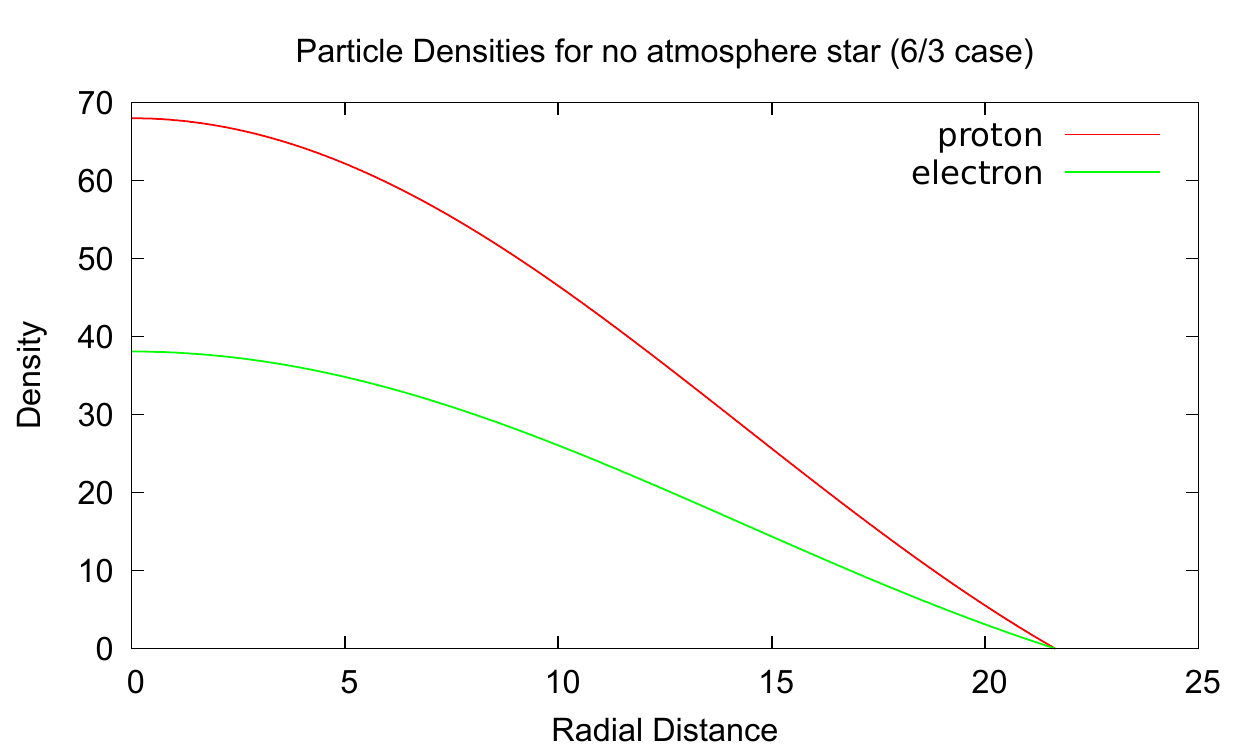} %{mpmeOFxiNOatmo.pdf} \vspace{-6truecm}
\caption{Shown are the density functions $\upos(\rho)$ and $\uneg(\rho)$ (units in sect.III, 1st paragraph), 
computed numerically for the 5/3 and analytically for the 6/3 model, of a star without atmosphere,
for SciFi values $G\mPR^2/e^2=\frac12$ and $\mEL/\mPR=\frac{1}{10}$. } \vspace{-.5truecm}
\label{53and63nuVSrhoNOatmo}
\end{figure}

 We begin with the distinguished pair of solutions consisting of the densities of a star without atmosphere, 
when both $\upos(\rho)$ and $\uneg(\rho)$ vanish at the same dimensionless bulk radius $\rho_0^{}$,
 We show the density functions of both the 5/3 and the 6/3 model; see Fig.~\ref{53and63nuVSrhoNOatmo}.
 The no-atmosphere solutions in the two models behave qualitatively similar; however, note the difference in the scales! 
 The no-atmosphere solutions of the 6/3 model with equal proton number $\Npos$ have a much
more spread-out bulk than those of the 5/3 model, and the central densities are much smaller in the 
6/3 model than in the 5/3 model.\vspace{-.5truecm}

 In the 6/3 model a star without an atmosphere has $\rho_0^{}=\pi/\kappa_t^{}$, as we discussed earlier,
and both densities then are scaled $n=1$ polytropes.
 For the proper 5/3 model it can also be shown that both densities are scaled polytropes, though for $n=3/2$
of course; as follows.

 Note that in the 6/3 model one has $\uneg(\rho) = \lambda \upos(\rho)$ when $\rho_0^{}=\pi/\kappa_t^{}$, 
with $\lambda$ given by the right-hand side of Eq.(\ref{eq:BnegOverBposTRIG}).
 Let's instead make the ansatz $\uneg(\rho) = \lambda \upos(\rho)$ in Eqs.(\ref{eq:PoissonMUp}) and (\ref{eq:PoissonMUe})
of the 6/3 model. 
 We then obtain two equations for $\upos(\rho)$ (say), and this generally overdetermines the problem.
 Their compatibility condition is the quadratic problem $a\lambda^2 +b\lambda +c =0$,
with $a= \left(1+ {G\mEL\mPR}/{e^2}\right)/\veps>0$,
$b= -\left(1 -{G\mPR^2}/{e^2}\right)/\veps + \left(1 - {G\mEL^2}/{e^2}\right)<0$,
and $c= - \left(1+ {G\mEL\mPR}/{e^2}\right)<0$.
 The ``quadratic formula'' yields two real solutions, 
\begin{equation}
\lambda_\pm^{} = -\tfrac{b}{2a}\left(1\pm\sqrt{1 - 4 \tfrac{ac}{b^2}}\right),
 \label{eq:lamdbaPM}
\end{equation}
one of which is positive and the other one negative.
 Now a particle density cannot be negative, so $\lambda=\lambda_+^{}$, and this 
is \emph{precisely} the right-hand side of Eq.(\ref{eq:BnegOverBposTRIG}).

 Similarly one can insert the ansatz $\nuneg(r) = \lambda \nupos(r)$ also into the equations of the 5/3 model,  
i.e. Eqs.(\ref{eq:PoissonMUpTHREEhalf}) and (\ref{eq:PoissonMUeTHREEhalf}), and now the compatibility condition 
is the vanishing of the degree-5 polynomial $a\eta^5 +b_e\eta^3 +b_{\mathrm{p}}\eta^2+d =0$, where $\eta :=\lambda^{1/3}$,
and  $a= \left(1+ {G\mEL\mPR}/{e^2}\right)/\veps>0$, 
$b_e= \left(1 - {G\mEL^2}/{e^2}\right)>0$, $b_{\mathrm{p}} = -\left(1 -{G\mPR^2}/{e^2}\right)/\veps <0$, 
and $c= - \left(1+ {G\mEL\mPR}/{e^2}\right)<0$.
 There generally does not exist a solution in closed form, but 
from the signs of the coefficients in this polynomial one can deduce right away that there exists a unique
positive solution $\eta_+$, say, very close to $1$, and for $\lambda=\eta_+^3$
both (\ref{eq:PoissonMUpTHREEhalf}) and (\ref{eq:PoissonMUeTHREEhalf}) reduce to the equation {(cf. \cite{HK})}
\begin{eqnarray}
\hspace{-15pt}
\veps \zeta \frac{1}{r^2}\!\!\left(r^2\nupos^{\frac23}{}^\prime(r)\right)^\prime\!\!
= \! \left[\!1 -\tfrac{G\mPR^2}{e^2} -\lambda\! \left(\!1\! +\!\tfrac{G\mPR\mEL}{e^2}\right)\! \right]\!
\nupos(r),
 \label{eq:PoissonMUpMUe}
\end{eqnarray}
which is equivalent (not identical) to the polytropic equation of index $n=3/2$, 
Eq.(\ref{eq:LaneEmdenTHREEhalfA}).
 Indeed, setting $\nupos^{\frac23}(r)=\nupos^{\frac23}(0)\theta(\xi)$ and scaling $r=C\xi$ appropriately
converts (\ref{eq:PoissonMUpMUe}) into the standardized format 
$-\frac{1}{\xi^2}\left(\xi^2\theta^{\prime}(\xi)\right)^\prime = \theta_+^{3/2}(\xi)$, cf. \cite{Emden}, \cite{Chandra},
and thus the no-atmosphere densities are obtained by rescaling the standardized $n=3/2$ polytrope.

 We remark that inserting the dimensionless bulk radius of a star without an atmosphere,
$\rho_0^{}=\pi/\kappa_t^{}$, into our formula for the $\rho_0^{}$-dependent
number of electrons per proton in the 6/3 model yields
\begin{equation}
\frac{\Nneg}{\Npos}
 = 
\frac{\lambda\left(1 -\frac{G\mPR^2}{e^2}\right)+ \veps\left(1 +\tfrac{G\mPR\mEL}{e^2}\right)}
{\lambda\left(1 +\tfrac{G\mPR\mEL}{e^2}\right)+ \veps\left(1 -\frac{G\mEL^2}{e^2}\right)},
% = \frac{\left(1 -\frac{G\mPR^2}{e^2}\right)+ \veps(\Bpos^t/\Bneg^t)\left(1 +\tfrac{G\mPR\mEL}{e^2}\right)}
% {\left(1 +\tfrac{G\mPR\mEL}{e^2}\right)+ \veps(\Bpos^t/\Bneg^t)\left(1 -\frac{G\mEL^2}{e^2}\right)},
 \label{eq:NnegOverNposNOatmoEXPLICIT}
\end{equation}
with $\lambda = \Bneg^t/\Bpos^t$ given by (\ref{eq:BnegOverBposTRIG}).
 Alternatively, knowing that $\uneg=\lambda\upos$ in this case, (\ref{eq:NnegOverNposNOatmoEXPLICIT})
follows directly from multiplying Eq.(\ref{eq:PoissonMUp}) by $\lambda 4\pi\rho^2$
and Eq.(\ref{eq:PoissonMUe}) by $\veps 4\pi\rho^2$, then integrating over $\rho$, then subtracting the
first result from the second, followed by simple algebra.

 Similarly the number of electrons per proton of the no-atmosphere solution of a 
failed white dwarf star as computed with the physical 5/3 model is obtained from 
Eqs.(\ref{eq:PoissonMUpTHREEhalf}) and (\ref{eq:PoissonMUeTHREEhalf}).
 With $\lambda=\eta_+^{3}$ one finds {\cite{HK}}
\begin{equation}
\frac{\Nneg}{\Npos}
 = 
\frac{\lambda^{2/3}\left(1 -\frac{G\mPR^2}{e^2}\right)+ \veps\left(1 +\tfrac{G\mPR\mEL}{e^2}\right)}
{\lambda^{2/3}\left(1 +\tfrac{G\mPR\mEL}{e^2}\right)+ \veps\left(1 -\frac{G\mEL^2}{e^2}\right)}.
 \label{eq:NnegOverNposNOatmoEXPLICITfor5third}
\end{equation}

\begin{figure}%[ht]
  \includegraphics[width = 8.2truecm,scale=1.1]{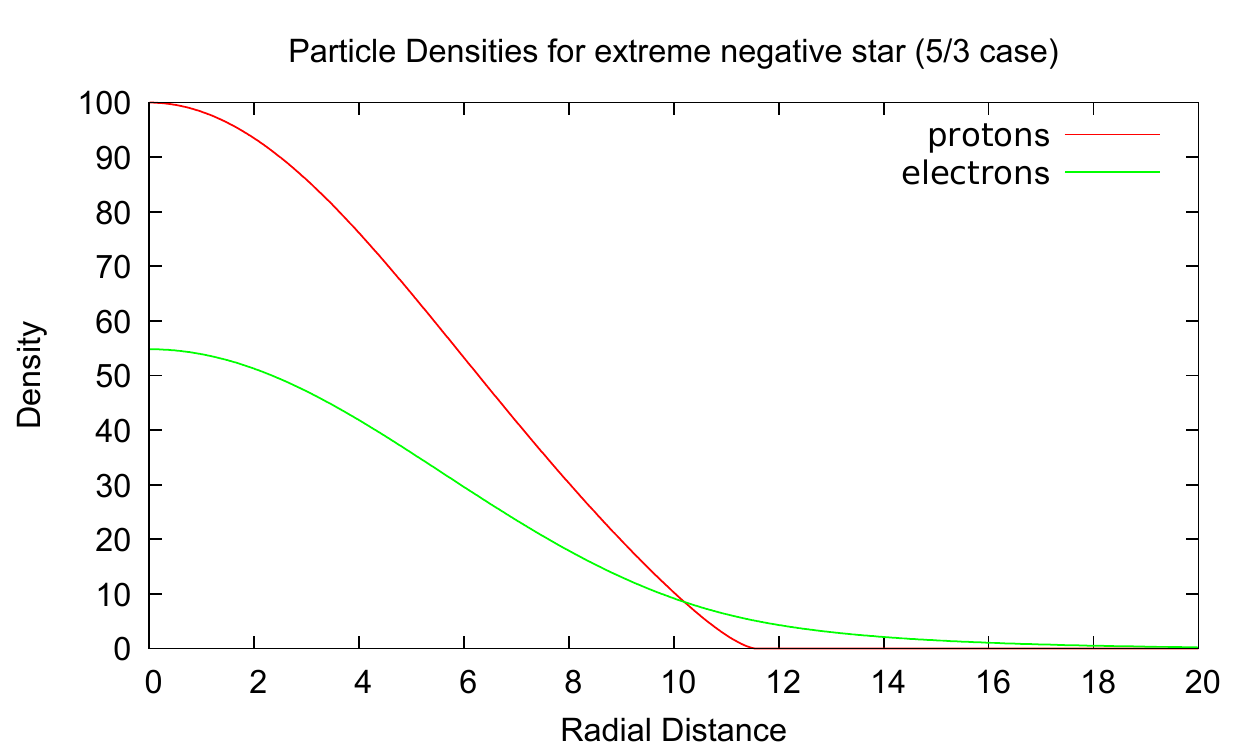} %{mpmeOFxiNEG.pdf} \vspace{-6.2truecm}
  \includegraphics[width = 8.2truecm,scale=1.1]{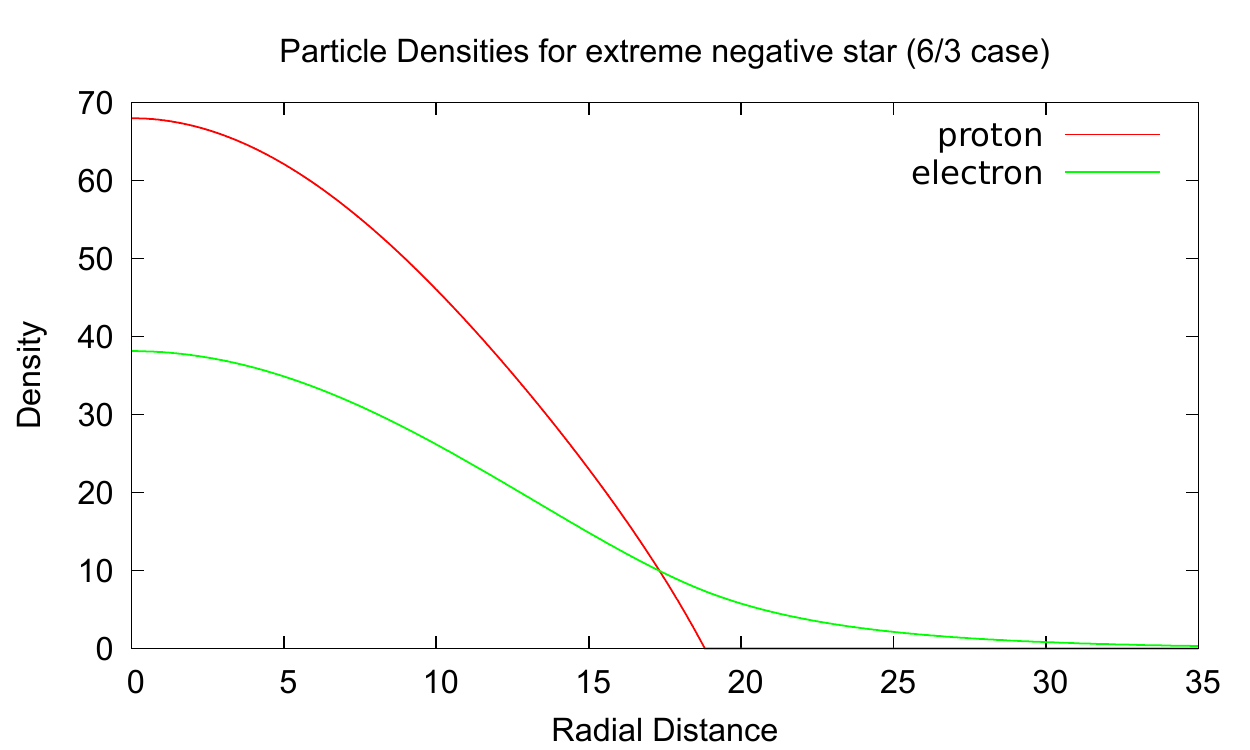} %{mpmeOFxiNEG.pdf} \vspace{-6.2truecm}
\caption{Shown are the density functions $\upos(\rho)$ and $\uneg(\rho)$ (for the units, see sect.III, first paragraph)
of the upper extreme ratio $\Nneg/\Npos = {( 1 + \tfrac{G\mPR\mEL}{e^2})}/{(1 - \tfrac{G\mEL^2}{e^2})}$,
 with science fiction values $G\mPR^2/e^2=1/2$ and $\mEL/\mPR=1/10$. 
The top graph is computed numerically, the bottom graph analytically.} \vspace{-.8truecm}
\label{53and63nuVSrhoNEGatmo}
\end{figure}

\begin{figure}%[ht]
  \includegraphics[width = 8.2truecm,scale=1.1]{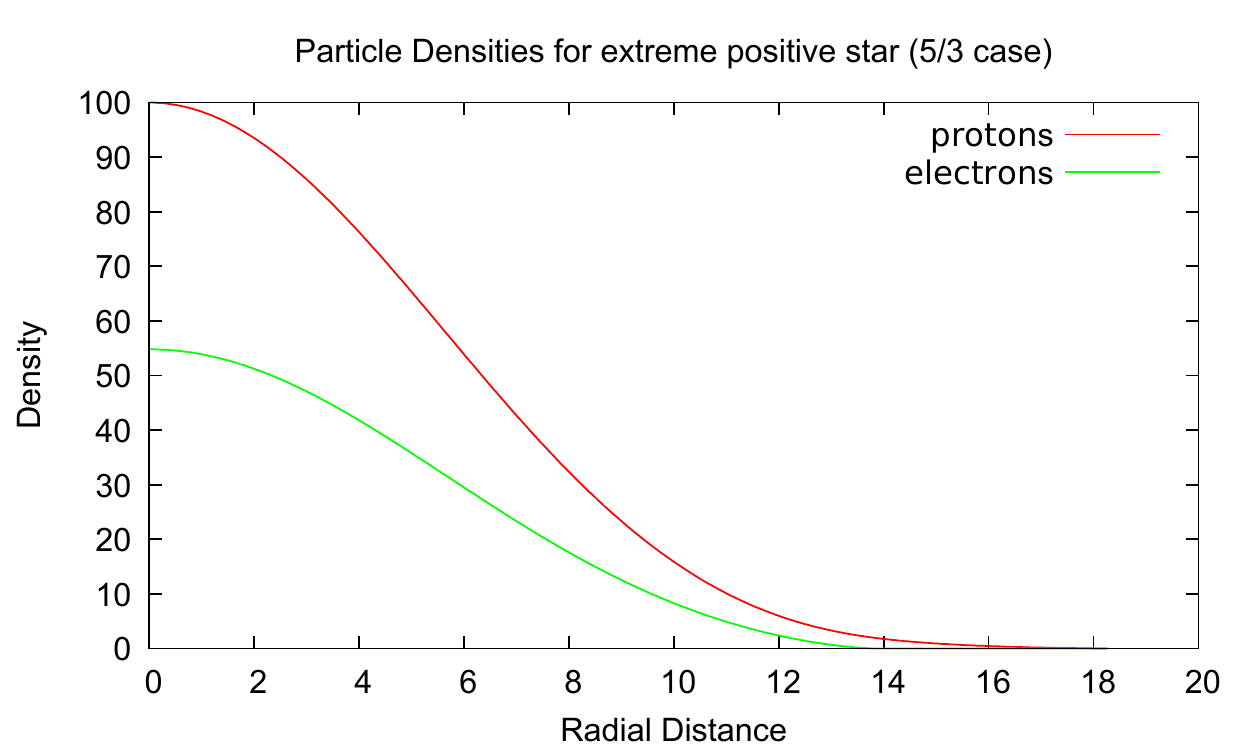} %{mpmeOFxiPOS.pdf} \vspace{-6truecm}
  \includegraphics[width = 8.2truecm,scale=1.1]{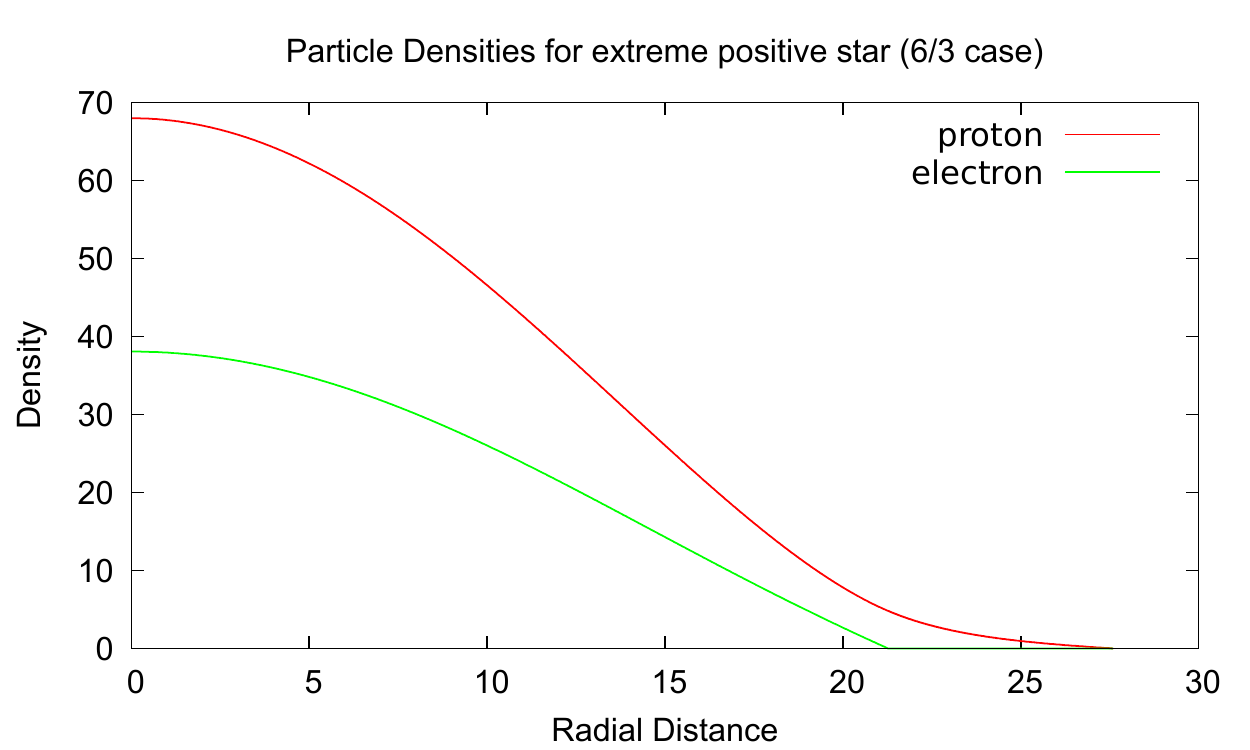} %{mpmeOFxiPOS.pdf} \vspace{-6truecm}
\caption{Shown are the density functions $\upos(\rho)$ and $\uneg(\rho)$ (for the units, see sect.III, first paragraph)
of the lower extreme ratio $\Nneg/\Npos = {(1 - \tfrac{G\mPR^2}{e^2} )}/{( 1 + \tfrac{G\mPR\mEL}{e^2})}$, 
with science fiction values $G\mPR^2/e^2=1/2$ and $\mEL/\mPR=1/10$. 
The top graph is computed numerically, the bottom graph analytically. } \vspace{-1truecm}
\label{53and63nuVSrhoPOSatmo}
\end{figure}

 Finally we turn to the extremely surcharged stars, whose densities are shown in Figs.~\ref{53and63nuVSrhoNEGatmo} 
and \ref{53and63nuVSrhoPOSatmo}.

 It is manifest that also the extreme solutions in the 6/3 model behave qualitatively similar to those in 
the 5/3 model.
 The extreme solutions of the 6/3 model have a much more spread-out bulk than those of the 5/3 model
with equal proton number $\Npos$, but their central densities are much smaller than those in the 5/3 model. 
 Interestingly, the ratio of the two central densities in the 6/3 model seems to roughly equal the one in the 5/3 model.

\begin{figure}[ht]
  \includegraphics[width = 10.2truecm, scale=2]{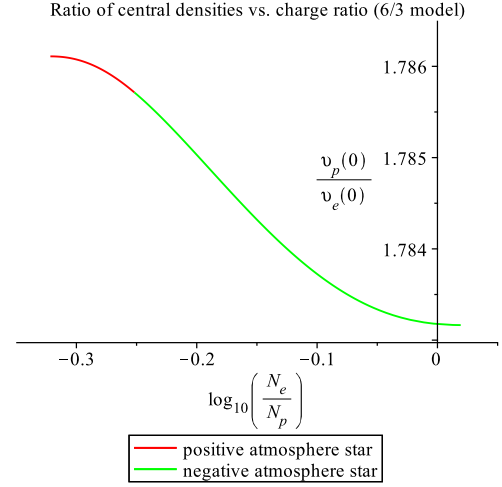} \vspace{-6truecm}
\caption{Shown is the ratio of the central proton density over central electron density 
vs. $\log_{10}$, computed analytically, of the number of electrons per proton, $\Nneg/\Npos$, covering the
full range of allowed $\Nneg/\Npos$ ratios, though for SciFi values $G\mPR^2/e^2=1/2$ and $\mEL/\mPR=1/10$. 
 Note that the central proton density is always larger than the central electron density. 
 Also this asymmetry is caused by the mass ratio $\mEL/\mPR < 1$.} %\vspace{-.5truecm}
\label{63npOneVsNeOverNp}
\end{figure}

 In all density function plots the central proton density is larger than the central electron density. 
 In Fig.~\ref{63npOneVsNeOverNp} we display the ratio of the central proton density over the central electron density
as a function of the ratio $\Nneg/\Npos$, or rather its decadic logarithm, in the 6/3 model. 
 Note the strong asymmetry caused by $\mEL/\mPR\ll 1$. 
 In the 5/3 model there will be such a curve for each value of $\Npos$ separately; all {such} curves 
coincide in the 6/3 model.\vspace{-1.2truecm}

\begin{figure}[ht]
  \includegraphics[width = 8truecm]{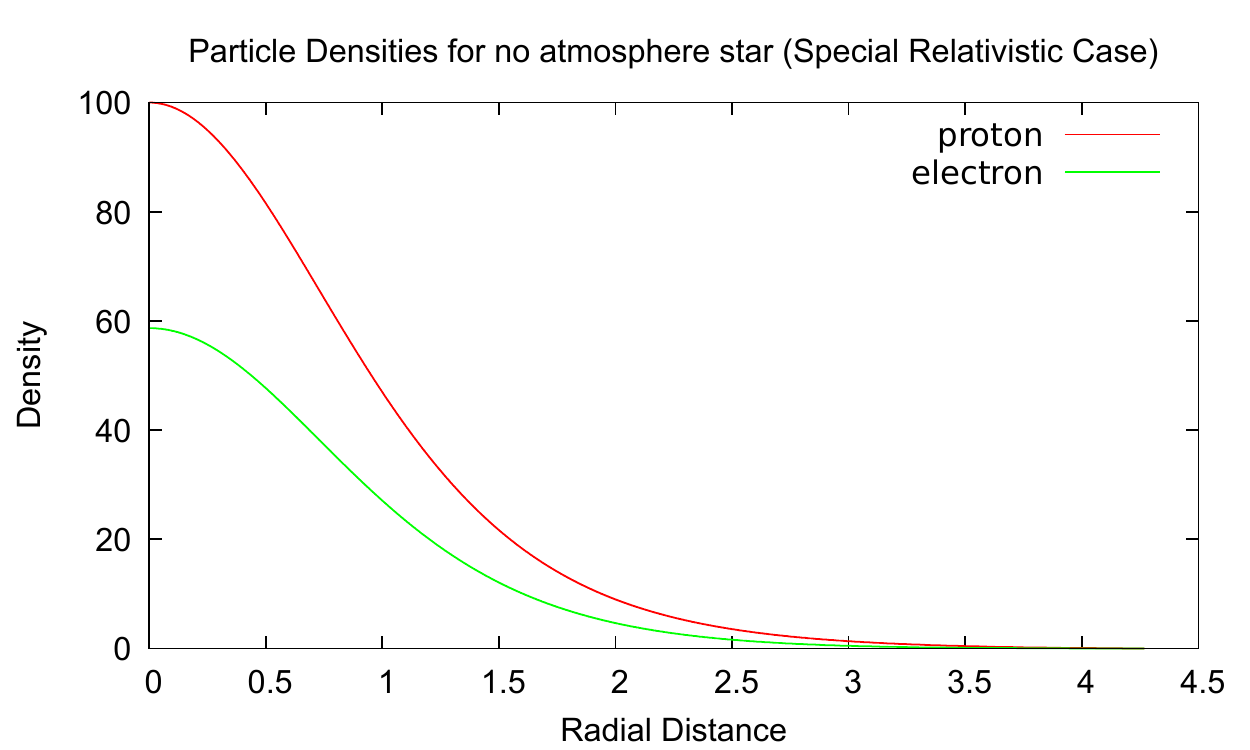} %\vspace{-7truecm}
\caption{Shown are the density functions $\upos(\rho)$ and $\uneg(\rho)$ (units in sect.III, 1st paragraph), computed numerically,
of a star without atmosphere, for SciFi values $G\mPR^2/e^2=1/2$ and $\mEL/\mPR=1/10$.}  \vspace{-.75truecm}
\label{Jfnoatmos}
\end{figure}

\begin{figure}[ht]
  \includegraphics[width = 8truecm]{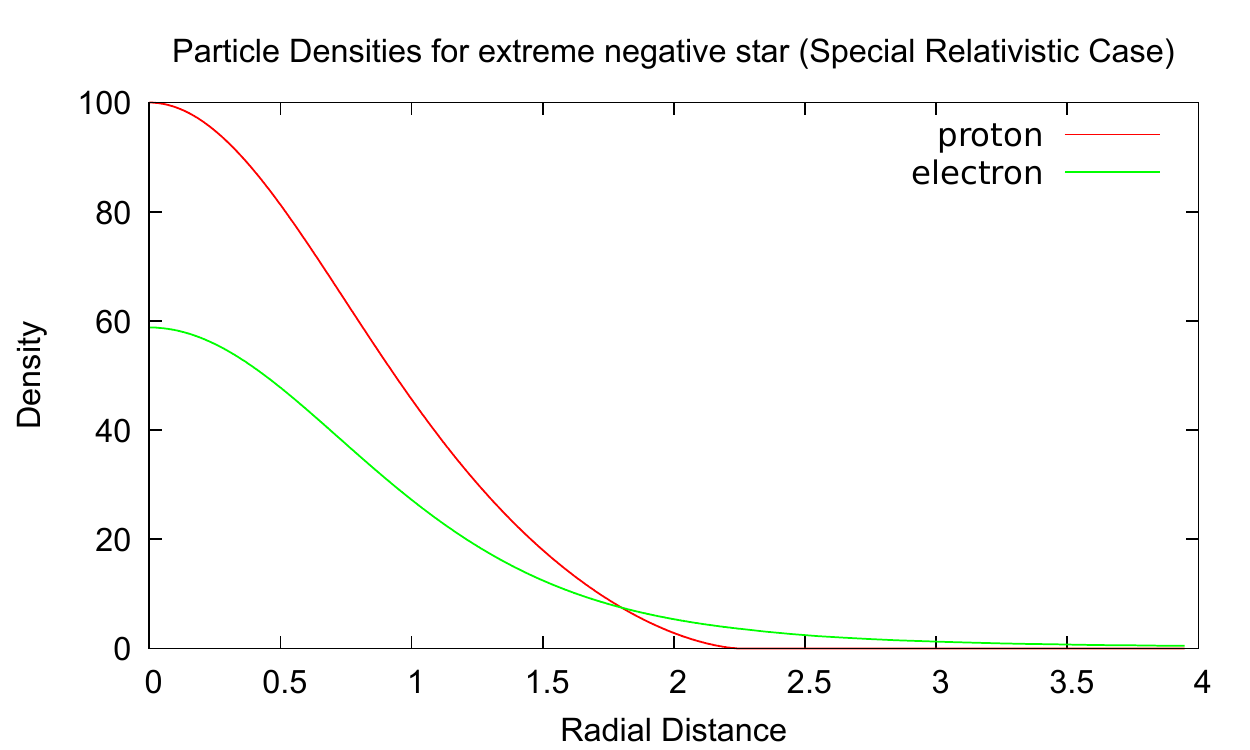}% \vspace{-7truecm}
\caption{Shown are the density functions $\upos(\rho)$ and $\uneg(\rho)$ (for the units, see sect.III, first paragraph), computed numerically,
of the upper extreme ratio $\Nneg/\Npos = {( 1 + \tfrac{G\mPR\mEL}{e^2})}/{(1 - \tfrac{G\mEL^2}{e^2})}$,
 with science fiction values $G\mPR^2/e^2=1/2$ and $\mEL/\mPR=1/10$.}  % \vspace{-.5truecm}
\label{Jfnegatmos}
%\end{figure}
%\begin{figure}[ht]
  \includegraphics[width = 8truecm]{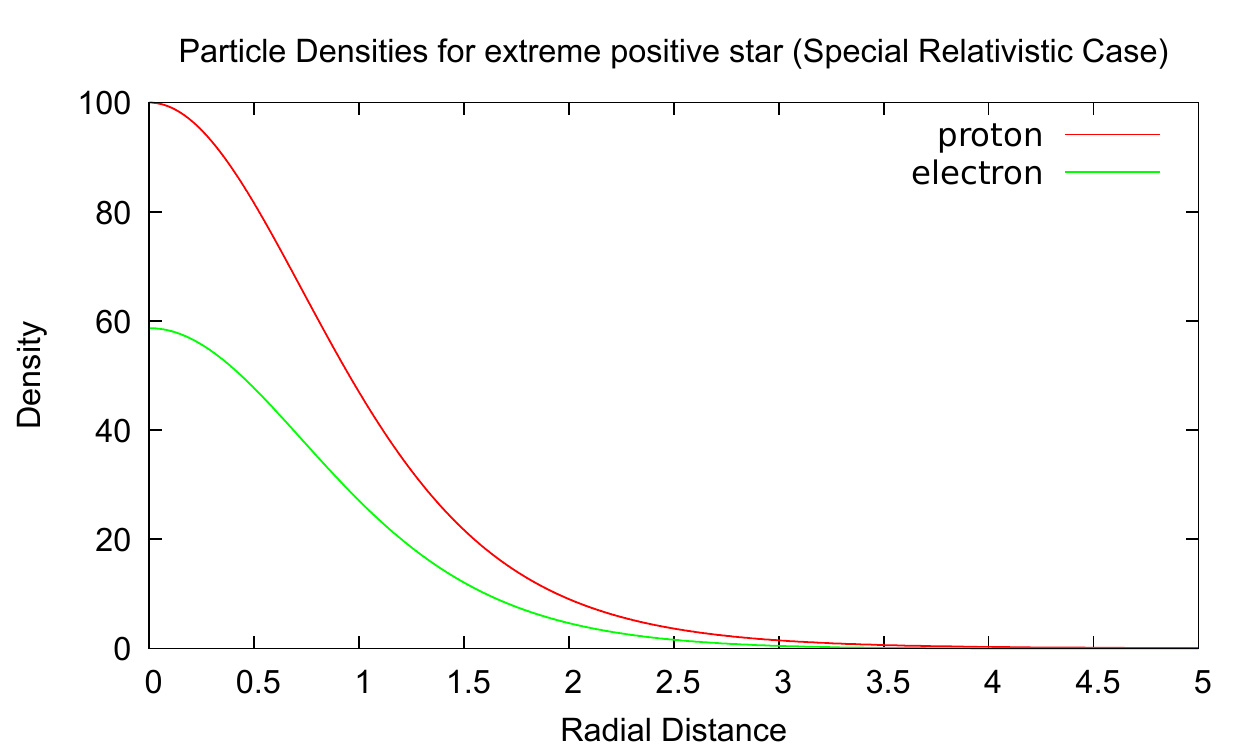} %\vspace{-7truecm}
\caption{Shown are the density functions $\upos(\rho)$ and $\uneg(\rho)$ (units in sect.III, 1st paragraph), computed numerically,
of the lower extreme ratio $\Nneg/\Npos = {(1 - \tfrac{G\mPR^2}{e^2} )}/{( 1 + \tfrac{G\mPR\mEL}{e^2})}$, 
with science fiction values $G\mPR^2/e^2=1/2$ and $\mEL/\mPR=1/10$.}  %\vspace{-.5truecm}
\label{Jfposatmos}
%\end{figure}
%\begin{figure}[ht]
  \includegraphics[width = 8truecm]{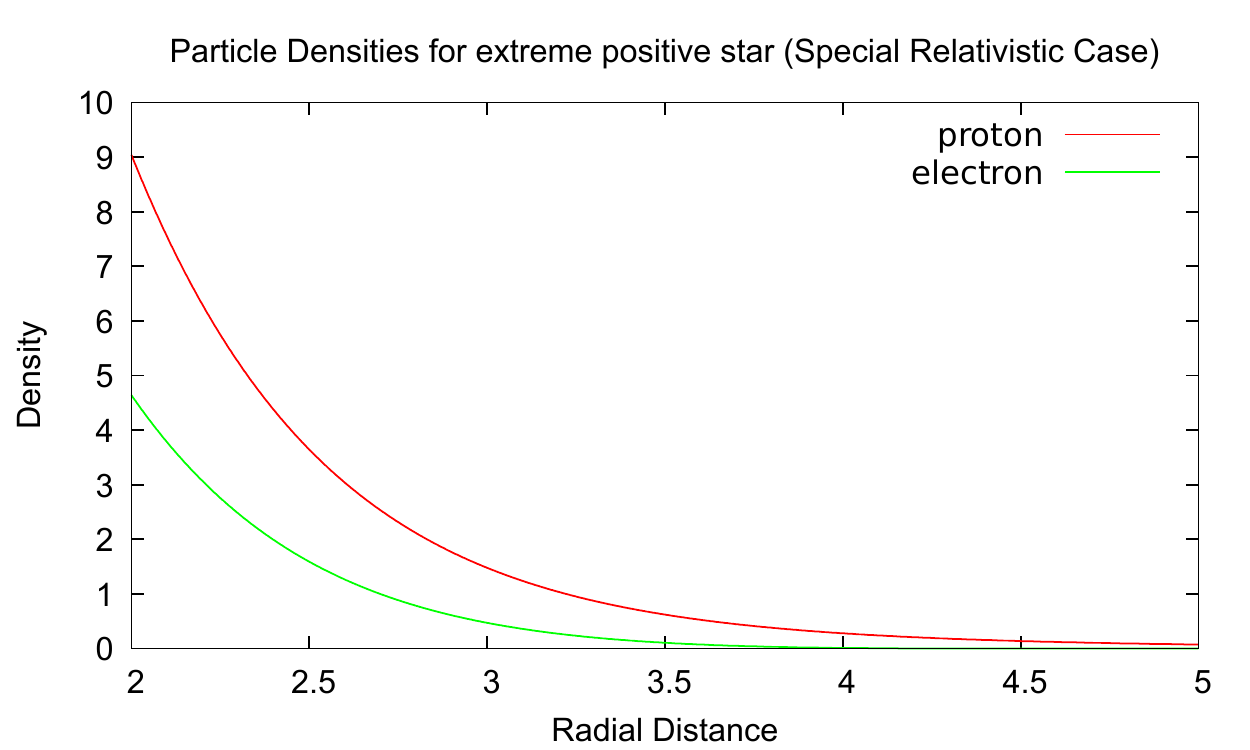} %\vspace{-7truecm}
\caption{Zoom-in of the density functions $\upos(\rho)$ and $\uneg(\rho)$ of Fig.~\ref{Jfposatmos},
revealing the positive atmosphere of the star.}  %\vspace{-1truecm}
\label{closeup}
\end{figure}

 Overall our algorithm worked sufficiently accurately so that we decided to trust it also 
for the special-relativistic model of the Chandrasekhar type.
 Figs. \ref{Jfnoatmos}, \ref{Jfnegatmos}, and \ref{Jfposatmos} are the Chandrasekhar-type special relativistic model 
equivalents of Figs. \ref{53and63nuVSrhoNOatmo}, \ref{53and63nuVSrhoNEGatmo}, and \ref{53and63nuVSrhoPOSatmo}; 
see \ref{sec:U} for a theoretical discussion and the presentation of the equations of this model.
 The figures indicate that the solutions of this model behave qualitatively in the same way as the 6/3 and 5/3 models, 
although we have done no rigorous analysis to estimate the similarity. It is difficult to see a difference between
 the plots with no atmosphere and with an extremely positive atmosphere when viewing the entire density curves.
 We have therefore included an extra plot of the positive atmospheric model at a smaller scale so that one can 
see that the behavior is indeed the same qualitatively as {in} the other models.\vspace{-.5truecm}
 \newpage

\begin{figure}[ht]
  \includegraphics[width = 8truecm]{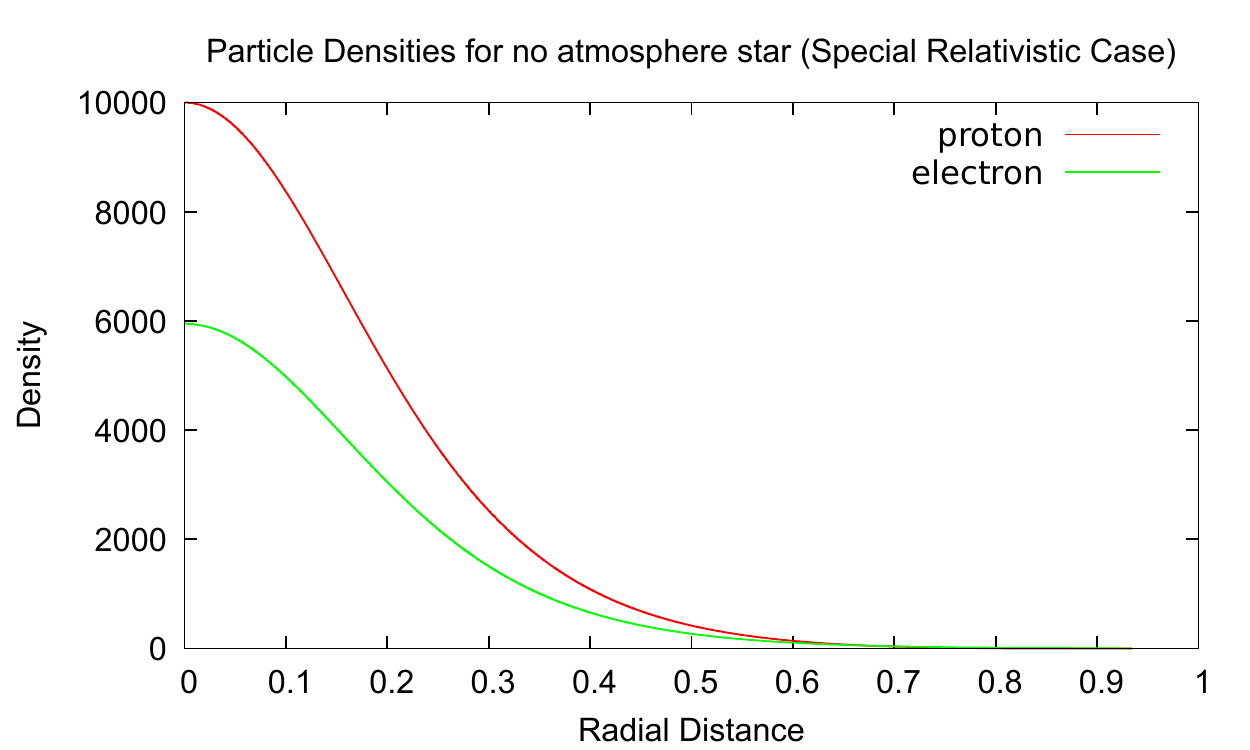} %\vspace{-7truecm}
\caption{Shown are the density functions $\upos(\rho)$ and $\uneg(\rho)$ (units in sect.III, 1st paragraph), computed numerically,
of a star without atmosphere and higher central densities, for science fiction values $G\mPR^2/e^2=1/2$ and $\mEL/\mPR=1/10$.} 
\vspace{-.5truecm}
\label{highdensity}
\end{figure}

 Here we can make some observations about the numerics of this model. 
As in the case of comparing the 6/3 to the 5/3 model, the special relativistic model has bulk radius smaller than the 5/3 model.
 This makes sense as the pressure law interpolates between 5/3 and 4/3. 
 
 Another characteristic of these solutions which cannot easily be shown in figures is that the decay rate of the extreme solutions 
is smaller than those of the 5/3 model. Again, this makes sense when one considers the exact atmospheric solution for the 5/3 model
 presented in the appendix. An ultra-relativistic 4/3 model would have an exact atmospheric solution with a 
(non-integrable) decay rate of $\eta=-3$ compared to $\eta=-6$ for the 5/3 model.

 A final interesting observation is that it appears that the possible masses for the special-relativistic model exists in a small 
band; that is, one cannot scale the solutions and increase the total number of proton and electrons as in the 6/3 or 5/3 models. 
Fig. \ref{highdensity} shows that as the central densities of the model are increased, the radius decreases. The result is that 
the total mass changes by a small amount, hardly at all. 
 
 Of course, these observations are only based on numerical results, and therefore need to be verified mathematically to make any 
definitive statement. This will be the subject of our future work. %\vspace{-.5truecm}
%\newpage

%$\phantom{ix}$ 
 \newpage

%%%%%%%%%%%%%%%%%%%%%%%%%%%%%%%%%%%%%%%%%%%%%%%%%%%%%%%%%%%%%%
%%%%%%%%%%%%%%%%%%%%%%%%%%%%%%%%%%%%%%%%%%%%%%%%%%%%%%%%%%%%%%
%%%%%%%%%%%%%%%%%%%%%%%%%%%%%%%%%%%%%%%%%%%%%%%%%%%%%%%%%%%%%%%%%%%%
\section{Can excess charge have a noticeable effect on the orbits of a binary system?}\vspace{-.5truecm}
%%%%%%%%%%%%%%%%%%%%%%%%%%%%%%%%%%%%%%%%%%%%%%%%%%%%%%%%%%%%%%
%%%%%%%%%%%%%%%%%%%%%%%%%%%%%%%%%%%%%%%%%%%%%%%%%%%%%%%%%%%%%%
%%%%%%%%%%%%%%%%%%%%%%%%%%%%%%%%%%%%%%%%%%%%%%%%%%%%%%%%%%%%%%%%%%%%

 As an application of our surcharge bounds (\ref{eq:NnegOverNposINTERVAL}), consider the following scenario.
 Suppose a maximal negatively charged and a maximal positively charged failed
white dwarf have formed a binary system of two equal mass components, each with mass $M$.
 The binary system is supposed to be sufficiently separated to vindicate the spherical approximation for their
shapes.
 Moreover, the atmospheric densities, which are rapidly decaying to zero, will be treated as having an effectively 
finite radius compared to the separation distance.
 The maximal charge imbalance is tiny, true, but since the microscopic electric coupling constants are so much stronger than 
the gravitational ones, it is in principle conceivable that even a tiny surcharge could be influencing the dynamics 
in a significant way.
 So let us find out by doing a calculation.

Note that $\Nneg\approx\Npos := N$ to high accuracy. 
 From (\ref{eq:NnegOverNposINTERVAL}) we obtain for the Coulomb coupling coefficient
of a maximal oppositely surcharged binary 
\begin{eqnarray}
- \tfrac{G\mPR\mEL}{e^2} \tfrac{G\mPR^2}{e^2} N^2e^2 \approx
- \tfrac{G\mPR\mEL}{e^2} GM^2,
 \label{eq:GbeatsE}
\end{eqnarray}
where we have used that the mass $M$ of each binary component is $\approx \mPR N$.
 Since $GM^2$ is the gravitational coupling coefficient between the two binaries, (\ref{eq:GbeatsE})
reveals that the electrical attraction between the two binary components 
is still $10^{-40}$ times smaller than their gravitational attraction.
 
 Thus astronomers can relax. 
 The validity of the determination of the masses of binary components based on their orbital data with the help of the
gravitational Kepler problem is not in question.% \vspace{-.5truecm}

% \newpage
%%%%%%%%%%%%%%%%%%%%%%%%%%%%%%%%%%%%%%%%%%%%%%%%%%%%%%%%%%%%%%
%%%%%%%%%%%%%%%%%%%%%%%%%%%%%%%%%%%%%%%%%%%%%%%%%%%%%%%%%%%%%%
%%%%%%%%%%%%%%%%%%%%%%%%%%%%%%%%%%%%%%%%%%%%%%%%%%%%%%%%%%%%%%%%%%%%
\section{Cosmic Censorship}\label{sec:Penrose}\vspace{-5pt}
%%%%%%%%%%%%%%%%%%%%%%%%%%%%%%%%%%%%%%%%%%%%%%%%%%%%%%%%%%%%%%%%%%%%
%%%%%%%%%%%%%%%%%%%%%%%%%%%%%%%%%%%%%%%%%%%%%%%%%%%%%%%%%%%%%%%%%%%%
%%%%%%%%%%%%%%%%%%%%%%%%%%%%%%%%%%%%%%%%%%%%%%%%%%%%%%%%%%%%%%%%%%%%

 Formula (\ref{eq:NnegOverNposINTERVAL}) is equivalent to the two inequalities
\begin{eqnarray}
\left(\Npos - \Nneg \right)e^2
\leq 
{G\left(\Npos\mPR + \Nneg\mEL\right)\mPR}
 \label{eq:QupperBOUND}
\end{eqnarray}
and
\begin{eqnarray}
\left(\Npos - \Nneg \right)e^2
\geq 
 - {G\left(\Npos\mPR + \Nneg\mEL\right)\mEL}
\;.
 \label{eq:QlowerBOUND}
\end{eqnarray}
 Noting that $\left(\Npos - \Nneg \right)e = Q$ is the net charge of the star 
and $\Npos\mPR + \Nneg\mEL=M$ its mass, these yield the interval 
\begin{eqnarray}
\boxed{
 - \frac{GM\mEL}{e}
\leq 
Q
\leq 
\frac{GM\mPR}{e}
      }\;
 \label{eq:Qinterval}
\end{eqnarray}
for the total charge a star made of electrons and protons can carry. 
 This interval is ``universal'' in the same sense as formula (\ref{eq:NnegOverNposINTERVAL}) is, 
recall section IV.
 Furthermore,  \cite{HKTFH}, the left inequality in \eqref{eq:Qinterval} is also ``universal'' in a wider sense, 
namely it holds also in a Thomas--Fermi--Hartree model for the ground state of a star which consists 
of electrons, protons, and several species of heavier nuclei that are bosons 
(recall our discussion in the introduction); the right inequality is possibly no longer true in the presence of bosons. 
 This of course does not follow from our derivation here, in which nuclei heavier than protons are absent.

 From \eqref{eq:Qinterval} we can derive an important inequality for $Q^2$, as follows.
 Considering first the left inequality in \eqref{eq:Qinterval},
we multiply through with $ - \Nneg e$, which yields $- \Nneg e Q \leq GM\Nneg \mEL$. 
  Considering next the right inequality in \eqref{eq:Qinterval},
we multiply through with $ \Npos e$, which yields $\Npos e Q \leq GM\Npos \mPR$. 
 Adding these then yields $(\Npos - \Nneg)eQ \leq GM(\Npos\mPR+\Nneg\mEL)$, viz.
\begin{eqnarray}
\boxed{
Q^2 
\leq 
GM^2
}\;.
 \label{eq:QsqrBOUND}
\end{eqnarray}
 Inequality \eqref{eq:QsqrBOUND}, here derived from a Thomas--Fermi model for the ground state of
a star made of protons and electrons, is also valid for a Reissner--Weyl--Nordstr\"om black hole. 
 A  Reissner--Weyl--Nordstr\"om spacetime which violates \eqref{eq:QsqrBOUND} features a naked
singularity, i.e. a singularity which is not hidden from an infinitely remote observer behind a 
closed event horizon. 
 In section IV.D we explained why our bounds (\ref{eq:NnegOverNposINTERVAL}) are to be expected to be
valid also when Newtonian gravity is replaced by Einsteinian gravity, hence a general-relativistic
treatment of a two-species Thomas--Fermi model of a failed white dwarf should also obey the bounds
\eqref{eq:QsqrBOUND} on the stellar charge. 
 This implies that if the quantum mechanical stabilization was magically turned off, such a star
could not collapse to a charged naked singularity but would turn into a charged black hole.

 Thus we have arrived at an important result in support of Penrose's weak cosmic censorship
hypothesis. 

%%%%%%%%%%%%%%%%%%%%%%%%%%%%%%%%%%%%%%%%%%%%%%%%%%%%%%%%%%%%%%
%%%%%%%%%%%%%%%%%%%%%%%%%%%%%%%%%%%%%%%%%%%%%%%%%%%%%%%%%%%%%%
%%%%%%%%%%%%%%%%%%%%%%%%%%%%%%%%%%%%%%%%%%%%%%%%%%%%%%%%%%%%%%%%%%%%
\section{Conclusions}\label{sec:CONCLUSIONS}\vspace{-5pt}
%%%%%%%%%%%%%%%%%%%%%%%%%%%%%%%%%%%%%%%%%%%%%%%%%%%%%%%%%%%%%%%%%%%%
%%%%%%%%%%%%%%%%%%%%%%%%%%%%%%%%%%%%%%%%%%%%%%%%%%%%%%%%%%%%%%%%%%%%
%%%%%%%%%%%%%%%%%%%%%%%%%%%%%%%%%%%%%%%%%%%%%%%%%%%%%%%%%%%%%%%%%%%%

In this paper we have presented the complete solution of an approximate model of a failed white dwarf star,
here for simplicity assumed to consist of electrons and protons only, in which the polytropic power $5/3$ of 
the pressure-density relation, predicted by non-relativistic quantum mechanics, is replaced with the nearby $6/3$, 
and which was introduced in \cite{KNY}.
 Based on the availability of the elementary exact solutions of this model we were able to discuss the whole
solution family thoroughly. 
 The model captures the qualitative behavior of the solutions of the physical $5/3$ and special relativistic models
correctly, and even gets the quantitative answer to the question of the maximal relative surcharge exactly
right, see (\ref{eq:NnegOverNposINTERVAL}); this we have shown in section \ref{sec:U} {(more on that below).}
 As was demonstrated with the 5/3 model, it can serve as a test case for computer algorithms which tackle 
physically more realistic many species models; see also the appendix for a brief discussion of how the 
{model generalizes} to more than two species.
  The approximate model also can easily be incorporated in an introductory astrophysics course which covers
the basic equations of stellar structure, in particular for white and brown dwarf stars. 

 In this vein we can compare the 6/3 model of \cite{KNY} with 
examples from statistical mechanics that come to mind.
 The most prominent ones, perhaps, are the two-dimensional Ising model, \cite{Onsager}, 
and the two-dimensional ice models \cite{Lieb}, \cite{Sutherland}, which have 
provided valuable qualitative insights into the behavior of the more realistic physical models
that require heavy use of numerical methods.

 The equations for two-species models corresponding to different polytropic laws than those discussed 
above can easily be written down, although the arguments presented here concerning the structure of solutions may not hold.
 For example, the reasoning behind  the saturation of the $N_{\mathrm{p}}/N_e$ bounds does not hold for polytropic laws of 
index larger than $n=3$, the ultrarelativistic white dwarf case. 
 But the $n=5$ case is exactly solvable and represents a border case of Lane-Emden equations, so it may 
be of some interest what can be said about the corresponding two species model. 
 While not related to the structure of white dwarfs, 
the $n=5$ polytrope does find uses in stellar dynamics and some fluid sphere solutions in general relativity. 
 We have included in the appendix an analytic solution of the two species model corresponding to the $n=5$ polytrope. 
Of course, this is just the beginning of a description of the solutions of such a model. 

 An interesting by-product of our investigation are the  $\hbar$- and $c$-independent 
bounds (\ref{eq:NnegOverNposINTERVAL}) on $\Nneg/\Npos$.
 We have presented compelling arguments for why our bounds (\ref{eq:NnegOverNposINTERVAL}) 
are the correct bounds for a failed white dwarf star made of electrons and individual protons, not only 
non-relativistically but also in the special- and general-relativistic theories, because one is far away 
from the Chandrasekhar mass.
 A more pedagogical account of these findings is presented in \cite{HK}.
 As we explained in section VIII, our bounds support Penrose's weak cosmic censorship hypothesis. 

 After submitting this paper we started to investigate more realistic stellar ground state models that,
inevitably, are no longer exactly solvable; cf. \cite{HKTFH} for a non-relativistic Thomas--Fermi--Hartree model
of electrons, protons, and several species of nuclei that are bosons (such as $\alpha$ particles).
 The bounds on $\Nneg/\Npos$ will then be replaced by bounds on the ratios $N_z/\Nneg$, where $N_z$ is the
number of nuclei in the star with $z$ elementary charges.
 In regard to the question of the maximal and minimal electric charges on a star, we found that the lower bound expressed in 
\eqref{eq:Qinterval} is valid also in the Thomas--Fermi--Hartree model; the upper bound in 
\eqref{eq:Qinterval} is modified, though.
 Thus our negative surcharge bound obtained in this paper with a highly simplified model is very robust.

 We also plan to investigate the special-relativistic formulation of the problem all the way up to near to
the critical Chandrasekhar mass, taking a mixture of different nuclei species into account with a mix of 
special-relativistic Thomas--Fermi and Hartree type equations. 
 We expect the left inequality in  \eqref{eq:Qinterval} to remain valid. 

 We also want to investigate the general-relativistic problem, with its effects on the critical mass;
an interesting question is whether \eqref{eq:QsqrBOUND} will hold.

 In the pursuit of more realism also the weak and strong nuclear forces should eventually
be taken into account if one considers stellar ground states with masses close to the 
above-mentioned critical Chandrasekhar mass, respectively the general-relativistic critical mass, for
then the central densities exceed the threshold for inverse $\beta$ decay, which will turn a certain
percentage of electrons and protons (bound in the nuclei) into neutrons, thus changing the composition
of the star and affecting  the critical mass. 
 Since inverse $\beta$ decay preserves the total charge involved in the process, it should not affect
the allowed surplus charge $Q$ on a star.

 So much on the excess charges of stellar ground states.
 We close this discussion by reminding the reader that the question of electrical surplus of charge
on a star is mostly meaningful for the ground state. 
 Real stars in the universe are estimated not yet to be in their ground state, and since finite temperature 
effects include the phenomenon of solar / stellar winds, real stars which constantly evaporate render 
the question of their electrical surcharge pointless.  

 Back to the exactly solvable model, very much of interest is to extend it, by including 
magnetism and rotation.
 This will complicate the problem considerably, for the spherical symmetry of the problem will be broken 
both by rotation, due to centrifugal effects (obviously), and by  magnetism's anisotropy, cf.
\cite{DM1}, \cite{DM2}, \cite{BB}, \cite{CFC}, \cite{CFD}. 

\vspace{-15pt}

%%%%%%%%%%%%%%%%%%%%%%%%%%%%%%%%%%%%%%%%%%%%%%%%%%%%%%%%%%%%%%%%%%%%%%%%%%%%%%%%%
%%%%%%%%%%%%%%%%%%%%%%%%%%%%%%%%%%%%%%%%%%%%%%%%%%%%%%%%%%%%%%%%%%%%%%%%%%%%%%%%%
%%%%%%%%%%%%%%%%%%%%%%%%%%%%%%%%%%%%%%%%%%%%%%%%%%%%%%%%%%%%%%%%%%%%%%%%%%%%%%%%%
\section*{Appendix} \vspace{-10pt}
%%%%%%%%%%%%%%%%%%%%%%%%%%%%%%%%%%%%%%%%%%%%%%%%%%%%%%%%%%%%%%%%%%%%%%%%%%%%%%%%%
%%%%%%%%%%%%%%%%%%%%%%%%%%%%%%%%%%%%%%%%%%%%%%%%%%%%%%%%%%%%%%%%%%%%%%%%%%%%%%%%%
%%%%%%%%%%%%%%%%%%%%%%%%%%%%%%%%%%%%%%%%%%%%%%%%%%%%%%%%%%%%%%%%%%%%%%%%%%%%%%%%%

%%%%%%%%%%%%%%%%%%%%%%%%%%%%%%%%%%%%%%%%%%%%%%%%%%%%%%%%%%%%%% 
%%%%%%%%%%%%%%%%%%%%%%%%%%%%%%%%%%%%%%%%%%%%%%%%%%%%%%%%%%%%%%
                \subsection{The 6/3 model for more than two species}\vspace{-10pt}
%%%%%%%%%%%%%%%%%%%%%%%%%%%%%%%%%%%%%%%%%%%%%%%%%%%%%%%%%%%%%%
%%%%%%%%%%%%%%%%%%%%%%%%%%%%%%%%%%%%%%%%%%%%%%%%%%%%%%%%%%%%%%

 The $6/3$ model can be easily generalized to
an arbitrary number of fermion species without affecting its exact solubility. 
 It is of course to be seen as a $5/3\to 6/3$ approximation to a more-than-two species $5/3$ model, 
and such a model has the physical deficiency that all species are fermions, whereas both the primordial 
nucleosynthesis \cite{Schramm} and also nuclear fusion in stars \cite{KippenhahnWeigert} essentially produces 
effectively bosonic heavier nuclei, in particular 
${}^4$He (both primordial nucleosynthesis and stellar fusion), ${}^{12}$C and ${}^{16}$O (the latter predominantly
only in stellar fusion).
 A multi-species fermion model is surely to be taken with some grain of salt. 
 All the same, the usual local neutrality approximation, traditionally used in astrophysical works on stellar 
structure, throws all these differences out the window also, so that a multi-species fermionic model is presumably 
not worse. 

 Of course, the combinatorial complexity increases, and for more than four species the $\kappa$ and
$\varkappa$ eigenvalues can no longer be expressed in closed form, but even for three and four species,
when one can, the closed form expressions are not very illuminating. 
 Fortunately this is not necessary, since the fantastic tininess of the ratios of the coupling constants of the
various species allow a very efficient evaluation with approximate expressions which are more accurate than
any typical numerical approximation on a machine. 

 Instead of presenting here the generalization to an arbitrary number of fermion species, we present
the three-species version, pretending that because of some unlikely fluke
the primordial nucleosynthesis \cite{Schramm} has, in some corner of the universe, produced a mix
of only protons and ${}^3$He, a spin-1/2 fermion with two elementary charges known as \emph{helion},
which together with the protons and the electrons now constitutes our failed white dwarf, if 
the total mass remains below $\approx 80$ Jupiter masses.
 One may also contemplate modelling a low mass white dwarf (no longer failed)
if the mass is a bit above the threshold for the onset of nuclear fusion, but not too high so that no fusion
into heavier nuclei than helium happened; one has to pretend that by some even more unlikely statistical fluke, 
also in the star only ${}^3$He is produced. 
 (Obviously, this narrative is not meant to be taken literally.)

Choosing ${}_h$ as subscript for ${}^3$He we now have the following coupled system of three linear 
second-order differential equations for the dimensionless density functions $\uHe$, $\upos$, and $\uneg$ in the bulk region,
\vfill
\begin{widetext}
\begin{eqnarray} 
- \veps_h \varsigma \frac{1}{\rho^2}\left(\rho^2\uHe^{\prime}(\rho)\right)^\prime
&\,\  =  - \left(4 -\tfrac{G\mHe^2}{e^2}\right) \uHe(\rho) - \left(2 -\tfrac{G\mHe\mPR}{e^2}\right) \upos(\rho) 
+ \left(2^{} +\tfrac{G\mHe{\mEL}_{}}{e^2}\right) \uneg(\rho),  \label{eq:PoissonMUdrei}\\
- \veps \varsigma \frac{1}{\rho^2}\left(\rho^2\upos^{\prime}(\rho)\right)^\prime
&\,\  = - \left(2 -\tfrac{G\mHe\mPR}{e^2}\right) \uHe(\rho) - \left(1 -\tfrac{G\mPR^2}{e^2}\right) \upos(\rho) 
+ \left(1 +\tfrac{G\mPR\mEL}{e^2}\right) \uneg(\rho),
 \label{eq:PoissonMUpNEU}\\
- \varsigma \frac{1}{\rho^2}\left(\rho^2\uneg^{\prime}(\rho)\right)^\prime
&\!\!  = 
\left(2 + \tfrac{G\mHe\mEL}{e^2}\right) \uHe(\rho) +
\left(1 + \tfrac{G\mPR\mEL}{e^2}\right) \upos(\rho) - \left(1 - \tfrac{G\mEL^2}{e^2}\right) \uneg(\rho),
 \label{eq:PoissonMUeNEU}
\end{eqnarray}
\end{widetext}
valid where $\upos(\rho)>0$, $\uneg(\rho)>0$, and $\uHe(\rho)>0$.   % ******
 Here, $\mHe\approx 3\mPR$ is the mass of ${}^3$He and $\veps_h:=\mEL/\mHe\approx \veps/3$. 
 
Sandwiched between the three-species bulk and the single-species atmosphere regions 
there is now an intermediate region where exactly one of the densities vanishes and two of the densities are non-zero. 
 The density functions in the intermediate region satisfy precisely the bulk equations of the two-species model,
except perhaps that we need to allow for the possibility that it is not a proton-electron system now but a ${}^3$He-electron
system, even though astrophysical stellar models suggest that the Helium zone resides inside the Hydrogen zone.

 Different from the discussion of the bulk region of the two-species model, though, the two-species intermediate zone does not
require using only $\sin(\kappa_t \rho)/\rho$ and $\sinh(\kappa_h\rho)/\rho$ linear combinations, because one stays away
from the center of the star.

 As for the three-species bulk region, it is clear that the ansatz $\upsilon_f(\rho) = A_f\frac{\exp(\kappa \rho)}{\rho}$
will once again lead to an eigenvalue problem for $\kappa$, this time it is a cubic equation in $\kappa^2$.
 Initial conditions at $\rho=0$ are posed, namely the vanishing of the radial derivatives, while the central densities
are to be chosen such as to satisfy the constraints that the particle densities integrate to $\Nhel$, $\Npos$, and $\Nneg$;
recall, $\Npos$ is the number of individual protons, not bound in nuclei with $Z>1$ elementary charges.
 At the interface between bulk and intermediate regions, the two non-vanishing densities of the intermediate region have to 
go over continuously differentiable into the bulk region, and at the intermediary-atmosphere interface, the earlier continuous
differentiability conditions for the atmospheric density is imposed. 

 It is clear that the combinatorial possibilities are already daunting for this three-species setup, but it is also
clear that it can be worked out completely, and analogously one can proceed with an arbitrary number of fermion
species, in principle at least. % \vspace{-15pt}
\newpage

%%%%%%%%%%%%%%%%%%%%%%%%%%%%%%%%%%%%%%%%%%%%%%%%%%%%%%%%%%%%%%
%%%%%%%%%%%%%%%%%%%%%%%%%%%%%%%%%%%%%%%%%%%%%%%%%%%%%%%%%%%%%%
                \subsection{Some exact solutions to related models}\vspace{-10pt}
%%%%%%%%%%%%%%%%%%%%%%%%%%%%%%%%%%%%%%%%%%%%%%%%%%%%%%%%%%%%%%
%%%%%%%%%%%%%%%%%%%%%%%%%%%%%%%%%%%%%%%%%%%%%%%%%%%%%%%%%%%%%%

%%%%%%%%%%%%%%%%%%%%%%%%%%%%%%%%%%%%%%%%%%%%%%%%%%%%%%%%%%%%%%
                \subsubsection{An elementary no-atmosphere solution for\\ the two-species 6/5 model}\vspace{-10pt}
%%%%%%%%%%%%%%%%%%%%%%%%%%%%%%%%%%%%%%%%%%%%%%%%%%%%%%%%%%%%%%

  As has been discussed, the multi-species 6/3 model is a generalization of the Lane--Emden equation of index $n=1$,
and its no-atmosphere solutions are obtained by rescaling the elementary solution of the Lane--Emden equation of index $n=1$.
 Aside from the less interesting index $n=0$ case, there is also an elementary no-atmosphere solution to the index $n=5$ case,
which corresponds to a polytropic pressure law with power $\gamma=6/5$.
  Following the procedure in \ref{5363}, we can derive the  two-species equivalent of (\ref{eq:PoissonMUpTHREEhalf}) and 
(\ref{eq:PoissonMUeTHREEhalf}) for the index $n=5$ case: 
\begin{widetext}
\begin{eqnarray}
- \veps \tau \frac{1}{\rho^2}\frac{d}{d\rho}\left(\rho^2\frac{d}{d\rho} \upos^{1/5}(\rho)\right)
&\,\  =  - \left(1 -\frac{G\mPR^2}{e^2}\right) \upos(\rho) + \left(1 +\frac{G\mPR\mEL}{e^2}\right) \uneg(\rho),
 \label{eq:6/5p}\\
- \tau \frac{1}{\rho^2}\frac{d}{d\rho}\left(\rho^2\frac{d}{d\rho} \uneg^{1/5}(\rho)\right)
&\!\!  = \left(1 + \frac{G\mPR\mEL}{e^2}\right) \upos(\rho) - \left(1 - \frac{G\mEL^2}{e^2}\right) \uneg(\rho);
 \label{eq:6/5e}
\end{eqnarray}
here, $\tau=\frac{\hbar c}{e^2}\frac{\pi^{1/3}3^{5/3}}{50}$, and these are equations valid where both densities are positive.
 Let $\theta_f(\rho)=\upsilon_f^{1/5}(\rho)$ on the set where $\upsilon_f$ is positive.
 Then these equations become  
\begin{eqnarray}
- \veps \tau \frac{1}{\rho^2}\frac{d}{d\rho}\left(\rho^2\frac{d}{d\rho} \theta_{\mathrm{p}}(\rho)\right)
&\,\  =  - \left(1 -\frac{G\mPR^2}{e^2}\right) \theta^5_{\mathrm{p}}(\rho) + \left(1 +\frac{G\mPR\mEL}{e^2}\right) \theta^5_e(\rho),
 \label{eq:6/5pNEW}\\
- \tau \frac{1}{\rho^2}\frac{d}{d\rho}\left(\rho^2\frac{d}{d\rho} \theta_e(\rho)\right)
&\!\!  = \left(1 + \frac{G\mPR\mEL}{e^2}\right) \theta^5_{\mathrm{p}}(\rho) - \left(1 - \frac{G\mEL^2}{e^2}\right) \theta^5_e(\rho).
 \label{eq:6/5eNEW}
\end{eqnarray}
Let us try $\theta_f(\rho)=\alpha_f(1+\kappa_f\frac{\rho^2}{3})^{-1/2}$, adapting the Lane--Emden index $n=5$
 solution similarly to how we adapted the Lane--Emden index $n=1$ solution for the $\gamma=6/3$ case. 
Then this system reduces to 
\begin{eqnarray}
\veps \tau \alpha_{\mathrm{p}} \kappa_{\mathrm{p}} \left(1+\kappa_{\mathrm{p}}\tfrac{\rho^2}{3}\right)^{-5/2}
&\,\  =  -\alpha_{\mathrm{p}} \left(1 -\frac{G\mPR^2}{e^2}\right)\left(1+\kappa_{\mathrm{p}}\frac{\rho^2}{3}\right)^{-5/2}  
+\alpha_e \left(1 +\frac{G\mPR\mEL}{e^2}\right) \left(1+\kappa_e\frac{\rho^2}{3}\right)^{-5/2},
 \label{eq:thetap}\\
\tau\alpha_e \kappa_e \left(1+\kappa_e\tfrac{\rho^2}{3}\right)^{-5/2} 
&\!\! = \alpha_{\mathrm{p}}\left(1 + \frac{G\mPR\mEL}{e^2}\right) \left(1+\kappa_{\mathrm{p}}\frac{\rho^2}{3}\right)^{-5/2} 
- \alpha_e\left(1 - \frac{G\mEL^2}{e^2}\right)\left(1+\kappa_e\frac{\rho^2}{3}\right)^{-5/2} .
 \label{eq:thetae}
\end{eqnarray}
\end{widetext}
To have equality, we then must take $\kappa_{\mathrm{p}}=\kappa_e=\kappa$. So as in the 6/3 case, we obtain a matrix problem: 
\begin{equation}
\hspace{-.2truecm}
\left(
\begin{array}{cc}
1 -\tfrac{G\mPR^2}{e^2} +  \kappa \veps\tau \;; &  - 1 -\tfrac{G\mPR\mEL}{e^2} \\
- 1 -\tfrac{G\mPR\mEL}{e^2}\quad\; ; & 1 -\tfrac{G\mEL^2}{e^2} +  \kappa \tau
\end{array}
\right)\!\!
\left(
\begin{array}{c}
\!\alpha_{\mathrm{p}}\!  \\
\!\alpha_e\!
\end{array}
\right)
=
\left(
\begin{array}{c}
\!0\!  \\
\!0\!
\end{array}
\right). \hspace{-.3truecm}
 \label{eq:matrixPROBLEM2}
\end{equation}
Again, the determinant must be zero, {and we obtain the following quadratic in $\kappa$, viz}: 
$a\kappa^2 +b\kappa +c =0$, with $a=\veps\tau^2>0$, $b= \tau\left(1+\veps -{G(\veps\mEL^2+\mPR^2)}/{e^2}\right)>0$,
and $c= - {G\left(\mEL + \mPR\right)^2}/{e^2}<0$.
 We find $\kappa_+\approx -1837$ and $\kappa_-\approx 3.2329\cdot 10^{-47}$.
 We cannot use $\kappa_+$ since $\theta_f(\rho)$ would have a singularity at $\sqrt{3/(-\kappa_{\mathrm{p}})}$,
 {and} so we let $\theta_f(\rho)=\alpha_f(1+\kappa_-\frac{\rho^2}{3})^{-1/2}$. 
The matrix (\ref{eq:matrixPROBLEM2}) also gives us that 
\begin{equation}\label{6/5prop}
    \alpha_{\mathrm{p}}=\frac{1+G\frac{m_em_{\mathrm{p}}}{e^2}}{1-\frac{Gm_{\mathrm{p}}^2}{e^2}+\kappa \veps\tau}\alpha_e
\end{equation}
We then obtain a one-parameter family of solutions to (\ref{eq:thetap}), (\ref{eq:thetae}) such that $\theta_{\mathrm{p}}$ 
and $\theta_e$ are in proportion as given by (\ref{6/5prop}) and have unbounded support. 
The same can therefore be said about $\upos$ and $\uneg$, solutions to (\ref{eq:6/5p}), (\ref{eq:6/5e}).\vspace{-.6truecm}

%%%%%%%%%%%%%%%%%%%%%%%%%%%%%%%%%%%%%%%%%%%%%%%%%%%%%%%%%%%%%%
                \subsubsection{An elementary solution for\\ the two-species isothermal model}\vspace{-10pt}
%%%%%%%%%%%%%%%%%%%%%%%%%%%%%%%%%%%%%%%%%%%%%%%%%%%%%%%%%%%%%%

 One of the earliest self-gravitating models, together with Homer Lane's polytropes, was Z\"ollner's isothermal 
self-gravitating ideal classical gas ball model.
 Its basic equations were later studied much more thoroughly by Emden \cite{Emden} and are nowadays named in his honor.
 We recall that the pressure-density relation of the isothermal ideal classical gas reads $p = \kB T\nu$, 
where $\kB$ is Boltzmann's constant. 
 Treating both electrons and protons as isothermal classical perfect gases, with equal temperature $T>0$,
yields the following system of nonlinear second-order differential equations for the density functions $\nupos$ and $\nuneg$,
valid wherever both  $\nupos(r)>0$ and $\nuneg(r)>0$:
\begin{widetext}
\begin{eqnarray}
- \varpi \frac{1}{r^2}\Ddr\left(r^2\Ddr \ln\nupos(r)\right)
&\,\  =  - \left(1 -\frac{G\mPR^2}{e^2}\right) \nupos(r) + \left(1 +\frac{G\mPR\mEL}{e^2}\right) \nuneg(r),
 \label{eq:PoissonMUpISOtherm}\\
- \varpi \frac{1}{r^2}\Ddr\left(r^2\Ddr \ln \nuneg(r)\right)
&\!\!  = \left(1 + \frac{G\mPR\mEL}{e^2}\right) \nupos(r) - \left(1 - \frac{G\mEL^2}{e^2}\right) \nuneg(r);
 \label{eq:PoissonMUeISOtherm}
\end{eqnarray}\vspace{-.5truecm}
\end{widetext}
here, $\varpi:= \kB T/ {4\pi e^2}$.
 Like the single-species Emden equation for the isothermal self-gravitating classical gas ball, also the system 
(\ref{eq:PoissonMUpISOtherm}), (\ref{eq:PoissonMUeISOtherm}) is not generally solvable in closed form. 
 However, following Z\"ollner's treatment of the single-species model, we can find one elementary solution to this 
two-species model by making the Ansatz $\nu_f(r) = 2\varpi A_f /r^2$ with $A_f >0$, and $_f$ standing for either $_{\mathrm{p}}$ or $_e$, 
as before.
 This Ansatz turns (\ref{eq:PoissonMUpISOtherm}), (\ref{eq:PoissonMUeISOtherm}) into 
\begin{eqnarray}
1 &\,\  =  - \left(1 -\frac{G\mPR^2}{e^2}\right) A_{\mathrm{p}} + \left(1 +\frac{G\mPR\mEL}{e^2}\right) A_e,
 \label{eq:PoissonMUpISOthermZ}\\
1 &\!\!  = \left(1 + \frac{G\mPR\mEL}{e^2}\right) A_{\mathrm{p}} - \left(1 - \frac{G\mEL^2}{e^2}\right) A_e.
 \label{eq:PoissonMUeISOthermZ}
\end{eqnarray}
 Thus
\begin{equation}
\hspace{-.2truecm}
\left(
\begin{array}{c}
\!\Apos\!  \\
\!\Aneg\!
\end{array}
\right)
=
\left(
\begin{array}{cc}
\quad - 1 +\tfrac{G\mPR^2}{e^2}\quad; &  1 + \tfrac{G\mPR\mEL}{e^2} \\
 1 + \tfrac{G\mPR\mEL}{e^2}\;\; ; & - 1 +\tfrac{G\mEL^2}{e^2} 
\end{array}
\right)^{-1}\!\!
\left(
\begin{array}{c}
\! 1\!
\\
\! 1\!
\end{array}
\right)\!\!,
 \label{eq:matrixAposAnegSOL}
\end{equation}
and the inverse matrix is the negative of (\ref{eq:matrixINV}), so
\begin{equation}
\left(
\begin{array}{c}
\!\Apos\!  \\
\!\Aneg\!
\end{array}
\right)
=
\frac{e^2}{G(\mPR+\mEL)^2}
\left(
\begin{array}{c}
2 + \tfrac{G(\mPR-\mEL)\mEL}{e^2} \\
2 - \tfrac{G(\mPR-\mEL)\mPR}{e^2} 
\end{array}
\right).
 \label{ApAeSOL}
\end{equation}
 Both $\Apos >0$ and $\Aneg>0$ thanks to the smallness of the ratio of gravitational to electric coupling constants,
hence we have found an exact solution pair to (\ref{eq:PoissonMUpISOtherm}), (\ref{eq:PoissonMUeISOtherm}).

 These $1/r^2$ densities are singular at the origin, but locally integrable.
 Of course they are not globally integrable, so $\Npos=\infty=\Nneg$. 
 Interestingly, though, the number of particles of species $_f$ inside a sphere of radius $r$, i.e.
$\cN_f(r) := 4\pi \int_0^r \nu_f(s)s^2 ds = (2\kB T/e^2) A_f r$, yields the $r$-independent ratio
\begin{equation}
\frac{\cN_e(r)}{\cN_{\mathrm{p}}(r)} = \frac{\Aneg}{\Apos} = 
\frac{2 - \tfrac{G(\mPR-\mEL)\mPR}{e^2}}{2 + \tfrac{G(\mPR-\mEL)\mEL}{e^2}}.
 \label{AeToApRATIO}
\end{equation}

 For general solution pairs of  (\ref{eq:PoissonMUpISOtherm}), (\ref{eq:PoissonMUeISOtherm}) one may proceed analogously.
 Since Emden's isothermal gas ball solutions all tend asymptotically for large $r$ to a $1/r^2$ 
behavior, we expect that $\lim_{r\to\infty}\cN_e(r)/\cN_{\mathrm{p}}(r)$ exists for each pair, and plays the role of $\Nneg/\Npos$
for such infinite-mass solutions. 
 Moreover, whenever $\lim_{r\to\infty}\cN_e(r)/\cN_{\mathrm{p}}(r)$ exists, a small modification of our arguments in section IV 
shows that the limit obeys the bounds (\ref{eq:NnegOverNposINTERVAL}) without saturation.
\vspace{-.6truecm}

%%%%%%%%%%%%%%%%%%%%%%%%%%%%%%%%%%%%%%%%%%%%%%%%%%%%%%%%%%%%%%
                \subsubsection{An exact atmospheric solution of\\ the two-species 5/3 model}\vspace{-5pt}
%%%%%%%%%%%%%%%%%%%%%%%%%%%%%%%%%%%%%%%%%%%%%%%%%%%%%%%%%%%%%%

 The nonlinearity of Eqs.(\ref{eq:PoissonMUpTHREEhalf}) and (\ref{eq:PoissonMUeTHREEhalf}) 
stands in the way of solving them generally in closed form, yet 
one atmospheric density solution actually can be obtained explicitly.
We show this for the negative atmosphere case.
 
Consider (\ref{eq:PoissonMUeTHREEhalf}) with $\upos(\rho)=0$ for $\rho>\rho_0^+$; it does not matter 
where $\rho_0^+$ is located, all we use is that it is a finite distance.
 We now make the ansatz $\uneg(\rho) = \Aneg \rho^\eta$ and find
$\eta = -6$ and $\Aneg = 12\zeta/\left(1 - {G\mEL^2}/{e^2}\right)$.
 While this is slower than the exponential decay to zero, it still is fast enough to be integrable at $\rho\to\infty$, 
viz. $\rho^2\uneg^\prime(\rho)\to 0$ as $\rho\to\infty$.
 This solution would still have to be matched to the bulk interior, which may or may not be possible!

\vspace{-10pt}

%%%%%%%%%%%%%%%%%%%%%%%%%%%%%%%%%%%%%%%%%%%%%%%%%%%%%%%%%%%%%% 
%%%%%%%%%%%%%%%%%%%%%%%%%%%%%%%%%%%%%%%%%%%%%%%%%%%%%%%%%%%%%%
                \subsection{The local neutrality approximation}\vspace{-5pt}
%%%%%%%%%%%%%%%%%%%%%%%%%%%%%%%%%%%%%%%%%%%%%%%%%%%%%%%%%%%%%%
%%%%%%%%%%%%%%%%%%%%%%%%%%%%%%%%%%%%%%%%%%%%%%%%%%%%%%%%%%%%%%

 So suppose temporarily that $\nupos(r)=\nuneg(r) =:\nu(r)$ for all $r$. 
 Then $\sigma = 0$ by Eq.(\ref{eq:chargedensity}), and Eq.(\ref{eq:PoissonC}) is then solved by $\phi_C^{}=0$.
 Moreover, by Eq.(\ref{eq:massdensity}) we now have $\mu(r) =(\mPR+\mEL)\nu(r)$.
 This is usually approximated
further by neglecting the electron mass versus the proton mass, yet technically this does not yield a simplification.

 A subtler step is the next one. 
 We still have to deal with Eqs.(\ref{eq:forceBALANCEpos}) and (\ref{eq:forceBALANCEneg}), but having set
 $\nupos=\nuneg =:\nu$, we then have two different equations for one unknown, $\nu(r)$, 
and this overdetermines the problem, strictly speaking.
 What this shows is that the strict local neutrality approximation cannot be exactly correct, but of course
it was never assumed to be exactly correct. 
 Therefore, to proceed in the spirit of the approximation, one needs to mold the two equations
 (\ref{eq:forceBALANCEpos}) and (\ref{eq:forceBALANCEneg}) into one. 
 This is done by replacing them by their sum, which in concert with $\phi_C^{}=0$ yields
the mechanical force balance equation
\begin{equation}
-\mu(r) \phi_N^{\prime}(r) %- \sigma(r) \phi_C^{\prime}(r) 
-  p^\prime(r) =0,
 \label{eq:forceBALANCE}
\end{equation}
where the pressure function $p(r) = \ppos(r) +\pneg(r)$ reads
\begin{equation}
p(r) = \hbar^2\left(\frac{1}{m_{\text{p}}}+\frac{1}{m_{\text{e}}}\right)\frac{(3\pi^{2})^{2/3}}{5}\nu^{5/3}(r).
 \label{eq:totalP}
\end{equation}
 This is usually approximated further by neglecting $1/\mPR$ versus $1/\mEL$, yet again technically this 
does not yield a simplification either.

 Since $\mu(r) =(\mPR+\mEL)\nu(r)$, Eq.(\ref{eq:forceBALANCE}) with $p(r)$ given by (\ref{eq:totalP})
can be integrated once to yield $\phi_N^{}$ as a function of $\nu$, which can be inverted to yield 
\begin{equation}
\nu(r) = 
\left(\frac{2}{(3\pi^{2})^{2/3}}\frac{m_{\text{p}}m_{\text{e}}}{\hbar^2}\left[\phi^*_N-\phi_N^{}(r)\right]_+^{}\right)^{3/2};
 \label{eq:nuOFphi}
\end{equation}
here, the notation $[g]_+$ means ``positive part,'' {i.e. $[g]_+(r)= g(r) >0$ for $0<r<R$, where $R$ is the
smallest $r$-value for which $g(r)=0$, and $[g]_+(r)= 0$ for $r\geq R$.
 Furthermore,}
$\phi_N^*$ is a constant of integration determined by $\int \nu(r) d^3r = \Npos$. 
 Inserting this relation into the Poisson equation (\ref{eq:PoissonN}) yields the familiar Lane--Emden
{equation of the polytropic gas ball for $\gamma=5/3$, equivalently of index $n:=1/(\gamma-1)=3/2$,
\begin{equation}
\frac{1}{r^2}\left(r^2\phi_N^{\prime}(r)\right)^\prime
  = C \left[\phi^*_N-\phi_N^{}(r)\right]_+^{3/2},
 \label{eq:LaneEmdenTHREEhalfA}
\end{equation}
\begin{equation}
C  = \frac{2^{7/2}}{3{\pi}}
\frac{G}{\hbar^3}\left(\mPR+\mEL\right)
\left({m_{\text{p}}m_{\text{e}}}\right)^{3/2};
 \label{eq:LaneEmdenTHREEhalfB}
\end{equation}
see \cite{Emden}, \cite{Chandra}, \cite{KippenhahnWeigert}.}
 By shifting and scaling, Eq.(\ref{eq:LaneEmdenTHREEhalfA}) can easily be brought into the dimensionless standardized format
$-\frac{1}{\xi^2}\left(\xi^2\theta^{\prime}(\xi)\right)^\prime = \theta_+^{3/2}(\xi)$, complemented with the 
initial conditions $\theta(0)=1$ and $\theta^\prime(0)=0$; cf. \cite{Emden}, \cite{Chandra}, \cite{SilbarReddy}. 
 The equations for the polytropic gas balls, or gas spheres as they are often called, have been studied
extensively in the astrophysical literature in dependence on their parameter $\gamma$, respectively $n$.
 For $\gamma=\infty$, $\gamma=2$, and $\gamma=6/5$ ($n=0$, $n=1$, and $n=5$)
the polytropic gas ball equation can be solved in terms of elementary functions, in all other cases the
equation itself defines the polytropic density functions. 
 In particular the case $\gamma=5/3$ has been studied thoroughly due to its importance in the theory of
white dwarf structure \cite{Chandra}. 

 For our purposes the case $\gamma=2$, viz. $n=1$, is of particular interest because of
our $5/3\to 6/3$ approximation.
 As a primer we briefly discuss this approximation in the context of the single-density model.
\vspace{-10pt}

%%%%%%%%%%%%%%%%%%%%%%%%%%%%%%%%%%%%%%%%%%%%%%%%%%%%%%%%%%%%%%
                \subsubsection{The $\frac53\to\frac63$ approximation in the single-density model}\vspace{-10pt}
%%%%%%%%%%%%%%%%%%%%%%%%%%%%%%%%%%%%%%%%%%%%%%%%%%%%%%%%%%%%%%

 We again set $r =: (\hbar/\mEL c)\rho$ and $\nu(r) =: (\mEL c/\hbar)^3\upsilon(\rho)$.
 Inserted into the formula for the degeneracy pressure, we find $p(r)\propto \upsilon(\rho)^{5/3}$, and 
since $\upsilon(\rho)$ is dimensionless, we may \emph{now} replace $\upsilon^{5/3}$ by $\upsilon^{6/3} (=\upsilon^2)$. 
  We also set $\phi_N^{}(r) =: c^2 \psi_N^{}(\rho)$ and
 proceed analogously to how we arrived at the polytropic equation with index $n=\frac32$, 
this time it's index $n=1$, except that there is little incentive now to invert the linear relationship between $\psi_N^{}$ and
$\upsilon$, which results from the force balance equation (\ref{eq:forceBALANCE}) wherever $\upsilon(\rho)>0$,
\begin{equation}
-\psi_N^\prime(\rho) =  % - \frac{2(3\pi)^{2/3}}{5}\frac{\mEL}{\mPR}\upsilon^{\prime}(\rho).
\veps K \upsilon^{\prime}(\rho).
 \label{eq:psiVSmu}
\end{equation}
Here we introduced $\veps:= {\mEL}/{\mPR} \approx 1/1836$ and $K := {2(3{\pi^2})^{\frac23}}/{5}$.
 We can even avoid the step of integrating (\ref{eq:psiVSmu}) and 
instead use it directly to eliminate $\psi_N^\prime(\rho)$ (viz. $\phi_N^\prime(r)$)
from Eq.(\ref{eq:PoissonN}) in favor of $\upsilon^\prime(\rho)$ to get
\begin{equation}
- \frac{1}{r^2}\left(r^2\upsilon^{\prime}(\rho)\right)^\prime
  = \kappa^2 \upsilon(\rho),
 \label{eq:LaneEmdenONEa}
\end{equation}
\begin{equation}
\kappa^2  = \frac{10}{3^{2/3}{\pi^{1/3}}
}\frac{G\mPR\left(\mPR+\mEL\right)}{\hbar c}.
 \label{eq:LaneEmdenONEb}
\end{equation}
 Note that the Lane--Emden equation of index $n=1$, Eq.(\ref{eq:LaneEmdenONEa}), is 
valid until $\upsilon(\rho)$ runs into its first zero.

 Several observations are in order. 

 First, we note that ${G\mPR\left(\mPR+\mEL\right)}/{\hbar c}\approx 6\times 10^{-39}$ is a gravitational 
analog of Sommerfeld's fine structure constant ${e^2}/{\hbar c}:= \alphaS \approx {1}/{137.036}$;
it is much much smaller, though.
 This means that to see any appreciable effect in a solution of Eq.(\ref{eq:LaneEmdenONEa}) the variable $\rho$
has to reach very large values. 
 But this is only to be expected, for our unit of length is the reduced Compton length of the electron, and
sure enough the structure of a star varies on scales which are gigantic in terms of these units. 

 Second, the Lane--Emden equation of index $n=1$, Eq.(\ref{eq:LaneEmdenONEa}), is not only linear, it is one of the three
special cases which can be solved in terms of elementary functions.
 It is a special case of a Bessel-type differential equation % (see \cite{Boyce} chpt. 5, sect. 7.),
and the solution relevant to our discussion is given by a spherical Bessel function, explicitly
\begin{equation}
\upsilon(\rho)  = B \frac{\sin\big(\kappa \rho\big)}{\rho},\qquad \rho\in(0,\pi/\kappa);
 \label{eq:LaneEmdenONEsol}
\end{equation}
the bulk amplitude $B$ is determined by $\int \upsilon(\rho)d^3\rho = \Npos^{}$. 

 Third, the radius of the star in this approximate single-density model is 
$R = \frac{\pi}{\kappa} \frac{\hbar}{\mEL c}$.
 Inserting the values for the physical and mathematical constants yields 
\begin{equation}\label{R}
 R \approx %1.5
2.2566\times 10^{19} \frac{\hbar}{\mEL c} \approx 8,714\; \mbox{km}, 
\end{equation}
 i.e. $\approx3/2$ earth radii, compatible with the accepted radius of white dwarf stars with half the mass of the sun.

 Fourth, note that $R$ is \emph{independent} of $\Npos$ (or $\Nneg$ for this matter).
 This of course is not physically reasonable. 
 However, we note that the physical range of acceptable values
for $\Npos$ (hence, $\Nneg$) is very narrow.
 Indeed, to have the interior of a gravitational object accurately modeled as an ideal Fermi gas, the mass needs  
to be sufficiently big, say $\Npos >1.5\cdot 10^{55}$ (13 Jupiter masses), and to be allowed to work with the non-relativistic
approximation, it can't be too big either, say $\Npos < 10^{57}$ (a solar mass). 
 Furthermore, we also assumed that the white dwarf failed to ignite, yet surely our sun did not. 
 This assumption reduces the allowed range of $\Npos$ to $\Npos<9\cdot 10^{55}$. 
 For such a narrow range of $\Npos$ values it is not too unrealistic to have the model predict an $\Npos$-independent
radius, and a central density which increases proportional to $\Npos$.
 
 The 5/3 model breaks the scaling invariance, and then the radii are $\Npos$-dependent, as visible in our
Figs.~\ref{53rhoVsNeOverNp} $\&$ \ref{53reVSrp}. % *** 
\vspace{-25pt}
\begin{widetext}
%%%%%%%%%%%%%%%%%%%%%%%%%%%%%%%%%%%%%%%%%%%%%%%%%%%%%%%%%%%%%%%%%%%%%%%%%%%%%%%%%%%
\section*{Acknowledgment}\vspace{-5pt}
 We thank Elliott H. Lieb for interesting discussions and encouragement.
 We also thank Andrey Yudin for pointing out \cite{KNY} and for helpful comments.
 Thanks are also extended to the referee for helpful suggestions.
%%%%%%%%%%%%%%%%%%%%%%%%%%%%%%%%%%%%%%%%%%%%%%%%%%%%%%%%%%%%%%%%%%%%%%%%%%%%%%%%%

\newpage
\end{widetext}

\section*{References} %\vspace{-20pt}  

\bibliographystyle{apsrev}

$\phantom{nix}$

ph325@math.rutgers.edu 

miki@math.rutgers.edu

\end{document}